%% file: VL_arXiv_v2.tex
\documentclass[a4paper,11pt]{article}
\pdfoutput=1 % if your are submitting a pdflatex (i.e. if you have
\usepackage{jheppub} % for details on the use of the package, please
\makeatletter
\DeclareRobustCommand*{\bfseries}{%
  \not@math@alphabet\bfseries\mathbf
  \fontseries\bfdefault\selectfont
  \boldmath
}
\makeatother

\usepackage{calc}
\usepackage{rotating}
\usepackage[english]{babel}
\usepackage{graphicx}
\usepackage{subfig}
\usepackage{float}
\usepackage{amsmath}
\usepackage{amssymb}
\usepackage{amsthm}
\usepackage{latexsym}
\usepackage{dcolumn}
\usepackage{hyperref}
\input macros_VL.tex
\restylefloat{figure}

\title{Muon $g-2$ and related phenomenology in constrained vector-like extensions of the MSSM}

\author[a,b]{Arghya Choudhury,}
\author[c]{Luc Darm\'e,}
\author[a,c]{Leszek Roszkowski,}
\author[c]{Enrico Maria Sessolo}
\author[c,d]{and Sebastian Trojanowski}

\affiliation{$^a$ Consortium for Fundamental Physics, Department of Physics and Astronomy,\\ University of Sheffield,
Sheffield S3 7RH, United Kingdom\\
$^b$ Consortium for Fundamental Physics, Department of Physics and Astronomy,\\ University of Manchester,
  Manchester, M13 9PL, United Kingdom\\
$^c$ National Centre for Nuclear Research,\\
Ho{\.z}a 69, 00-681 Warsaw, Poland\\
$^d$ Department of Physics and Astronomy, University of California,\\
Irvine, California 92697, USA} 

\emailAdd{a.choudhury@sheffield.ac.uk}
\emailAdd{luc.darme@ncbj.gov.pl}
\emailAdd{leszek.roszkowski@ncbj.gov.pl}
\emailAdd{enrico.sessolo@ncbj.gov.pl}
\emailAdd{sebastian.trojanowski@uci.edu}

\abstract{We analyze two minimal supersymmetric constrained models with low-energy vector-like matter preserving gauge coupling unification. In one we add to the MSSM spectrum a pair $\mathbf{5}+\mathbf{\bar{5}}$ of $SU(5)$, in the other a pair $\mathbf{10}+\mathbf{\overline{10}}$. We show that the muon $g-2$ anomaly can be explained in these models while retaining perturbativity up to the unification scale, satisfying electroweak and flavor precision tests and current LHC data. We examine also some related phenomenological features of the models, including Higgs mass, fine-tuning, dark matter and several LHC signatures. We stress that, at least for the $\mathbf{5}+\mathbf{\bar{5}}$ model, the parameter space consistent with $g-2$ is entirely in reach of the LHC with a moderate increase in luminosity with respect to the current data set.}

\dedicated{UCI-HEP-TR-2016-27}

\begin{document}
\maketitle
%\flushbottom
%%%%%%%%%%%%%%%%%%%%%%%%%%%%%%%%%%%%%%%%%%%%%%%%%%%%%%%%%%%
%%%%%%%%%%%%%%%%%%%%%%%%%%%%%%%%%%%%%%%%%%%%%%%%%%%%%%%%%%%%%%%%%%%%%%%%%%%%%%%%%%%%%
\section{Introduction}

The lack of convincing signals of beyond the Standard Model (BSM)
physics at the LHC has severely constrained many scenarios for new
physics. More precisely, the idea, mainly motivated by the hierarchy
problem, that BSM physics should be found around or just above the
electroweak symmetry breaking (EWSB) scale seems to be now under
strain in many frameworks, including low scale supersymmetry (SUSY).
Direct searches for the sparticles of the Minimal Supersymmetric
Standard Model (MSSM) by
CMS\cite{CMS-PAS-SUS-16-028,CMS-PAS-SUS-16-014,CMS-PAS-SUS-16-015,CMS-PAS-SUS-16-016}
and
ATLAS\cite{ATLAS-CONF-2016-078,ATLAS-CONF-2016-050,ATLAS-CONF-2016-054,ATLAS-CONF-2016-052}
have now pushed the gluino mass bound to 1.7--1.9\tev\ for most
choices of spectrum and decay cascade,\footnote{For the most recent
  interpretation of the ATLAS direct search results in the framework
  of the phenomenological MSSM~(pMSSM) see
  Ref.\cite{Kowalska:2016ent}.} each of the light generation
squarks to $\sim1\tev$ and above, and the lightest stop to
700--800\gev\ and above.

On the other hand, SUSY masses at, or actually above, the 1\tev\ range
show the greatest consistency with the Higgs boson mass at 125\gev,
especially in models defined at the scale of Grand Unification (GUT)
and motivated by supergravity, like the Constrained MSSM (CMSSM) and
the Non-Universal Higgs Mass (NUHM) model. In these models the favored
parameter space shows sparticles in the range of a few~TeV (see,
e.g.,\cite{Roszkowski:2014wqa}), somewhat decoupled from the EWSB
scale, so that all precision observables are expected to yield values
in agreement with the Standard Model (SM) within the present
experimental sensitivity.  As a bonus, one obtains a naturally
embedded dark matter (DM) candidate, the lightest neutralino, which
can easily satisfy the relic density constraint and yields signatures
in reach of present and future direct and indirect DM searches.

By and large precision observables and rare meson decays have been
measured in recent years to be in good or even excellent agreement
with the SM.  However, there exist some long-standing anomalies that
point to the existence of BSM physics close to the EWSB scale.  The
most outstanding and thoroughly studied among them is arguably the
anomalous magnetic moment of the muon, \gmtwo, which shows a deviation
from the SM value at more than
$3\sigma$\cite{Bennett:2006fi,Miller:2007kk}.  The anomaly will soon
be either confirmed or falsified by the New Muon g-2 experiment at
Fermilab\cite{Grange:2015fou,Chapelain:2017syu}, which is projected to
reach a sensitivity of $7\sigma$ to possible BSM effects.

In SUSY, deviations from the SM value of \gmtwo\ are mainly due to the
contributions of smuon-neutralino and sneutrino-chargino loops, and
require these states to be relatively light. Direct
searches for these particles, at LEP first and now at the LHC, constrain
them above the few hundred GeV range, but even when recent direct LHC
bounds are taken into account, the \gmtwo\ anomaly can be easily
explained in the framework of the
MSSM\cite{Endo:2013bba,Fowlie:2013oua,Chakraborti:2014gea,Das:2014kwa,Lindner:2016bgg}.
It is much harder, however, to accommodate the discrepancy in
GUT-constrained models.  In particular, the bounds from direct squark
and gluino searches at the LHC already exclude\cite{Fowlie:2012im} the
parameter space that would lead to the correct value of \gmtwo\ in the
CMSSM and the NUHM. The simplest, although at the same time the
least motivated, way out in such models would be to disunify slepton
and squark masses. 
A more motivated solution is to relax the assumption of
a universal gaugino mass, as was shown in,
e.g.,\cite{Mohanty:2013soa,Akula:2013ioa,Chakrabortty:2013voa,Kowalska:2015zja,Chakrabortty:2015ika,
Belyaev:2016oxy,Okada:2016wlm,Fukuyama:2016mqb}.

As an alternative, one can resolve the \gmtwo\ discrepancy by
extending the particle content of the MSSM with vector-like (VL)
matter, as investigated, e.g.,
in\cite{Endo:2011xq,Endo:2011mc,Dermisek:2013gta,Dermisek:2014cia,Gogoladze:2015jua,Aboubrahim:2016xuz,Nishida:2016lyk,Higaki:2016yeh,Megias:2017dzd}.
The introduction of VL superfields in the
superpotential brings along extra degrees of freedom without spoiling
the successful unification of gauge interactions at the GUT
scale\cite{Martin:2009bg}.  Extra VL matter, moreover, has been
recently considered in the context of several long-standing
theoretical issues related to BSM models, and has been shown to be
able to provide the effective couplings needed to reconcile some of
the other few discrepancies from the SM that have been recently
reported by experimental collaborations.

Besides \gmtwo, it has been for instance pointed out that VL colored sparticles provide extra
contributions to the Higgs boson mass\cite{Graham:2009gy,Martin:2009bg,Faroughy:2014oka,Lalak:2015xea,Nickel:2015dna,Barbieri:2016cnt}, that VL quarks could possibly explain\cite{Angelescu:2015kga} the recently emerged $ttH$ 
anomaly\cite{ATLAS-CONF-2015-044}, and that VL superfields might ameliorate to some extent the fine tuning associated with large stops 
with respect to the MSSM\cite{Dermisek:2016tzw}. On the observational side, it has been shown that signatures of extra VL matter 
can be tested in the next generation of experiments probing lepton flavor violating decays\cite{Ibrahim:2015hva},
electric and chromoelectric dipole moments\cite{Aboubrahim:2015nza,Aboubrahim:2015zpa}, 
flavor violating Higgs decays\cite{Fathy:2016vli} and rare meson decays.

In this paper we perform a detailed investigation of \gmtwo, taking
into account the dark
matter, Higgs mass, and other constraints in two of the simplest VL
extensions of the CMSSM. These are constructed by introducing at the
GUT scale either a pair of multiplets in the
$\mathbf{5}+\mathbf{\bar{5}}$ representation of $SU(5)$, or a pair
$\mathbf{10}+\mathbf{\overline{10}}$ of $SU(5)$.  We show that within
these frameworks one can manage to maintain a reasonable level of
simplicity and be able to explain the \gmtwo\ anomaly. At same time
one can retain a good DM candidate without violating any of the
constraints from the LHC direct SUSY searches, Higgs measurements,
flavor sector, perturbativity in the renormalization group evolution
(RGE), and overall consistency with the GUT picture.  We provide
projections for possible direct signals in the next run of the LHC and
we present some comments on issues related to fine tuning, flavor
observables, and the $ttH$ anomaly.

The paper is organized as follows. We first present in \refsec{sec:models} the models along with their boundary conditions at the GUT scale. We then focus in \refsec{sec:pheno} on the low-energy phenomenology of our models and the corresponding bounds on the parameter space. Section~\ref{gm2analytic} presents a detailed description of the mechanisms increasing the value of 
\gmtwo\ in SUSY models with VL matter and provides analytical formulas for the effect. 
Finally we show in \refsec{numerics} our numerical results, and conclude in \refsec{sec:summary}. 
The appendices contain more information on the soft SUSY-breaking Lagrangian, the most relevant mass matrices, some useful calculations, and a detailed analysis of the collider constraints.

%%%%%%%%%%%%%%%%%%%%%%%%%%%%%%%%%%%%%%%%%%%% 
\section{The models\label{sec:models}}
%%%%%%%%%%%%%%%%%%%%%%%%%%%%%%%%%%%%%%%%%%%%%

We consider in this work models with new VL fields that are consistent
with perturbative gauge coupling unification.  The unified models
presented here are inspired both by ideas of GUT, based
on the $SU(5)$ gauge group, and by expectations of minimality. We
therefore do not include additional singlets and focus on simply
adding a pair $\mathbf{5}+\mathbf{\bar{5}}$ or a pair
$\mathbf{10}+\mathbf{\overline{10}}$ to the MSSM, the VL pair nature
of the new fields allowing as usual to give them a superpotential
mass. Similarly, we will not suppose any additional discrete symmetry
preventing direct mixing between the new fields and the MSSM
ones. Finally, let us recall that all of our new fields are charged
under lepton number.

In what follows we systematically use small letters for MSSM fields and indicate the $SU(3)\times SU(2)\times U(1)$ 
quantum numbers in parentheses. With this choice of notation, the MSSM fields are
\begin{align}
q&=(\mathbf{3},\mathbf{2},1/6)&  H_u&=(\mathbf{1},\mathbf{2},1/2) &  l&=(\mathbf{1},\mathbf{2},-1/2) \nonumber \\
u&=(\mathbf{\bar{3}},\mathbf{1},-2/3)&  H_d&=(\mathbf{1},\mathbf{2},-1/2) & e&=(\mathbf{1},\mathbf{1},1)  \\ 
d&=(\mathbf{\bar{3}},\mathbf{1},1/3) \,. &  & & & & & \nonumber
\end{align}
The MSSM part of the superpotential is
\be
W=\mu\,H_u H_d-Y_d\,q H_d d-Y_e\,l H_d e+Y_u\,q H_u u\,,
\ee
where $\mu$ is the Higgs/higgsino mass parameter, the $Y$ Yukawa couplings are to be understood as $3\times 3$ matrices in flavor space, 
and we have suppressed generation and isospin indices from the notation.

%%%%%%%%%%%%%%%%%
\subsection{The 5-plet LD model}
%%%%%%%%%%%%%%%%%%

For the first model we consider, which, following the convention of\cite{Martin:2009bg}, 
we refer to as \textbf{LD}, we add to the MSSM spectrum a VL pair $\mathbf{5}+\mathbf{\bar{5}}$  of $SU(5)$, corresponding to the following new fields:
\begin{align*}
D&=(\mathbf{\bar{3}},\mathbf{1},1/3)& D'&=(\mathbf{3},\mathbf{1},-1/3)\nonumber\\
L&=(\mathbf{1},\mathbf{2},-1/2)& L'&=(\mathbf{1},\mathbf{2},1/2)\,.
\end{align*}
Hence, with respect to the MSSM, there is one extra quark with charge $-1/3$ (and its antiparticle), one extra charged lepton (and its antiparticle), and 2 extra massive neutrinos. 
Correspondingly, there are two more squarks, two more sleptons, and two more sneutrinos.

Additional trilinear and bilinear terms are now allowed in the superpotential,
\be
W \supset -\lam_D\,q H_d D-\lam_L\,L H_d e+M_D D D'+M_L L L'+\widetilde{M}_L\,lL'+\widetilde{M}_D\,dD'\,,\label{superpot5}
\ee
where the new Yukawa couplings $\lam_L$ and $\lam_D$ and masses $\widetilde{M}_L$ and $\widetilde{M}_D$ 
responsible for the mixing with the SM fields are intended as 3-dimensional arrays spanning the SM generations. 

For the fields characterized by the same quantum numbers ($d,D$ and
$l,L$) it is possible to choose a basis such that the mixing mass
terms are rotated away. This amounts to a redefinition of the other
free parameters in the superpotential.  However, if this choice is
made at the GUT scale, the RGE will in fact regenerate these mixing
terms at the SUSY scale.\footnote{The respective 1-loop beta
  functions, $\beta_{\widetilde{M}_L}$ for $\widetilde{M}_L$ and
  $\beta_{\widetilde{M}_D}$ for $\widetilde{M}_D$, contain $M_L
  Y_e^\dagger \lam^*_L$ and $2 M_D Y_d \lam^*_D$, which ensure that
  even fixing $\widetilde{M}_L=\widetilde{M}_D=0$ at the GUT scale
  will nonetheless lead to their non-zero values at the SUSY scale.\label{foot}}  Not
including them would therefore amount to tuning the GUT-scale
parameters to ensure their subsequent vanishing at the SUSY scale.
Since such tuning is not well-motivated and furthermore would break the
universality assumption in our boundary conditions, we have chosen to
maintain in \refeq{superpot5} the most general form, which includes
explicit mass mixing.

The soft SUSY-breaking Lagrangian features additional terms with respect to the MSSM (see Appendix~\ref{app:soft} for the full expression),
\bea
-\mathcal{L}_{\textrm{soft}}&\supset&\left[m_L^2|\tilde{L}|^2+m_{L'}^2|\tilde{L}'|^2
+m_D^2|\tilde{D}|^2+m_{D'}^2|\tilde{D}'|^2+\left(\widetilde{m}_L^2\,\tilde{l}^{\dag}\tilde{L}
+\widetilde{m}_D^2\,\tilde{d}^{\dag}\tilde{D}+\textrm{h.c.}\right)\right]\nonumber\\
 & & +\left(B_{M_L}\tilde{L}\tilde{L}'+B_{\widetilde{M}_L}\tilde{l}\tilde{L}'
+B_{M_D}\tilde{D}\tilde{D}'+B_{\widetilde{M}_D}\tilde{d}\tilde{D}'+\textrm{h.c.}\right) \nonumber \\ 
& &-\left(T_D\,\tilde{q}H_d\tilde{D}^{\dag}
+T_L\,\tilde{L}H_d\tilde{e}^{\dag}+\textrm{h.c.}\right),\label{soft5}
\eea
where the terms mixing VL and MSSM matter, $\widetilde{m}_L^2$, $\widetilde{m}_D^2$, $T_L$, $T_D$, $B_{\widetilde{M}_L}$, and $B_{\widetilde{M}_D}$ are, again, to be understood as 3-dimensional arrays.

Since the mixing between the new VL fields and the MSSM ones will be a crucial part of the phenomenology of our model, 
the explicit form of the fermion and lepton mass matrices will often be very useful. We have therefore included them in Appendix~\ref{app:soft}. 

%%%%%%%%%%%%%%%%%%%%%%%%%%
\subsection{The 10-plet QUE model}\label{sec:10model}
%%%%%%%%%%%%%%%%%%%%%%%%%%

The second model that we consider in this work is obtained by the addition to the MSSM spectrum of a VL pair 
of fields in a $\mathbf{10}+\mathbf{\overline{10}}$ representation of $SU(5)$. We call it the \mtenm\ model. The quantum numbers of the new fields are:
\begin{align}
Q&=(\mathbf{3},\mathbf{2},1/6)& Q'&=(\mathbf{\bar{3}},\mathbf{2},-1/6)\nonumber\\
U&=(\mathbf{\bar{3}},\mathbf{1},-2/3)& U'&=(\mathbf{3},\mathbf{1},2/3)\nonumber\\
E&=(\mathbf{1},\mathbf{1},1)& E'&=(\mathbf{1},\mathbf{1},-1)\,.
\end{align}
With respect to the MSSM, the $\mathbf{10}+\mathbf{\overline{10}}$ 
model therefore features two extra quarks with charge $2/3$ (and their antiparticles), one extra 
quark with charge $-1/3$ and its antiparticle, and one extra lepton with its antiparticle. Correspondingly, there are 
four extra up-type squarks, two extra down-type squarks and two extra sleptons in the spectrum, with their respective antiparticles.  

The additional terms in the superpotential are given by \bea
W&\supset& \lambda_{Qu}\,Q H_u u-\lambda_{Qd}\,Q H_d d+\lambda_U\,q H_u U-\lambda_E\,l H_d E+Y_{10}\,Q H_u U-Y_{10}'\,Q' H_d U'\nonumber\\
& &+M_Q Q Q'+M_U U U'+M_E E E'+\widetilde{M}_Q\,q
Q'+\widetilde{M}_U\,u U'+\widetilde{M}_E\,e E',\label{superpot10} \eea
where, again, all mixing trilinear and mass terms are understood as
3-dimensional arrays spanning the SM generations.

The additional soft terms and the mass matrices can be obtained in a fashion similar to \refeq{soft5} and following. 
We leave their explicit form for Appendix~\ref{app:soft}. 

%%%%%%%%%%%%%%%%%%%%%%
\subsection{Boundary conditions}
%%%%%%%%%%%%%%%%%%%%%%

Besides the usual parameters of the CMSSM, \mzero, \mhalf, \azero, \tanb, and \signmu, in the \mfivem\ model there will be 
quantities parameterizing the additional terms given in Eqs.~(\ref{superpot5}) and (\ref{soft5}).
Since with a greater number of parameters it becomes more likely to miss possibly fine-tuned regions in a numerical scan,
we will try to strike a balance between thoroughness and economy, driven also by the expectation that the parameters 
sharing a common origin at the GUT scale should be unified.

Thus, we introduce a common VL superpotential mass value at the GUT scale, $M_V\equiv M_L=M_D$. 
We extend the definition of $m_0^2$ to include the GUT-scale value of $m_L^2$,  $m_{L'}^2$, $m_D^2$, $m_{D'}^2$ in \refeq{soft5}, 
and we use \azero\ to define $T_{L,D}=\lam_{L,D}\azero$ at the GUT scale. 
We also introduce a parameter $B_0$ such that, for example, $B_{M_L}=B_0 M_L$ at the GUT scale and similarly for all other B terms
in \refeq{soft5}.   

On the other hand, the flavor structure of the extra terms in the superpotential and soft SUSY-breaking Lagrangian 
is largely unknown and subject to model-building assumptions. 
As the scope of this analysis is phenomenological, we refrain from making any specific assumption on the flavor UV completion, but rather focus on reasonably wide regions of the parameter space in agreement with flavor constraints. 

To maximize the impact of our choice of parameters on the \gmtwo\ observable, which involves the second generation leptons, 
and at the same time minimize flavor-changing effects involving the first and third generation we assume that the GUT/Planck scale UV completion defines the following boundary conditions for the extra Yukawa couplings:
\begin{align}\label{5_lam_inv}
\lam_L=\lam_D =\begin{pmatrix}
                0 \\ \lambda_{5} \\ \epsilon\lambda_{5}
               \end{pmatrix},
\end{align}
where $\lam_5$ is a unified Yukawa coupling, $\epsilon$ is a
parameter smaller than 1, and the first generation Yukawa mixing is set to zero for practicity, but is to be rather intended 
as a parameter small enough to satisfy all bounds from first to second generation conversion in the lepton and quark sectors.  

The explicit mass mixing terms in Eqs.~(\ref{superpot5}) and (\ref{superpot10}) can be rotated away at the GUT scale, but are subsequently generated radiatively (see Footnote~\ref{foot}). This implies that they should feature roughly the same flavor structure as in 
\refeq{5_lam_inv}, with a highly suppressed first genetration mixing, and their size be in the few-GeV range. We will therefore choose $\widetilde{M}_L= \widetilde{M}_D=(0,\widetilde{M},\alpha\widetilde{M})$, expressed in terms of a unified GUT-scale value $\widetilde{M}$ and a parameter $\alpha$ 
smaller than 1. As we shall see below, $\widetilde{M}$ is also constrained to the few-GeV range by phenomenological bounds, 
so that it does not play a significant role in obtaining \gmtwo\ and other relevant signatures.

Conversely, the texture of the soft mass matrices in \refeq{soft5} does play an important role for the phenomenology. 
Since these terms are subject to largely the same flavor constraints as the 
Yukawa couplings that mix VL matter and the MSSM fields we assume that, while their diagonal part 
is set universally by \mzero, as is usually the case in GUT-constrained SUSY models, the off-diagonal mixing terms 
$\widetilde{m}^2_{L,D}$ follow a structure similar to \refeq{5_lam_inv}. We also parametrize them as
\begin{align}\label{5_mass_inv}
\widetilde{m}_L^2= \widetilde{m}_D^2=\begin{pmatrix}
                0 \\ \widetilde{m}^2 \\ \alpha\widetilde{m}^2 
               \end{pmatrix},
\end{align}
in terms of a unified GUT-scale value $\widetilde{m}^2$ and the parameter $\alpha$ smaller than 1.
We point out as a sidenote that a different treatment of the diagonal and off-diagonal elements of the soft matrices is not unreasonable, but rather 
typical of flavor models inspired by the Froggatt-Nielsen mechanism\cite{Froggatt:1978nt}, where the elements of the soft mass
matrices are generated proportionally to the \textit{difference} of the charges assigned to different generations (see, e.g.,\cite{Nelson:1997bt}).   

We finally adopt similar boundary conditions for the \mtenm\ model. 
We introduce the parameters $\lam_{10}$, $\epsilon$, $\widetilde{M}$, $\widetilde{m}^2$, and $\alpha$ as before, so that
\begin{align}\label{10_lam_hier}
\lam_{Qu(d)}=\lam_U=\lam_E =\begin{pmatrix}
                0 \\ \lambda_{10} \\ \epsilon\lambda_{10}
               \end{pmatrix},
\end{align}
and equations similar to \refeq{5_mass_inv} apply for the mass mixing terms. 
Additionally, we treat the purely VL Yukawa couplings as unified at the GUT scale, $Y_{10}=Y'_{10}$, and we scan this
independently.

Even if the boundary conditions outlined above favor mixing between the
second generation and the VL particles, in the following sections we will
comment on phenomena that involve predominantly third-generation
effects, like electroweak (EW) fine tuning, or fits to the $ttH$
anomaly, and that might favor models characterized by boundary
conditions different from Eqs.~(\ref{5_lam_inv}) and
(\ref{10_lam_hier}), for example, Yukawa couplings of the form
$\lam_U=(0,\epsilon\lam_{10},\lam_{10})$.

%%%%%%%%%%%%%%%%%%%%%%%%%%%%%%%%%%%%%%%%%%%%%%%%%%%%%%%%
\section{Low energy phenomenology\label{sec:pheno}}
%%%%%%%%%%%%%%%%%%%%%%%%%%%%%%%%%%%%%%%%%%%%%%%%%%%%%%%%

\subsection{Gaugino and scalar mass spectra}

The phenomenological properties of GUT-unified SUSY models enriched
with VL matter has been investigated, e.g., in\cite{Martin:2009bg}.
We briefly recall here some of the main characteristics, focusing in
particular on the differences with the CMSSM. 
For our numerical results we use the SPheno\cite{porod_spheno_2003,porod_spheno_2012} code generated by 
\SARAH~(see Refs.\cite{staub_sarah_2008,staub_automatic_2011,staub_superpotential_2010,staub_sarah_2013,staub_sarah_2014}). 
The Higgs pole mass is obtained at 2 loops in the effective potential approach and all other masses are given at one loop. 
The unified values of gaugino and scalar soft masses are given at
$\mgut\approx 10^{16}\gev$ and run down to \msusy, defined as usual as 
the geometrical mean of the physical stop masses. Typical
low-energy spectra are determined by the form of the RGE, which is
modified by the presence of extra matter fields.  

Particularly consequential are modifications to the running of the gluino mass,
which result in a reshaping of the GUT-scale parameter space with respect
to the CMSSM.
Note that, as the dimension of the VL representation increases, the
1-loop beta function becomes less negative. At 1 loop the gluino beta
function, which in the CMSSM is $\beta_{M_3}^{\textrm{MSSM}}=
-6g_3^2 M_3$, is modified in the \mfivem\ model to
$\beta_{M_3}^{\textrm{LD}}= -4g_3^2 M_3$, which results in a
gentler slope when running to low energies. In the \mtenm\ model,
where one obtains $\beta_{M_3}^{\textrm{QUE}}= 0$ at one loop, 2-loop effects make the
gluino mass actually run to smaller values at low energies.

%%%%%%%%%%%%%%%%%%%%%%%%%%%%%%%%
\begin{figure}[t]
\centering
\includegraphics[width=0.60\textwidth]{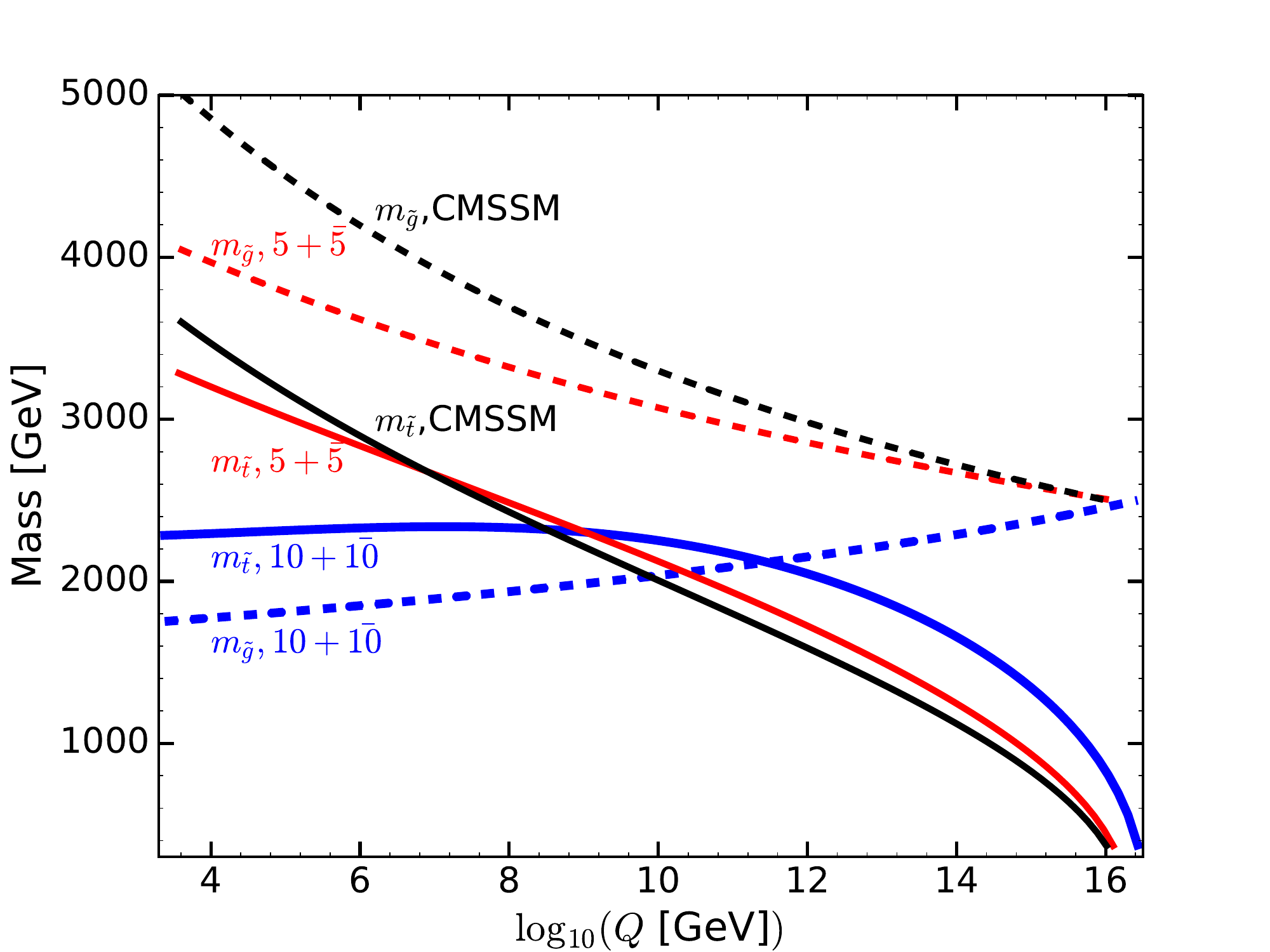}
\caption{The dashed lines show the running of the gluino soft mass,
  $ M_3\approx\mglu$, in the CMSSM (black), the \mfivem\ model
  (red), and the \mtenm\ model (blue).  The running of the
  right-chiral stop soft mass, $m_{\tilde{t}}$, is plotted with solid
  lines in the same color code. We have considered $\mhalf = 2.5 \tev, \mzero=600 \gev$ and $\azero=2.5 \tev$ in all cases and set the VL Yukawa couplings to zero. Note that the GUT scale in the \mfivem\ and \mtenm\ models is slightly lower than in the CMSSM.}
\label{fig:run}
\end{figure}
%%%%%%%%%%%%%%%%%%%%%%%%%%%%%%%%

We illustrate this behavior in \reffig{fig:run}, where the 2-loop running of the gluino mass in the CMSSM 
is indicated with a black dashed line, and is compared to the running of the gluino mass in the \mfivem\ 
model (red dashed line), and in the \mtenm\ model (blue dashed). 
  
When \mzero\ is not too large, which is the case of interest for explaining the anomalous magnetic moment of the muon, 
the running of the gluino mass provides the leading term to the low-scale renormalization of sparticle masses for the color sector and renders their
SUSY-scale value not much dependent on the initial choice of \mzero. 
The solid lines in \reffig{fig:run} follow the running of the lightest soft stop mass, $m_{\tilde{t}}$\,, in the 
CMSSM (black), in the \mfivem\ model (red), and in the \mtenm\ model (black). 
Note that, while in the CMSSM and in the \mfivem\ model the stops end up being lighter than the gluino at the low scale, 
in the \mtenm\ model they become heavier than the gluino, independently of \mzero.  As a consequence, given the LHC bounds on the gluino mass, $m_{\tilde{g}}\gsim 1.8\tev$, 
the stops must always be heavier than $\sim 2 \tev$ in the \mtenm\ model.

%\es{Modified:}  As is generally the case in GUT-scale SUSY models, the gluino mass running provides 
% the leading term to the low-scale renormalization of sparticle masses for the color sector, making their
% SUSY-scale value not much dependent on the initial choice of \mzero. 
% The solid lines in \reffig{fig:run} follow the running of the lightest stop soft mass, $m_{\tilde{t}}$\,, in the 
% CMSSM (black), in the \mfivem\ model (red), and in the \mtenm\ model (black). 
% Note that, while the stops end up lighter than the gluinos at the low scale 
% in the CMSSM and in the \mfivem\ model, they become heavier in the \mtenm\ model, and this is always true,
% independently of \mzero.  As a consequence, given the LHC bounds on gluinos, $m_{\tilde{g}}\gsim 1.8\tev$, 
% the stops must always be heavier than $\sim 2 \tev$ in the \mtenm\ model.

\begin{figure}[t]
\centering
\subfloat[]{%
%\label{fig:c}%
\includegraphics[width=0.33\textwidth]{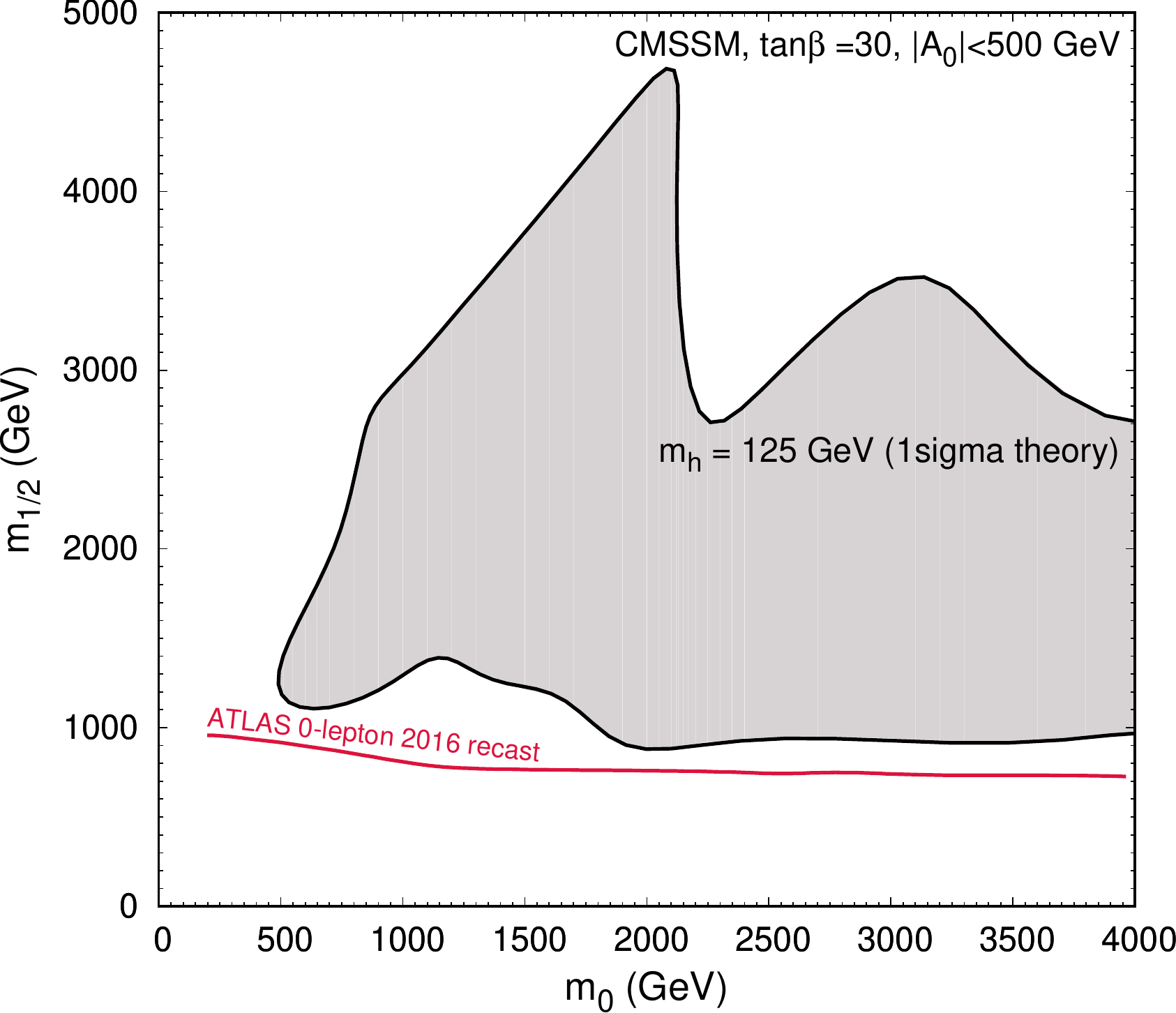}
}%
%\hspace{0.02\textwidth}
\subfloat[]{%
%\label{fig:a}%
\includegraphics[width=0.33\textwidth]{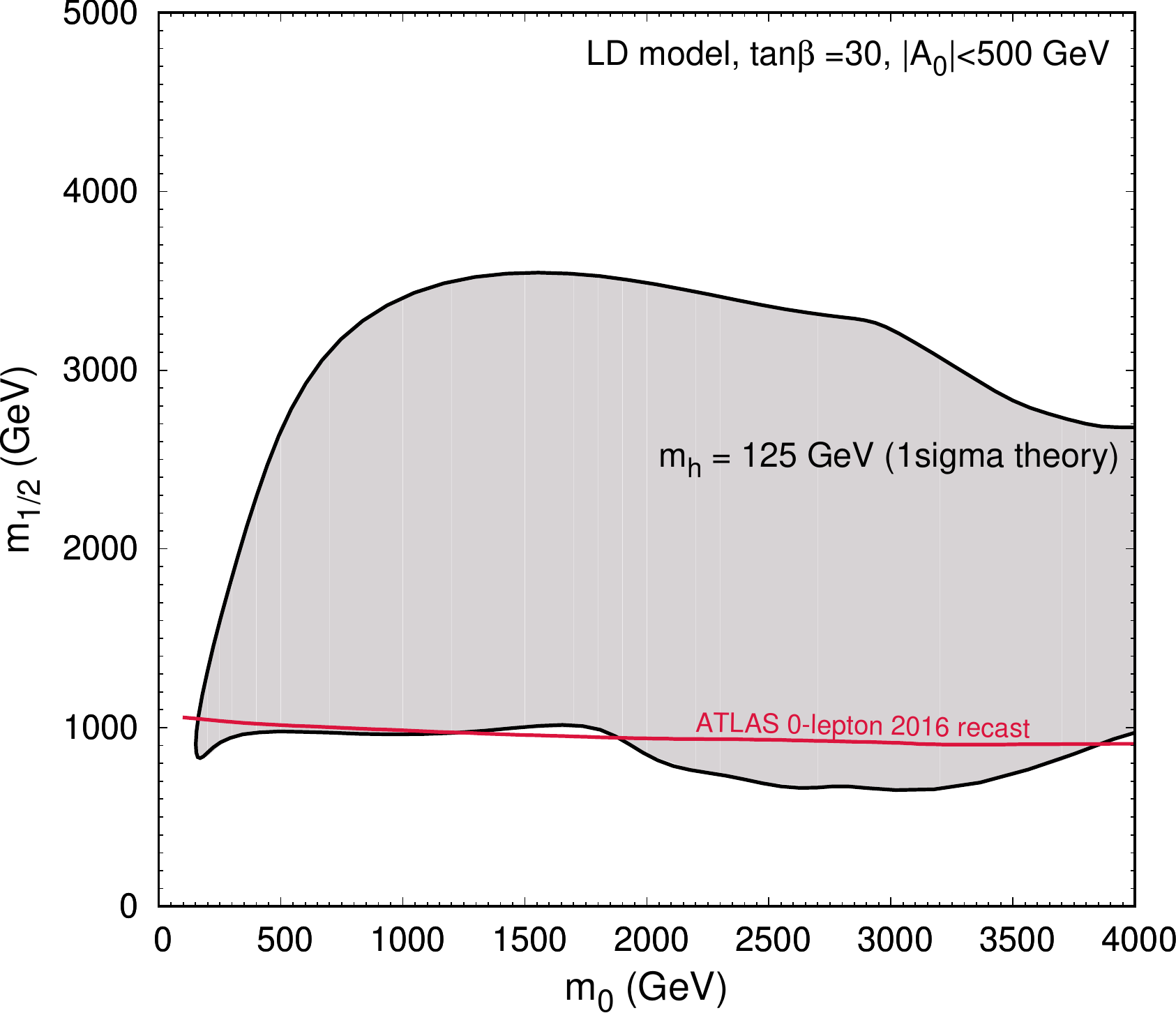}
}%
%\hspace{0.02\textwidth}
\subfloat[]{%
%\label{fig:b}%
\includegraphics[width=0.33\textwidth]{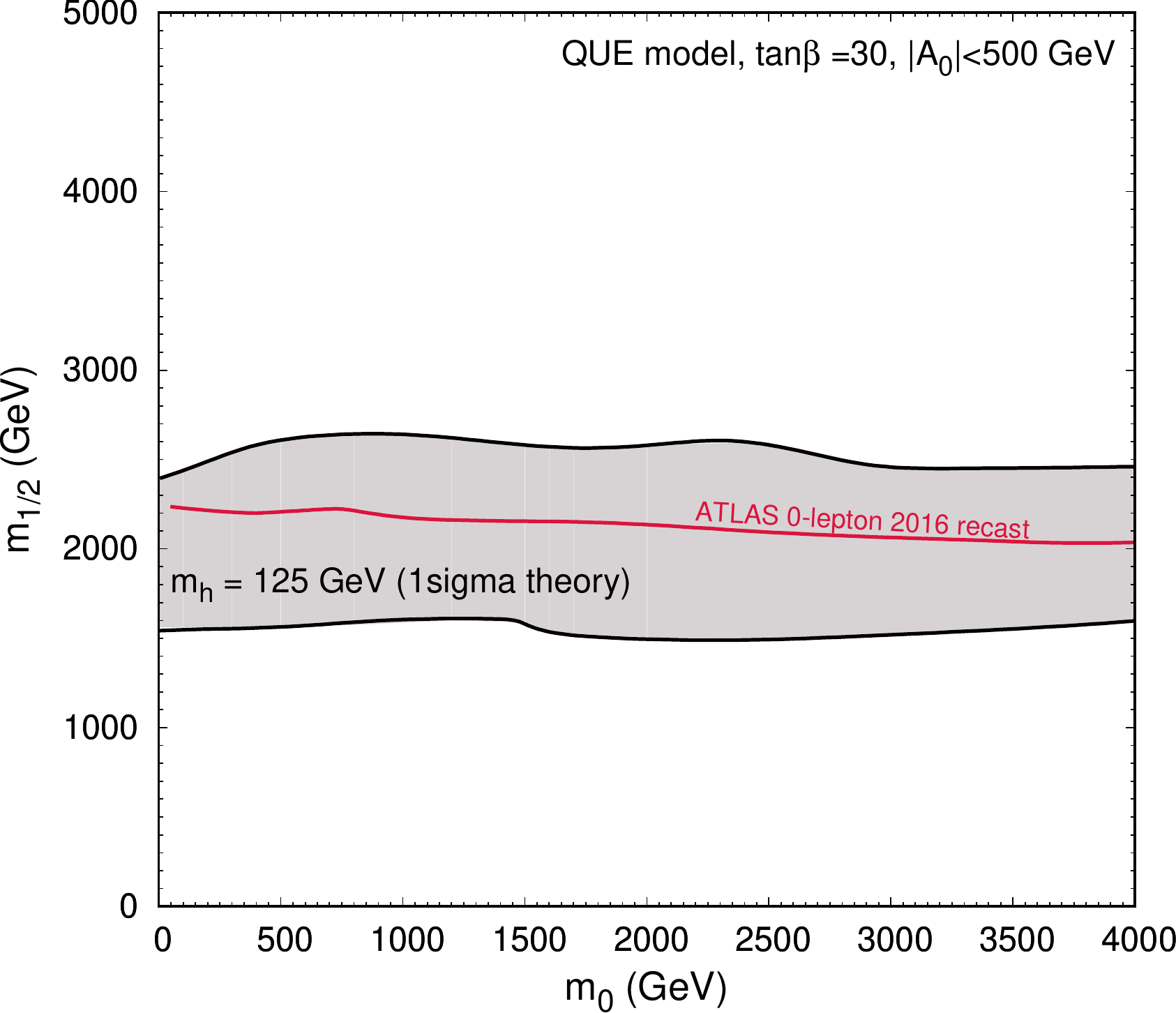}
}%
\caption{The red solid line marks the ATLAS 0-lepton search lower
  bound\cite{ATLAS-CONF-2016-078}.  The region of correct Higgs mass
  in the (\mzero, \mhalf) plane for $\tanb\approx 30$ and
  $|\azero|<500\gev$ is shown in pencil gray shaded in: (a) CMSSM, (b)
  the \mfivem\ model and (c) the \mtenm\ model.}
\label{fig:parspace}
\end{figure}

More generally, the modifications to the running of gaugino and scalar masses have the effect of shifting the GUT-scale 
parameter space of models with VL matter to larger values for a given LHC mass bound at the low scale. 
To give an example, in \reffig{fig:parspace} we show as a solid red line the current ATLAS 0-lepton direct bound on the mass of squarks and 
gluino\cite{ATLAS-CONF-2016-078} in the CMSSM and the two VL models analyzed in this work. 
The bound has been recast using the code of\cite{Kowalska:2016ent} and is shown in the (\mzero, \mhalf) plane of the 
CMSSM in \reffig{fig:parspace}(a) (solid red line). From Figs.~\ref{fig:parspace}(b) and \ref{fig:parspace}(c) one can infer that 
the direct bound bites increasingly into larger GUT-scale parameters in the models with additional matter. 

%%%%%%%%%%%%%%%%%%%%%%%%%%%%%%%%
%%%%%%%%%%%%%%%%%%%%%%%%%%%%%%%%

%%%%%%%%%%%%%%%%%%%%%%%%%%%%%%%%%%
\subsection{Higgs sector\label{sec:higgs}}
%%%%%%%%%%%%%%%%%%%%%%%%%%%%%%%%%%

As mentioned above, and in some contrast to most previous studies considering the Higgs mass in SUSY theories with new VL matter, 
in this work we include the 2-loop corrections arising from the new particles (notable exceptions 
are\cite{Nickel:2015dna,Staub:2016dxq}).

Additional VL fields can modify the Higgs pole mass in two ways. 
First, by adding extra loop corrections. This was thoroughly investigated in Ref.\cite{Martin:2009bg}, and it was shown
there that, in case of a large hierarchy between the fermionic and scalar components of the VL fields, 
corrections up to 15\gev\ could be obtained. 
Second, \mhl\ can be modified by altering the MSSM Yukawa couplings (and in particular the top Yukawa) 
through the mixing of the new VL fields with the MSSM ones.

As already pointed out in\cite{Martin:2009bg}, the \mfivem\ model offers small mass improvements compared to the 
MSSM. Essentially, the largest effect originates from the RGE modifications, which lead to trilinear couplings being 
typically more negative than in the CMSSM and thereby increase the Higgs mass for 
equivalent boundary conditions at the GUT scale.

In \reffig{fig:parspace}(b) one can see the region of correct Higgs mass within an assumed 
$1\sigma\approx 3\gev$ theoretical error in the (\mzero, \mhalf) plane of the \mfivem\ model.
It should be compared with the equivalent plane in the CMSSM, \reffig{fig:parspace}(a). We assume in the plots that
$\tanb\approx 30$ and $|\azero|<500\gev$. Note that the region of correct Higgs mass in the \mfivem\ model 
is characterized by a slightly smaller size of the GUT-scale soft masses than in the CMSSM. The difference is not, however, 
very dramatic.

\begin{figure}[t]
\begin{center}
\includegraphics[width=0.60\textwidth]{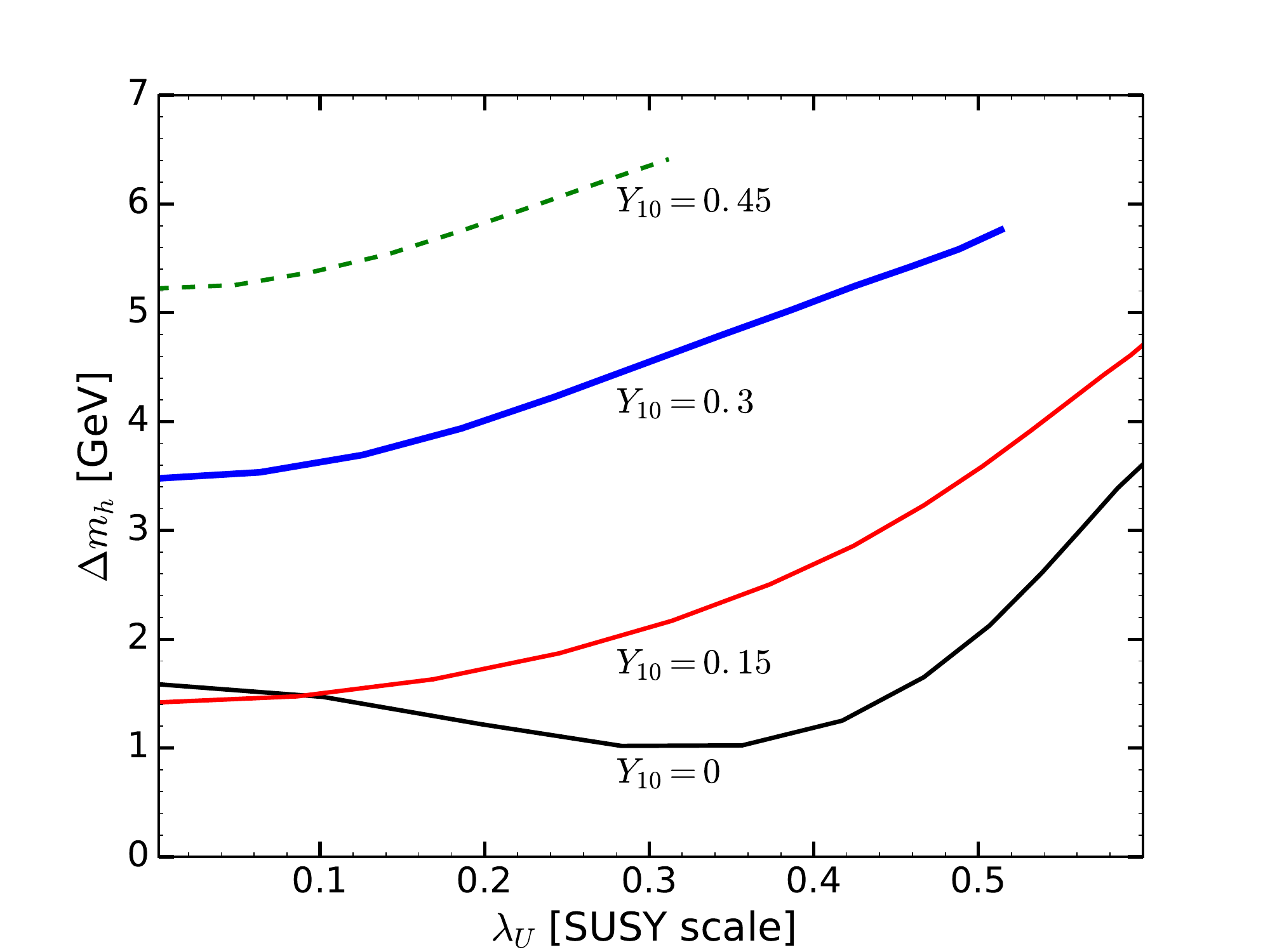}
\caption{Higgs mass difference in the \mtenm\ model with respect to
  the CMSSM as a function of $\lambda_{U}$ (SUSY scale) for various
  values of $Y_{10}$ at the GUT scale. From bottom to top
  $Y_{10}=0,\,0.15,\,0.3,\,0.45$. These values correspond roughly to
  0, 0.5, 0.75, and 1 at the SUSY scale. The values of \mzero, \mhalf,
  \azero, and \tanb\ are fixed as in the \mtenm\ benchmark point, see \refsec{sec:bench}.}
\label{fig:mhl10Y10}
\end{center}
\end{figure}

The \mtenm\ model differs more from the CMSSM. This is not only because 
of more substantial modifications to the RGE, but also because of the loop corrections involving the extra Yukawa couplings in \refeq{superpot10}, which
have the effect of giving a significant increase to the Higgs mass.
We give an example of this in \reffig{fig:mhl10Y10}, where we show the relative increase (with respect to the CMSSM) of the Higgs mass, $\Delta\mhl$, in the \mtenm\ model, as a function of the SUSY-scale value of $\lam_U$ for different choices of $Y_{10}$ at the GUT scale.
Note that the curves are obtained for fixed values of \mzero, \mhalf, \azero, and \tanb, set as in the \mtenm\ benchmark point presented in \refsec{sec:bench}. Note that, in the limit of zero VL Yukawa couplings, there remains a residual Higgs mass difference with the CMSSM which is due to different RGE in both models.

%%%%%%%%%%%%%%%%%%%%%%%%%%%%%%%%
%%%%%%%%%%%%%%%%%%%%%%%%%%%%%%%%%%

As a consequence, in the \mtenm\ model one obtains generally a good
Higgs mass value in the parameter space that is already being tested
at the LHC, as \reffig{fig:parspace}(c) shows.  Additionally, as the
Higgs mass can easily become too heavy, imposing $\mhl\approx125\gev$
often produces an upper bound on the extra Yukawa couplings given a
specific choice of \msusy\ and \tanb.

%%%%%%%%%%%%%%%%%%%%%%%%%%%%%%%%%%
\begin{figure}[t]
\begin{center}
\includegraphics[width=0.6\textwidth]{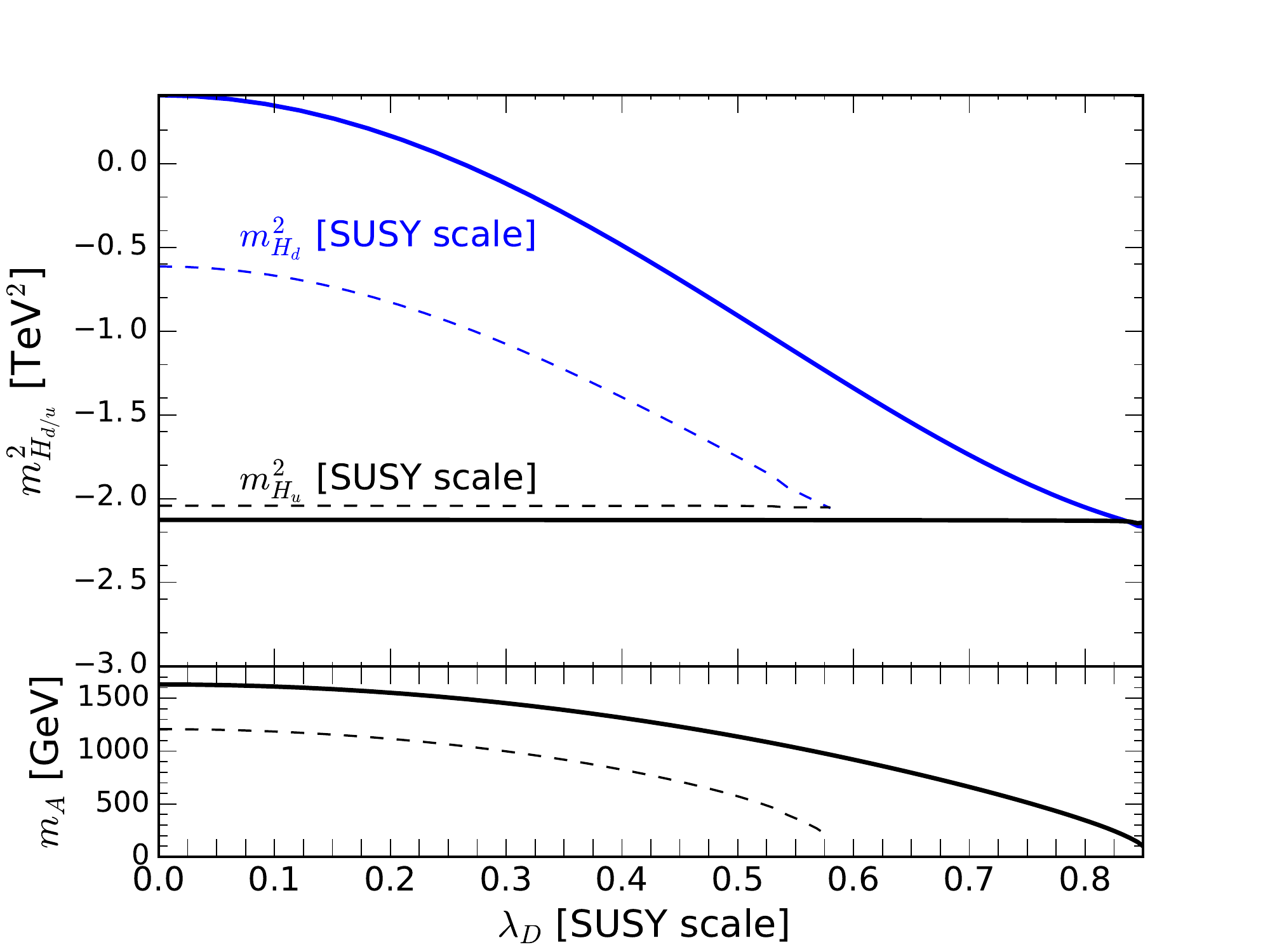}
\caption{Soft SUSY-breaking masses \mhusq\ and \mhdsq\ in the \mfivem\
  model, as well as the pseudoscalar Higgs pole mass $m_A$ as a
  function of the Yukawa coupling $\lambda_D$ at the SUSY scale.  We
  fix $\mzero=300\gev$ and $\mhalf =1500\gev$. We assume $\tanb=10$
  for the thick solid lines and $\tanb=40$ for the thin dashed lines.}
\label{fig:l5M5Mu}
\end{center}
\end{figure}
%%%%%%%%%%%%%%%%%%%%%%%%%%%%%%%%%%%

An interesting difference with the CMSSM pertains to the size of the
heavy Higgs boson masses in the \mfivem\ model.  The presence of the
$D$ superfield in \refeq{superpot5} modifies the running of \mhdsq, as
the relative beta function picks up a $6 \lambda_D^2$ contribution,
while the RGE for \mhusq\ remains unchanged.  We show in
\reffig{fig:l5M5Mu} the low scale values of \mhusq\ and \mhdsq\ as
functions of the $\lambda_D$ coupling at the SUSY scale.  As the Higgs
soft masses increasingly approach each other, the heavy Higgs bosons
become lighter (for instance, the tree-level form of the pseudoscalar
mass reads $m_{A,\textrm{tree}} \sim | m^2_{H_d} -
m^2_{H_u}|/\sqrt{1-\sin^2 2 \beta}$).

One of the main consequences of this effect is that, in the \mfivem\
model, the pseudoscalar mass $m_A$ can almost be traded for $\lam_D$
as a free parameter, thus opening up additional parameter space when
it comes to obtaining the correct DM relic density.  We will come back
to this point in \refsec{sec:dm}.  Note that the same freedom is not
seen in the \mtenm\ model, as in that case the beta function of
\textit{both} the \mhdsq\ and \mhusq\ soft terms are modified by large
contributions.

%%%%%%%%%%%%%%%%%%%%%%%%%%%%%%%%%%%%
\subsection{Fine tuning}
%%%%%%%%%%%%%%%%%%%%%%%%%%%%%%%%%%%%%

It was recently suggested in\cite{Dermisek:2016tzw}, but previously
emerged indirectly in\cite{Martin:2009bg} as well, that the presence
of new VL colored particles mixing with the third generation squarks
can lead to possible reductions in the fine tuning of the \mzero\
parameter with respect to the MSSM. This effect is possibly observed
particularly in the \mtenm\ model, which features some new couplings
involving the third generation squarks.

One can, as Ref.\cite{Dermisek:2016tzw} suggests, suppress
the Yukawa interactions between the VL fields and the Higgs sector
to effectively decouple the $H_u$ superfield from the
color fields responsible for setting the SUSY scale. If these Yukawa cuplings are not forbidden by some symmetry, however, 
one is more likely to see that for selected values
of $\lam_{10}$ and $Y_{10}$ additional ``focus point''
behavior is induced, as for some choices of the new Yukawa couplings the
SUSY-scale value of $\mhusq$ becomes less sensitive to the initial
value, \mzero. In this case, then, one has to take into account the fine tuning due to the chosen value of the Yukawas, 
which can be significant.
Note, however, that, as is usually the case in GUT-constrained models, the dominant source of fine tuning comes from the 2-loop effects on the renormalization of \mhusq\ driven by the gluino mass. 
This implies that the fine tuning is not significantly altered in our models with respect to the CMSSM.
 
%To give a quantitative feel for the fine tuning, we recall that the 1-loop beta function of \mhusq -- which in the MSSM is driven by
%$d\mhusq/dt\sim 6y_t^2 (m_{\tilde{q}_3}^2+m_{\tilde{t}_R}^2+\mhusq)$ -- is now modified by additional contributions:
%$\delta(d\mhusq/dt)\sim 6\lam_{Qu}^2 (m^2_Q+m_{\tilde{t}_R}^2+\mhusq)+6\lam_U^2 (m^2_U+m_{\tilde{q}_3}^2+\mhusq)+...$. Upon integrating, one obtains a modified dependence of the SUSY-scale value of \mhusq\ on the parameters of the GUT scale.

We can calculate \mhusq\ at the low scale as an approximate function of the GUT-scale parameters (in the range $\mzero,\mhalf\lesssim 3.5\tev$ and $\tanb\approx 30$). One finds 
\bea
\mhusq(\msusy)&\simeq&0.07\,m_0^2-0.92\,\mhalf^2-0.11\,\azero^2+0.31\,\azero\mhalf\,\,\textrm{ (CMSSM)}\\
\mhusq(\msusy)&\simeq&0.10\,m_0^2 -0.99\,\mhalf^2 -0.11\,\azero^2 +0.35\,\azero \mhalf\,\,\textrm{ (\mfivem)}\\
 \mhusq(\msusy)&\simeq&0.18\,m_0^2-1.02\,\mhalf^2 -0.13\,\azero^2 +0.49\,\azero \mhalf\,\,\textrm{ (\mtenm), }
\eea
for zero GUT-scale values of all new Yukawa couplings, and
 \bea
\mhusq(\msusy)&\simeq&0.08\,m_0^2-0.97\,\mhalf^2 -0.14\,\azero^2 +0.32\,\azero \mhalf\textrm{ (\mfivem)}\\
\mhusq(\msusy)&\simeq&-1.11\,\mhalf^2 -0.11\,\azero^2 +0.46\,\azero \mhalf  \textrm{ (\mtenm),}\label{eq:FT10}
\eea
for selected GUT-scale values of the Yukawas: $\lam_5\approx 0.1$, $\lam_{10}\approx 0.05$, $Y_{10}\approx 0.05$.
In all of the models the coefficient regulating the dependence on the unified gaugino mass, \mhalf, remains of order 1.
As a consequence, \mhalf\ and $\mu$ provide the main source of fine tuning, particularly when 
the constraints from the Higgs mass and LHC direct SUSY searches are taken into account.

We have calculated the fine tuning of our models with SPheno. 
The program calculates numerically for all input parameters $p_i$ the Barbieri-Giudice 
measure\cite{Ellis:1986yg,Barbieri:1987fn}, 
\be
FT(p_i)=\left|\frac{\partial\log M_Z^2}{\partial\log p_i}\right|,
\ee
where the pole mass $M_Z$ is calculated including the 1-loop tadpole corrections to the scalar potential, using the same procedure as in\cite{Ross:2017kjc}.

%%%%%%%%%%%%%%%%%%%%%%%%%%%%%%%%
\begin{figure}[t]
\centering
\subfloat[]{%
%\label{fig:c}%
\includegraphics[width=0.47\textwidth]{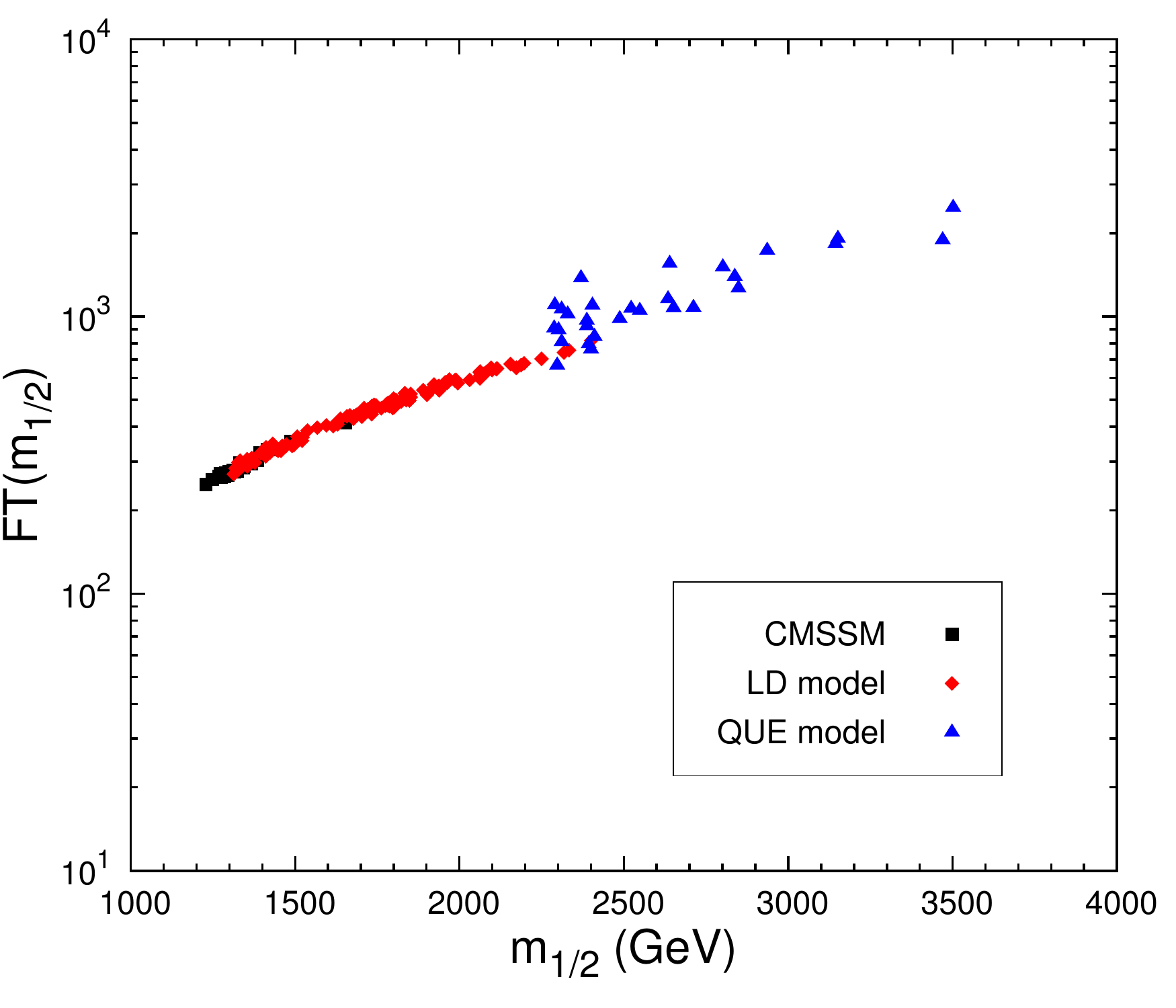}
}%
\hspace{0.02\textwidth}
\subfloat[]{%
%\label{fig:a}%
\includegraphics[width=0.47\textwidth]{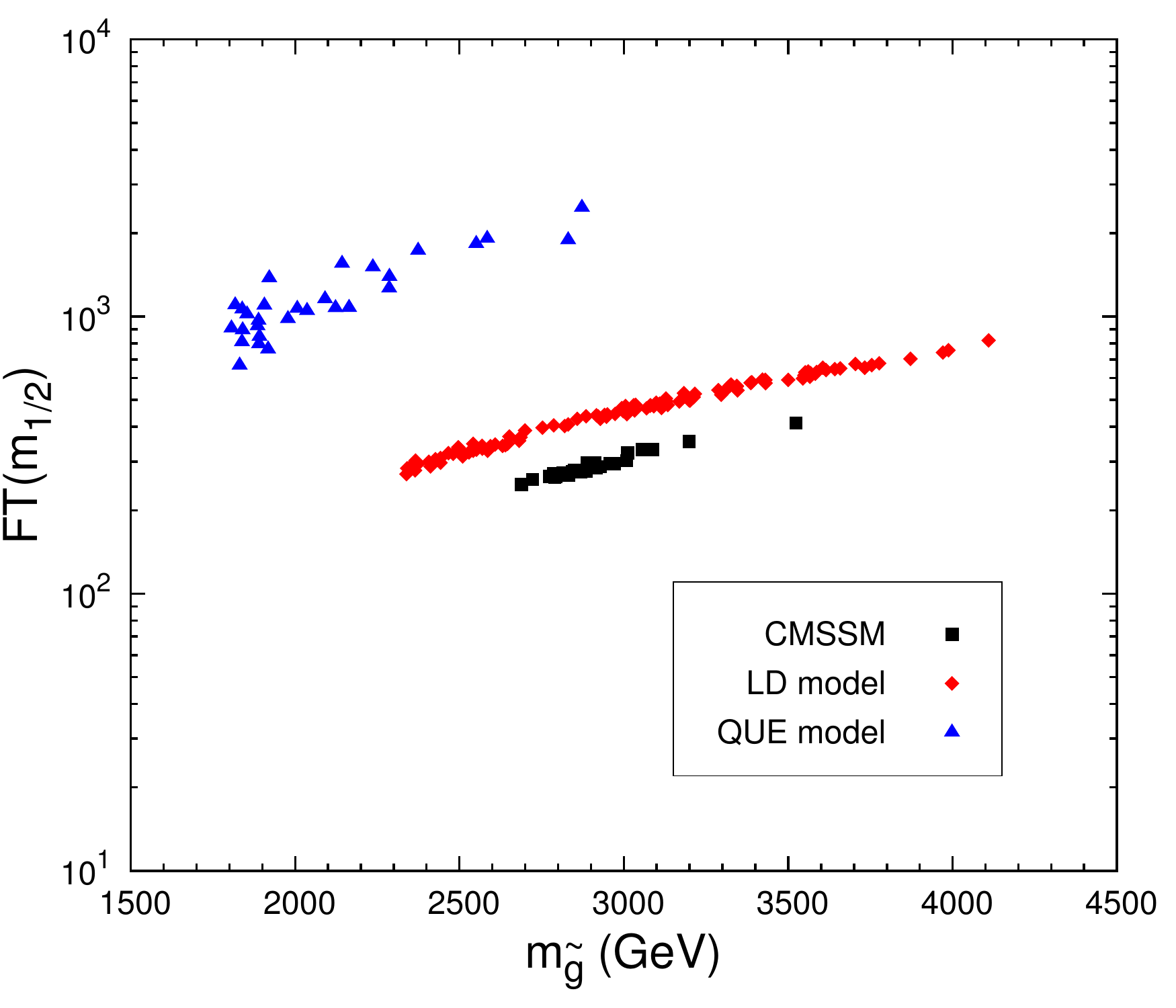}
}%
\caption{(a) The tuning of the GUT-scale parameter \mhalf\ as a function of \mhalf\ itself, 
for the CMSSM (black squares), the \mfivem\ model (red diamonds), and the \mtenm\ model (blue triangles), 
where we fixed $\mzero\approx1\tev$ and $|\azero|<500\gev$. (b) Same as (a), but as a function of the gluino pole mass $\mglu$\,.}
\label{fig:VL5tuning}
\end{figure}
%%%%%%%%%%%%%%%%%%%%%%%%%%%%%%%%

In \reffig{fig:VL5tuning}(a) we show the fine tuning of the \mfivem\ and \mtenm\ models, as a function of \mhalf, 
compared to the CMSSM, for a region of the parameter 
space in agreement with the constraints listed in \refsec{numerics}. 
The 5-plet model is thus currently 
tuned at the level of one part in $10^2-10^3$, not dissimilarly from the CMSSM, whereas the 10-plet model suffers from 
requiring larger GUT-scale values of the parameters, given an equivalent phenomenology. 

We present the dependence on \mglu\ in \reffig{fig:VL5tuning}(b), which shows that a given gluino mass leads to a 
$\sim50\%$ increase in the tuning of the \mfivem\ model with respect to the CMSSM,
due to a different RGE running.  
By the same token, the fine tuning associated with the \mtenm\ model increases even more drastically, 
as the physical SUSY masses are there associated with overall larger values of the GUT-scale input parameters.     

%%%%%%%%%%%%%%%%%%%%%%%%%%%%%%
\subsection{Bounds from perturbativity and physicality\label{sec:pert}}
%%%%%%%%%%%%%%%%%%%%%%%%%%%%%

Perturbativity is a key requirement that ends up placing constraints on the new Yukawa 
couplings introduced in \refeq{superpot5} and \refeq{superpot10}. 
A comprehensive study of the infra-red fixed points of the Yukawa couplings of some VL extensions of the MSSM 
was presented in\cite{Martin:2009bg}. In this subsection, we limit ourselves to discussing the 
bounds that apply to our specific models, the 5-plet and the 10-plet. 

Note, as a starting point, that the fact that the new Yukawa couplings are fixed at the GUT scale implies that they are always 
safe from Landau poles. Indeed, as a larger number of Yukawa couplings increase the beta function, 
choosing one sizable coupling at the GUT scale 
will simply induce smaller values at the SUSY scale for the other Yukawa couplings. Therefore, the only possible source of 
non-perturbative behavior are the MSSM Yukawa couplings, which are fixed at the EWSB scale by the experimental value of the fermion masses. 

In the \mfivem\ model, the problematic Yukawa is the one of the bottom
quark, $y_b$\,, as $\lambda_D$ and $\lambda_L$ introduce only a small
contribution to the top Yukawa RGE. Supposing for simplicity that the mixing terms involve the second generation only,
$\lam_{D}\equiv\lam_{D,2}$ and $\lam_{L}\equiv\lam_{L,2}$, one has
\begin{align}
\beta_{y_{b}} & \sim  y_b \Big(   3 |\lambda_{D}|^2 + |\lambda_{L}|^2 \Big), \label{eq:pert}
\end{align}
which can be used to impose an upper bound on $\lam_D$ and $\lam_L$. 
The ratios between the low- and high-scale values of these couplings are approximately 
$\lam_D\approx 1.9\,\lam_L\approx 2.3\,\lam_5$.
From preventing a Landau pole in $y_b$ one gets the \tanb-dependent bound 
\bea
\lam_D\lesssim~ \begin{cases}~1 \qquad \phantom{.7} (\tanb=5)\\ ~0.7 \qquad (\tanb=60) \end{cases}.\label{eq:pbound}
\eea

In the \mtenm\ model, the problematic coupling is the top Yukawa coupling $y_t$\,, for which new contributions to the beta function read
\begin{align}
\beta_{y_t} & \sim -6 y_t\left( |\lambda_{U,2}  |^2+2|\lam_{U,3}|^2\right) +3 {Y_{10}} \left( |\lambda_{U,3}|^2  + {Y_{10}} y_t \right)
\end{align}
in the limit $\lam_{Qu}=\lam_{U}$.

When $\lam_{U,3}$ is negligible, we get the bounds 
\bea
\lam_{U,2}\lesssim~ \begin{cases}~0.65 \qquad (Y_{10}=0.2)\\ ~0.45 \qquad (Y_{10}=0.6) \end{cases}.\label{eq:pbound2}
\eea
Conversely, the bounds on $\lam_{U,3}$ are even stronger: $\lam_{U,3}\lesssim 0.45$ with $Y_{10}=0$ and $\widetilde{M}=0$.

Interestingly, however, the bounds on the couplings of the \mfivem\
model are actually more severe than in \refeq{eq:pbound}, due to the requirement of
physical values for the masses of the heavy Higgs bosons, see
\refsec{sec:higgs}. This, coupled to the requirement that the top Yukawa
remains perturbative (which sets a lower bound on $\tan \beta$), yields the bound 
\bea \lam_D\lesssim~ \begin{cases}~0.9 \qquad (\tanb=5)\\ ~0.4
  \qquad (\tanb=60) \end{cases}.\label{eq:pbound3} \eea 
Note that these bounds depend on \mzero\,: while \refeq{eq:pbound3} has been determined for $\mzero \approx 0.5 \tev$, it becomes $\lam_D\lesssim~ 0.8$ ($0.1$) for $\tanb=5$ ($60$) at $\mzero \approx 2.5 \tev$.
\medskip

\textbf{Comments on $ttH$.} 
We conclude this section with a few comments on the possibility of enhancing the $t\bar{t}h$ production mode in the \mtenm\ model, 
while keeping the gluon-fusion production rate around its measured value, as was suggested in\cite{Angelescu:2015kga}.
Although this possibility remains enticing in generic VL scenarios, the anomaly cannot be explained in models 
constrained at the GUT scale once all phenomenological bounds are taken into account.
In light of the above discussion this is to be expected, as the values for the new Yukawa couplings considered 
in\cite{Angelescu:2015kga} are not compatible with our 
assumption of perturbativity up to the GUT scale.

%%%%%%%%%%%%%%%%%%%%%%%%%%%%%%
\subsection{Bounds from electroweak precision tests}
%%%%%%%%%%%%%%%%%%%%%%%%%%%%%

The mixing between the new VL and the SM leptons is strongly
constrained by precision tests of the EW theory.  A detailed study of
the constraints from these observables is beyond the scope of this
paper, but we will derive in this section rough bounds on the
parameters relevant in our models. Note that further bounds related to flavor and EW precision tests are directly implemented 
numerically and will be described in \refsec{numerics}.

We assume for simplicity that only the mixing between the second generation and VL particles is present.
The second-lightest charged lepton and neutrino mass eigenstates, $e_2$ and $\nu_2$, contain a fraction of the VL lepton fields.
Their gauge coupling to the $Z$ and $W$ bosons are thus
\begin{align}
 \scr{L} \supset  Z_{\mu} \bar{e}_{2} \gamma^\mu (P_L \glzmu + P_R \grzmu) e_2 + \left[ W_\mu \bar{\nu}_{2} \gamma^\mu P_L \glwmu e_2 +\textrm{h.c.} \right],
\end{align}
where the couplings for our models are given in detail in Appendix~\ref{app:prec}. 
The absence of a VL left-handed $SU(2)$-singlet in the \mfivem\ model as well as of a VL right-handed $SU(2)$-doublet in the \mtenm\ 
model limits the corrections to these couplings, as we shall see. 

We assume here that the following mass hierarchy holds:
\begin{align}
 m_\mu \ll \lambda_L v_d,\lambda_E v_d,\widetilde{M}_L,\widetilde{M}_E \ll M_L, M_E \ .
 \label{eq:masshierc}
\end{align}
The most stringent bounds on these parameter originate from two observables. 
First, the couplings to the $Z$ boson are strongly constrained by the $Z \rightarrow \mu^+ \mu^-$ 
branching ratio (see, e.g.,\cite{Olive:2016xmw}), which imposes a constraint on the modified couplings $\delta g_{L,R}^{Z\mu\mu}\equiv g_{L,R}^{Z\mu\mu}-g_{L,R,\textrm{SM}}^{Z\mu\mu}$\,:
\begin{align}
 \frac{ \delta  \glzmu }{\glzmuSM}\,\,,\,\,\frac{\delta  \grzmu }{\glzmuSM} \lesssim 0.1 \%\,,\label{prec_bound}
\end{align}
leading to 
\begin{align}
 \frac{ \lam_L v_d}{M_L}\,\,,\,\,\frac{\lam_E v_d}{M_E} \lesssim 2 \%\,.\label{prec_bound3}
\end{align}

Second, the measurements of the Fermi constant, using the muon lifetime, constrains the coupling with the $W$ boson 
(see, e.g.,\cite{Dermisek:2015oja}) as
\begin{align}
 \frac{\delta \glwmu }{\glwmuSM}\,\,,\,\,\frac{\delta  \grwmu}{\grwmuSM} \lesssim 0.05 \%\,,\label{prec_bound2}
\end{align}
thus producing a stronger bound on the coupling than the direct
measurements.

Note, however, that in the \mfivem\ model these
couplings are only generated at order $(\widetilde{M}_L/M_L)^4$ in the limit of mixing between the second generation and 
VL particles only. This leads to a mild constraint $\widetilde{M}_L/M_L\lesssim 0.1$ (cf. Appendix~\ref{app:prec}).
In the \mtenm\ model the equivalent constraint is weaker than the one
from $\textrm{BR}(Z \rightarrow \mu^+ \mu^-)$, \refeq{prec_bound3}.

%%%%%%%%%%%%%%%%%%%%%%%%%%%%%%%%%%%%%%%%%%%%%%%%%%%%%%%%%%%%
\section{The {\boldmath \gmtwo } anomaly in VL models \label{gm2analytic}}
%%%%%%%%%%%%%%%%%%%%%%%%%%%%%%%%%%%%%%%%%%%%%%%%%%%%%%%%%%%%%

There is a long-standing discrepancy, at the $3\sigma$ or more level, 
between the value of the anomalous magnetic moment of the muon, 
$a_{\mu} =\gmtwo/2$, measured at Brookhaven\cite{Bennett:2006fi,Hagiwara:2011af} and the SM expectation.

A recent update\cite{Davier:2016iru} of the lowest order hadronic contributions to the calculation of $a_{\mu}$ in the SM places the discrepancy at\,\footnote{An older estimate\cite{Davier:2010nc} places the value of \deltagmtwomu\ at $(28.7\pm 8.0)\times 10^{-10}$, whereas the estimate provided in\cite{Hagiwara:2011af} leads to $(26.1\pm 8.0)\times 10^{-10}$.}
\be
\deltagmtwomu \equiv a_\mu(\textrm{SM}) - a_\mu(\textrm{exp}) = (27.4 \pm 7.6 ) \times 10^{-10}.
\ee

The anomaly, if real, provides a clear hint for new physics not far
from the EWSB scale.  On the experimental side, the New Muon g-2
experiment at Fermilab\cite{Grange:2015fou,Chapelain:2017syu} will
soon probe the discrepancy at an unprecedented $7\sigma$ level,
which is bound to revive the interest of the particle physics
community in the subject.

It has been long known that, while the excess can be easily explained in the framework of the MSSM 
even after the most recent LHC bounds for direct SUSY searches are taken into account\cite{Endo:2013bba,Fowlie:2013oua,
Chakraborti:2014gea,Das:2014kwa,Kowalska:2015zja,Lindner:2016bgg}, 
the same bounds and the Higgs mass value prevent a good fit in the simplest 
constrained models, like the CMSSM and the NUHM\cite{Bechtle:2012zk,Fowlie:2012im,Buchmueller:2013rsa}.
The tension can be ameliorated if one relaxes the assumption of gaugino and/or squark universality, 
as shown for instance in\cite{Akula:2013ioa,Kowalska:2015zja}.

In this section we show that, as an alternative, the tension can be resolved by maintaining universal boundary 
conditions at the GUT scale, and considering instead additional VL matter, as the new sleptons can contribute to loop corrections 
and give rise to phenomenological features different from the MSSM. 
Note that solutions to the \gmtwo\ anomaly employing VL fermions have also been considered 
in\cite{Endo:2011xq,Endo:2011mc,Aboubrahim:2016xuz,Nishida:2016lyk},
in general by postulating the existence of a full new generation, which leads to new fermionic contributions in loops
involving the $W$ and $Z$ bosons, or in the framework of the MSSM with parameters 
defined at the SUSY scale. 

In the CMSSM-like VL extensions that we consider here we introduce $\mathbf{5}+\mathbf{\bar{5}}$, 
or $\mathbf{10}+\mathbf{\overline{10}}$ multiplets of $SU(5)$, as described in \refsec{sec:models}.
The complete one-loop corrections in the mass-eigenstate basis are 
well-known\cite{Moroi:1995yh,Martin:2001st} and already implemented in many codes, 
including the SARAH-generated SPheno routines that we use to find the regions of the parameters space that 
are in agreement with the measured value 
of \deltagmtwomu, the relic abundance, and the other constraints defined in \refsec{numerics}.
It is worth, however, first taking a look to the parametric dependence of \deltagmtwomu\ in our models.

As we explain in more detail in \refsec{sec:dm}, for the neutralino mass range considered here one 
obtains the correct value of the relic density 
in the slepton-coannihilation and $A$-funnel regions of the parameter space, which are both characterized by a mostly bino-like neutralino.
In the MSSM with bino-like DM, the dominant contributions to \deltagmtwomu\ are due to the well known neutralino/smuon and 
chargino/sneutrino loops, and are approximately of comparable strength.

The former, $\Delta a_{\mu}^{\chi^0}$, can be expressed, following\cite{Martin:2001st}, as a sum over smuon and neutralino 
mass eigenstates, $\tilde{\mu}_i$ and $\chi_m^0$:
\be
\Delta a_{\mu}^{\chi^0}\approx \frac{m_{\mu}}{48\,\pi^2}
\sum_{i,m} \left[\frac{m_{\chi^0_m}}{m^2_{\tilde{\mu}_i}}\,\textrm{Re}\left(n_{im}^L n_{im}^R\right) \mathcal{F}_N(x_{im})\right]\,,\label{eq:amuneut}
\ee
where $m_{\mu}$ is the muon mass, the loop function $\mathcal{F}_N$ takes the form
\be
\mathcal{F}_N(x) = \frac{ 3 }{ (1-x)^3 } 
\left(
1 - x^2 + 2 x \ln x
\right),
\ee
and $x_{im}\equiv m^2_{\chi_m^0}/m^2_{\tilde{\mu}_i}$.
The effective couplings $n_{im}^L$ and $n_{im}^R$ parameterize the interaction of the physical smuons with the neutralinos and 
with left-handed and right-handed muons, respectively. They can be expressed explicitly in terms of the eigenvectors of the 
neutralino and smuon mass matrices and can be found, e.g., in\cite{Martin:2001st}. 

Equivalently, the dominant chargino/sneutrino contribution, $\Delta a_{\mu}^{\chi^{\pm}}$, reads\cite{Martin:2001st}
\be
\Delta a_{\mu}^{\chi^{\pm}}\approx \frac{m_{\mu}}{24\,\pi^2}
\sum_{j,k} \left[\frac{m_{\chi^{\pm}_k}}{m^2_{\tilde{\nu}_{\mu,j}}}\,\textrm{Re}\left(c_{jk}^L c_{jk}^R\right) \mathcal{F}_C(z_{jk})\right]\,,\label{eq:amuchar}
\ee
where the loop function $\mathcal{F}_C$ is given by
\be
\mathcal{F}_C(x) = -\frac{ 3 }{2 (1-x)^3 } 
\left(
3-4x+x^2 + 2\ln x
\right),
\ee
and $z_{jk}\equiv m^2_{\chi_k^{\pm}}/m^2_{\tilde{\nu}_{\mu,j}}$.
Again, $c_{jk}^L$ and $c_{jk}^R$ are the effective couplings of the physical muon sneutrinos 
(of which there is one in the MSSM with minimal flavor violation) to the charginos and 
left-handed and right-handed muons. 

In the limit of an almost pure bino LSP -- roughly the case for the
$A$-funnel region of the CMSSM, but not necessarily for the
stau-coannihilation region, in which \deltagmtwomu\ features non
negligible contributions from diagrams involving heavier higgsino-like
neutralinos -- \refeq{eq:amuneut} takes the simple
form\cite{Martin:2001st} \be \Delta
a_{\mu}^{\chi^0}\approx\frac{g_1^2}{48\pi^2}\frac{m_{\mu}^2M_1(\mu\tanb-A_{\mu})}{m_{\tilde{\mu}_2}^2-m_{\tilde{\mu}_1}^2}
\left[\frac{\mathcal{F}_N(x_{11})}{m_{\tilde{\mu}_1}^2}-\frac{\mathcal{F}_N(x_{21})}{m_{\tilde{\mu}_2}^2}\right],\label{eq:bino_neut}
\ee where the smuon mixing term in the numerator, which depends
linearly on $\mu$ and $A_{\mu}$, provides the main source of chirality
flip in the loop. Under the same assumptions, the parametric form of
\refeq{eq:amuchar} can also be derived (see Appendix~\ref{app:gm2}),
and reads \be \Delta
a_{\mu}^{\chi^{\pm}}\approx\frac{g_2^2\,m_{\mu}^2}{24\,\pi^2}\frac{\mu\,M_2\tanb}{m_{\chartwo}^2-m_{\charone}^2}
\left[\frac{\mathcal{F}_C(z_{11})-\mathcal{F}_C(z_{12})}{m_{\tilde{\nu}_{\mu}}^2}\right].\label{eq:bino_char}
\ee

%%%%%%%%%%%%%%%%%%%%%%%%%%%%%%%%
\begin{figure}[t]
\centering
\subfloat[]{%
%\label{fig:c}%
\includegraphics[width=0.47\textwidth]{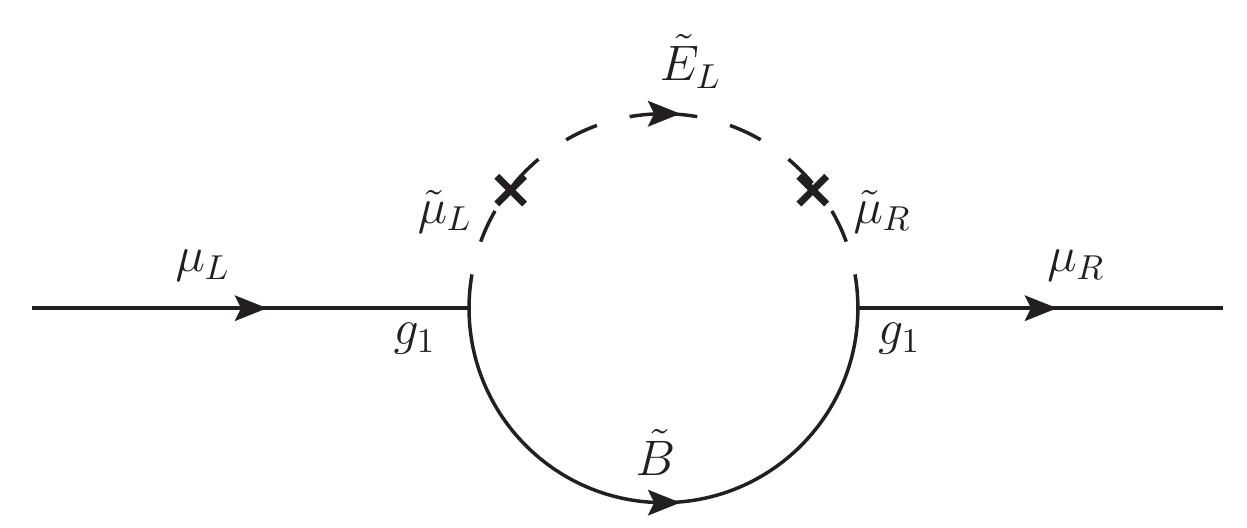}
}%
\hspace{0.02\textwidth}
\subfloat[]{%
%\label{fig:a}%
\includegraphics[width=0.47\textwidth]{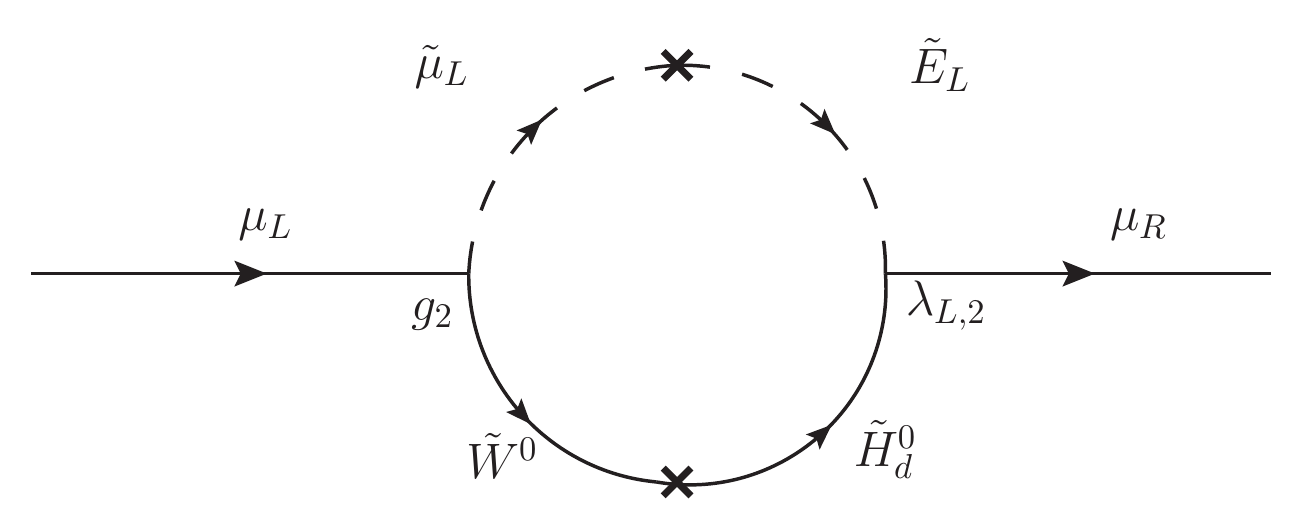}
}%
\\
\subfloat[]{%
%\label{fig:b}%
\includegraphics[width=0.47\textwidth]{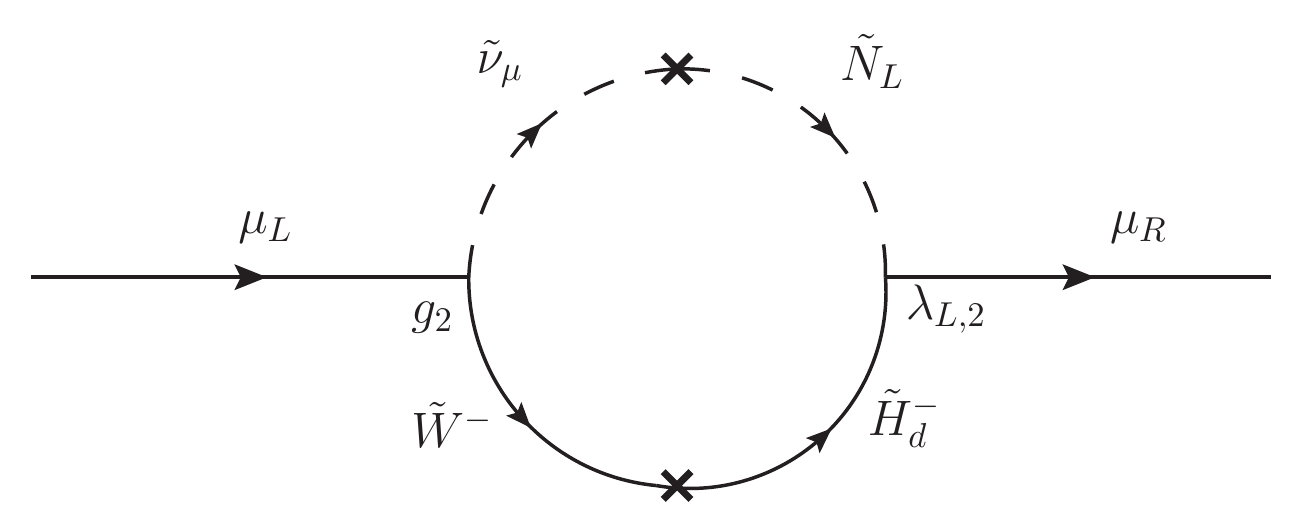}
}%
\caption{Main loops contributing to \deltagmtwomu\ in the \mfivem\ model. The photon line attached to one of the charged legs is implicit. We have explicitly written the doublet $L$ as $L=(N_L, E_L)^T$. 
}
\label{fig:g2loop5}
\end{figure}
%%%%%%%%%%%%%%%%%%%%%%%%%%%%%%%%

The presence of the new VL sector introduces new contributions to \gmtwo\ in two ways, 
which are summarized for the \mfivem\ model in \reffig{fig:g2loop5}:
\begin{itemize}
\item There are new sources of smuon mixing, as the second-generation
  sleptons are mixed with new VL matter, see \reffig{fig:g2loop5}(a).
\item There are new Yukawa couplings, $\lambda_L$ and $\lam_E$, that
  are a priori unconstrained and can be much larger than the muon
  Yukawa, see Figs.~\ref{fig:g2loop5}(b) and
  \ref{fig:g2loop5}(c). Note, however, that for models with
  perturbativity assumed up to the GUT scale the absence of Landau
  poles does constrain these couplings, see \refsec{sec:pert}.
\end{itemize}

In the \mfivem\ model, extra contributions to \refeq{eq:bino_neut} are provided by larger mixing between the smuons.
The loop correction depicted in \reffig{fig:g2loop5}(a) modifies \refeq{eq:bino_neut} in a non-trivial way, 
as the smuon mass matrix that must be 
diagonalized is now $4\times 4$, see \refeq{eq:smumass} in Appendix~\ref{app:soft}. The physical smuon mixing now depends 
on the new Yukawa coupling, $\lam_L$, and the superpotential and soft-SUSY breaking mixing terms, $\widetilde{M}_L$ and 
$\widetilde{m}^2_L$.

The loop correction of \reffig{fig:g2loop5}(c) affects instead the form of the chargino/sneutrino contribution, \refeq{eq:bino_char}, 
which is now modified by an additional term
\be
\Delta a_{\mu,\textrm{tot}}^{\chi^{\pm}}=
\Delta a_{\mu}^{\chi^{\pm}}+\Delta a_{\mu,\textrm{VL}}^{\chi^{\pm}},\label{eq:VL5tot}
\ee
where $\Delta a_{\mu,\textrm{VL}}^{\chi^{\pm}}$ is expressed in terms of the sneutrino mass squared matrix, 
\refeq{eq:snumass}, and reads (see Appendix~\ref{app:gm2})
\begin{multline}\label{eq:VL5char}
\Delta a_{\mu,\textrm{VL}}^{\chi^{\pm}}\approx\frac{g_2\,m_{\mu}M_W}{12\sqrt{2}\,\pi^2}\frac{\mu\,M_2\sin\beta}{m_{\chartwo}^2-m_{\charone}^2}\frac{\lam_L(M_L\widetilde{M}_L+\widetilde{m}_L^2)}{m^2_{\tilde{\nu}_{\mu,2}}-m^2_{\tilde{\nu}_{\mu,1}}}\\
\times\left[\frac{\mathcal{F}_C(z_{21})
-\mathcal{F}_C(z_{22})}{m_{\tilde{\nu}_{\mu,2}}^2}-\frac{\mathcal{F}_C(z_{11})-\mathcal{F}_C(z_{12})}{m_{\tilde{\nu}_{\mu,1}}^2}\right].
\end{multline}

%%%%%%%%%%%%%%%%%%%%%%%%%%%%%%%%
\begin{figure}[t]
\centering
\subfloat[]{%
%%\label{fig:b}%
\includegraphics[width=0.47\textwidth]{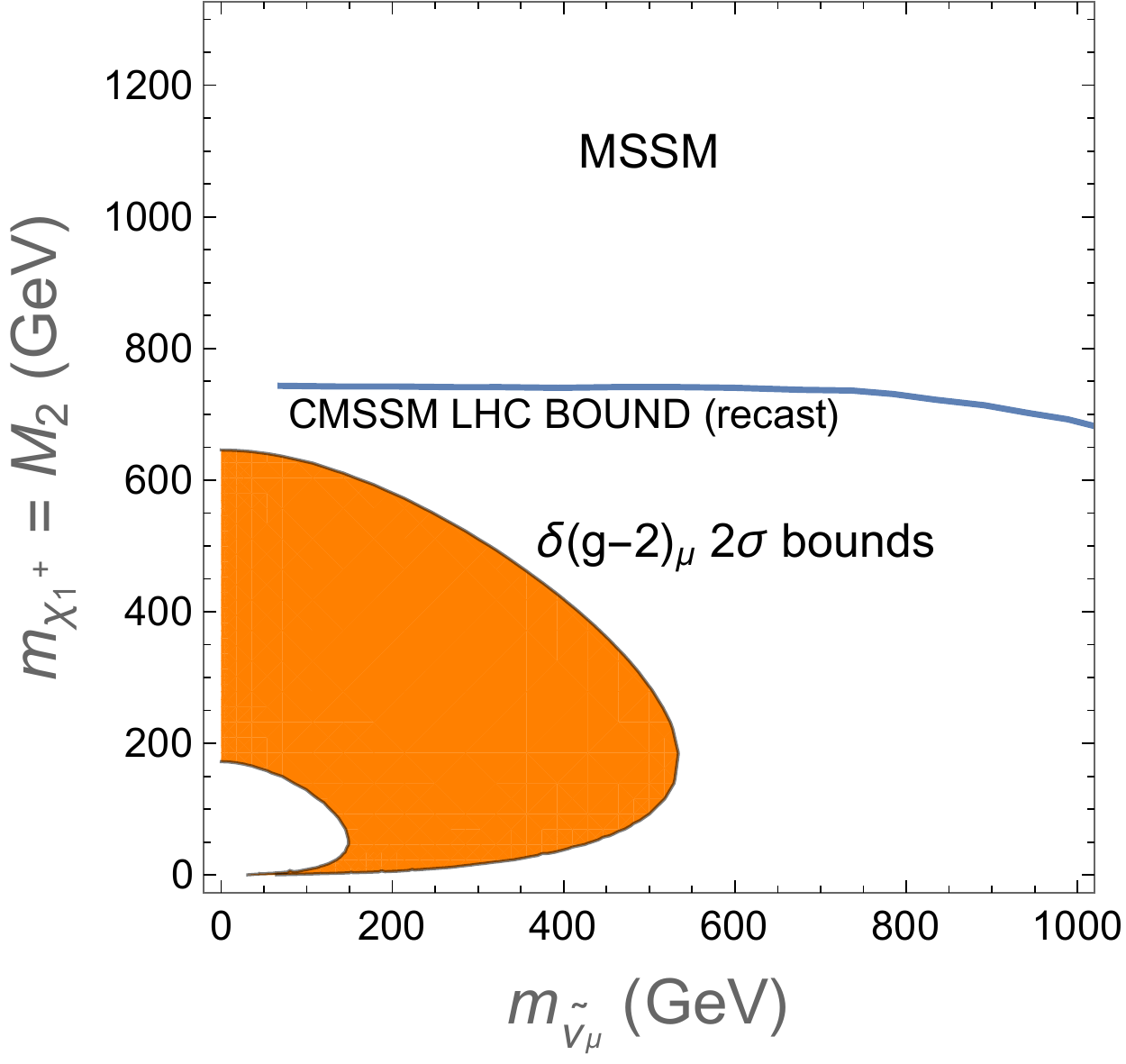}
}%
\hspace{0.02\textwidth}
\subfloat[]{%
%\label{fig:c}%
\includegraphics[width=0.47\textwidth]{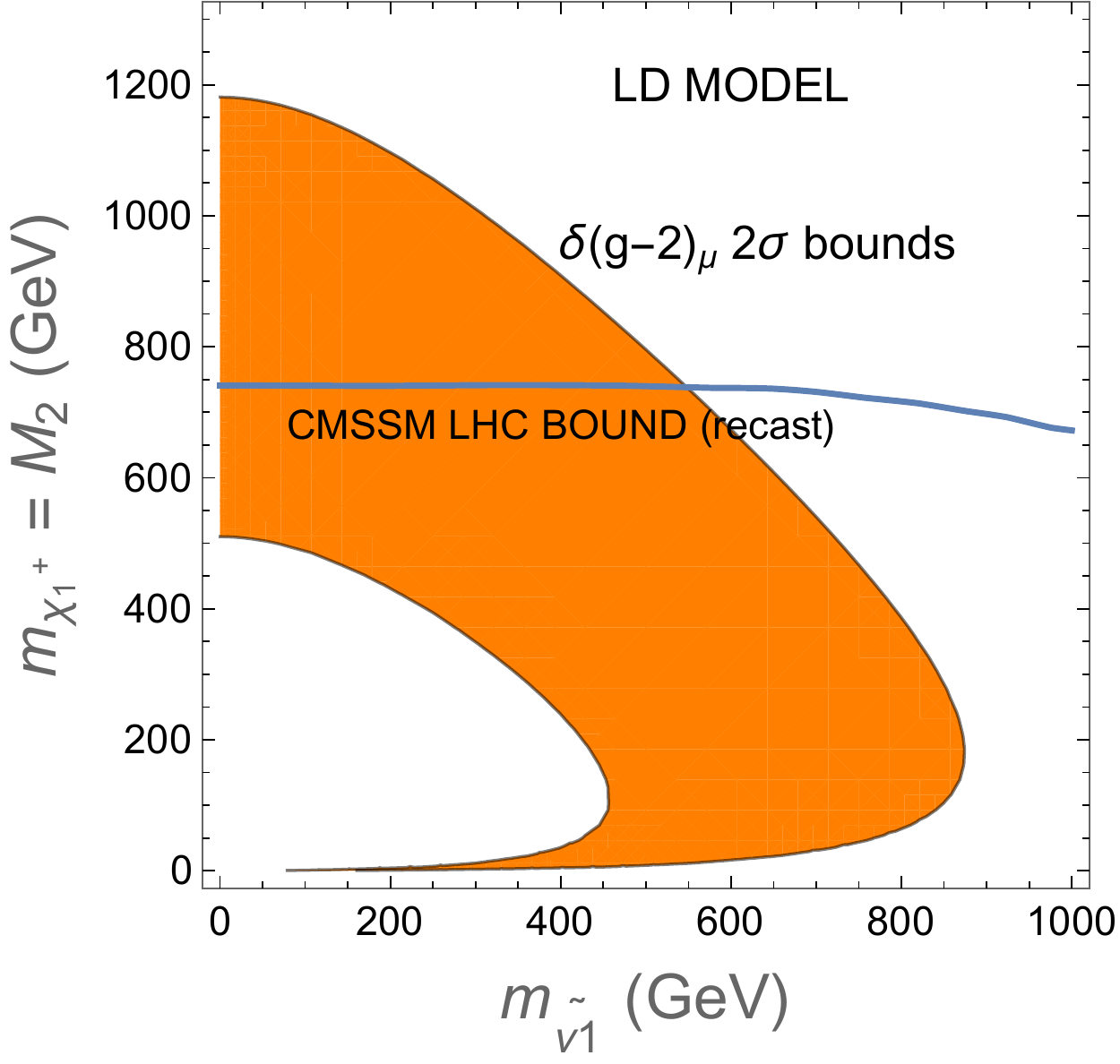}
}%
\caption{(a) We show in orange the $2\sigma$ region of \deltagmtwomu\
  from \refeq{eq:bino_char} in the ($m_{\tilde{\nu}_{\mu}}$, $M_2$)
  plane of the MSSM. We assume $\tanb=30$ and $\mu=1200\gev$.  The LHC
  lower bound from squark and gluino
  searches\cite{ATLAS-CONF-2016-078} projected to the
  ($m_{\tilde{\nu}_{\mu}}$, $M_2$) plane after CMSSM-like boundary
  conditions are applied is shown as a solid gray line.  (b) The
  $2\sigma$ region of \deltagmtwomu\ from \refeq{eq:VL5tot} in the
  ($m_{\tilde{\nu}_1}$, $M_2$) plane.  We assume $M_L=600\gev$,
  $\lam_L=0.25$, $\widetilde{M}_L=10\gev$,
  $\widetilde{m}^2_L=-(500\gev)^2$, and $m^2_L\approx
  m^2_{\tilde{\mu}_L}$, with all parameters defined at the SUSY
  scale.}
\label{fig:g2New}
\end{figure}
%%%%%%%%%%%%%%%%%%%%%%%%%%%%%%%% 

In \reffig{fig:g2New}(a) we derive the approximate $2\sigma$ bounds in the plane of the chargino mass versus 
muon sneutrino mass of the MSSM, ($m_{\tilde{\nu}_{\mu}}$, $M_2$), using the largest contribution, \refeq{eq:bino_char}. 
We superimpose a CMSSM recast of the current LHC bounds from direct squark and gluino searches\cite{ATLAS-CONF-2016-078}
obtained using the code of Ref.\cite{Kowalska:2016ent}.
The plot confirms that the region of the CMSSM favored by \gmtwo\ data is excluded.  

We use \refeq{eq:VL5tot} to show in \reffig{fig:g2New}(b) that VL contributions allow one to extend the available parameter space
within the $2\sigma$ bounds of \deltagmtwomu, so to evade the current LHC limit.
Selected values of the VL parameters are given in the caption.
Note that the enhanced value with respect to the SM is here due to an entirely supersymmetric effect, unlike
the enhancement obtained, e.g., in\cite{Aboubrahim:2016xuz}, which is instead due to loop contributions involving new neutrinos from VL matter belonging to larger representations than the ones considered in this work.

%%%%%%%%%%%%%%%%%%%%%%%%%%%%%%%%
\begin{figure}[t]
\centering
\subfloat[]{%
%\label{fig:c}%
\includegraphics[width=0.47\textwidth]{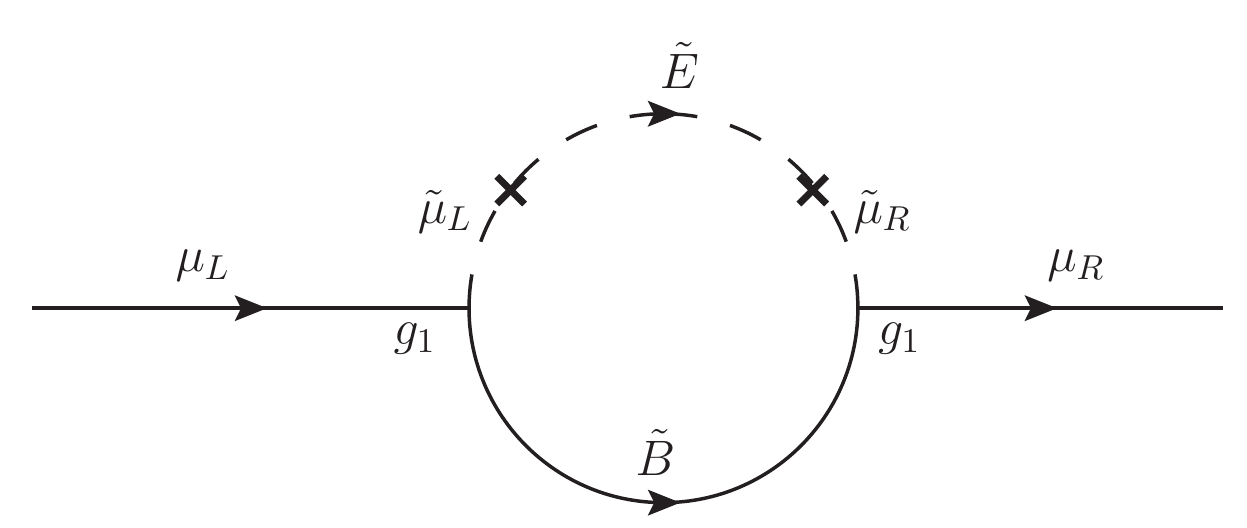}
}%
\hspace{0.02\textwidth}
\subfloat[]{%
%\label{fig:a}%
\includegraphics[width=0.47\textwidth]{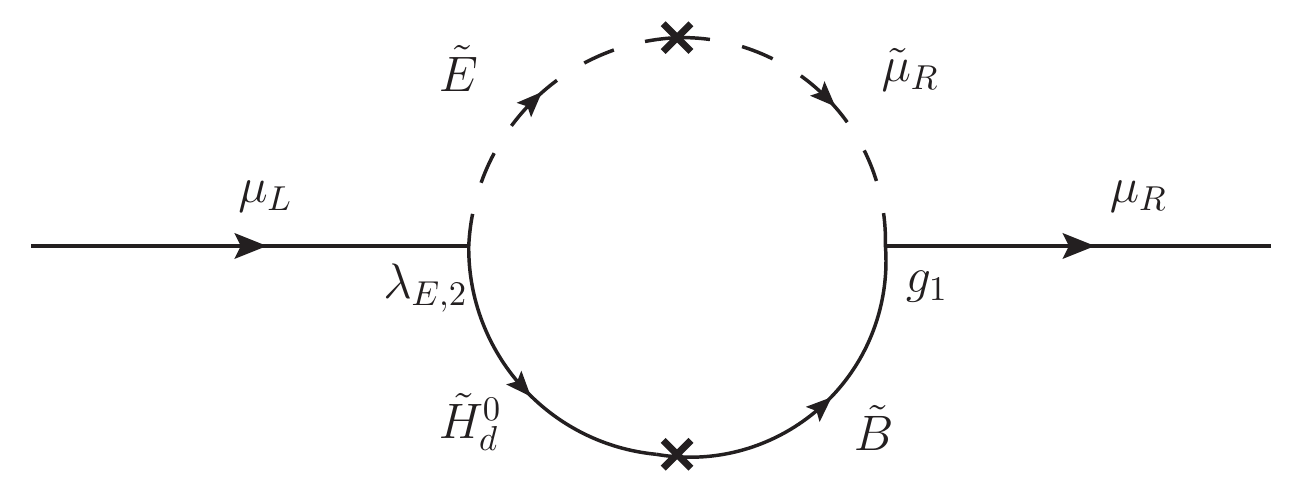}
}%
%\\
%\subfloat[]{%
%\label{fig:b}%
%\includegraphics[width=0.47\textwidth]{Plots/HinoNeutrino.eps}
%}%
\caption{
Main loops contributing to \deltagmtwomu\ in the \mtenm\ model. The photon line attached to one of the charged legs is implicit. We have explicitly written the scalar component of the supermultiplet $E$ as $\tilde{E}$.
}
\label{fig:g2loop10}
\end{figure}
%%%%%%%%%%%%%%%%%%%%%%%%%%%%%%%%

Similar considerations apply to the \mtenm\ model, with the obvious difference that only neutralino/smuon loops will be enhanced,
as there are no extra sneutrinos with respect to the MSSM. The most important contributions due to VL matter 
are shown in \reffig{fig:g2loop10}. In particular, the dominant one in the case with bino-like DM, \reffig{fig:g2loop10}(a), 
introduces a modification to $\Delta a^{\chi^0}_{\mu}$ equivalent to the contribution present in the LD model, 
where in this case one must use the elements of the mass squared matrix given in \refeq{eq:smumass2}.

%%%%%%%%%%%%%%%%%%%%%%%%%%%%%%%%%%%%%%%%%%%%%%%%%%%%%
\section{Numerical results\label{numerics}}
%%%%%%%%%%%%%%%%%%%%%%%%%%%%%%%%%%%%%%%%%%%%%%%%%%%%

We use \texttt{MultiNest}\cite{Feroz:2008xx} to direct the scanning procedure and we interface it with various publicly available codes. 
We use the \SARAH-produced SPheno code as our spectrum generator. 
The flavor related observables are obtained using the \SARAH-package FlavorKit\cite{Porod:2014xia}. 
Dark matter observables, \abundchi\ and \sigsip,
are computed with $\tt micrOMEGAs\ v.3.5.5$\cite{Belanger:2013oya}. 
The scan prior ranges we adopt for the parameters of the \mfivem\ and \mtenm\ models are shown in Appendix~\ref{app:scan}.

We build a global likelihood function using the constraints, central values, 
theoretical and experimental uncertainties shown in \reftable{tab:exp_constraints}. 
The Higgs sector is additionally constrained using $\tt HiggsSignals$\cite{Bechtle:2013xfa} and $\tt HiggsBounds$\cite{Bechtle:2008jh,Bechtle:2011sb,Bechtle:2013wla}. 
These codes ensure that our Higgs sector is in proper agreement with the most recent LHC bounds, despite possible 
modifications to the Yukawa couplings that originate from our mixing terms.
We also impose a hard cut on \sigsip\ from the latest LUX data\cite{Akerib:2016vxi}.

%************************************************************************************
\begin{table}[t]
\begin{center}
\begin{tabular}{|c|c|c|c|c|}
\hline
Constraint & Mean & Exp. Error & Th. Error & Ref. \\
\hline
Higgs sector & See text. & See text. & See text. & \cite{Bechtle:2013xfa,Bechtle:2008jh,Bechtle:2011sb,Bechtle:2013wla} \\
\hline
\sigsip\ & See text. & See text. & See text. & \cite{Akerib:2016vxi}\\
\hline
\abunchi\ & 0.1188 & 0.0010 & 10\% & \cite{Ade:2015xua}\\
\hline
$\brbxsgamma\times 10^4$ & 3.32 & 0.16 & 0.21 & \cite{Amhis:2016xyh} \\
\hline
$\brbutaunu \times 10^4$ & 0.72 & 0.27 & 0.38 & \cite{Adachi:2012mm} \\
\hline
\delmbs\ & 17.757~ps$^{-1}$ & 0.021~ps$^{-1}$ & 2.400~ps$^{-1}$ & \cite{Olive:2016xmw} \\
\hline
$\Delta \rho \times 10^4$ & $3.7$ & $2.3$ & $0.5$ & \cite{Olive:2016xmw} \\
\hline
$\brbsmumu \times 10^9$ & 2.9 & $0.7$ & 10\% & \cite{Aaij:2013aka,Chatrchyan:2013bka} \\
\hline
$\brtaumugamma \times 10^{8}$ & $<4.4$ & $0$ & $0$ & \cite{Aubert:2009ag} \\
\hline
\end{tabular}
\caption{The experimental constraints included in the likelihood function.}
\label{tab:exp_constraints}
\end{center}
\end{table}%
%********************************************************************************************************

An interesting consequence of the presence in the \mtenm\ model of a right-handed VL up-type quark
is the possible enhancement of the decay \brbutaunu. We have therefore included this observable in the likelihood of our scans. 
The presence of new down-type VL quarks in both the \mfivem\ and \mtenm\ model can also modify the flavor-changing neutral current $b\rightarrow s$.
We have consequently included in the likelihood of our scans the experimental values for \brbxsgamma\ and \brbsmumu, 
calculated at one-loop using FlavorKit. We additionally include all the bounds discussed in \refsec{sec:pheno}.

Finally, the constraints from the correction $\Delta \rho$ to the Veltman $\rho$-parameter are calculated at one-loop by SARAH and have been included in the likelihood of our scans.

%%%%%%%%%%%%%%%%%%%%%%%%%%%%%%%%%%%%%%%%%%%
\subsection{Muon g-2  benchmark points \label{sec:bench}}
%%%%%%%%%%%%%%%%%%%%%%%%%%%%%%%%%%%%%%%%%%%

We present in \reftable{tab:benchm} benchmark points for the models
\mfivem\ and \mtenm\ satisfying all the previous constraints including
\deltagmtwomu. In the \mfivem\ benchmark point the muon sneutrino is
light thanks to the mixing with the VL sneutrino, and gives the
greatest contribution to \deltagmtwomu. Conversely, the benchmark
point for the \mtenm\ model relies on a light slepton to generate a
sizable \deltagmtwomu.  Note the large splitting between the first
slepton mass eigenstate (which is mixed smuon/VL) and second slepton
eigenstate (which is the usual right-handed stau). Furthermore, as
could be inferred in \refsec{gm2analytic}, in order to have a positive
contribution to \deltagmtwomu\ the sign of the new Yukawa couplings
and of the new mixing soft terms should preferably be opposite.

%%%%%%%%%%%%%%%%%%%%%%%%%%%%%%%%%%%%%%%%%%
\begin{table}[htbp]
	\begin{center}
		\begin{tabular}{c|c|c|c}
			\hline
			\hline
			\rule{0pt}{2.5ex} \small{}  &       Parameter          &  \mfivem & \mtenm             \\
			\hline
			\hline
			\rule{0pt}{2.5ex}  & $\mzero$   & $600$ & $565$ \\
			& $\mhalf$   & $1550$    & $2500$   \\
			& $\azero$   & $100$    & $2500$   \\
			& $M_V$   & $1000$    & $250$   \\
			\small{\textbf{GUT inputs}}  & $B_0$   & $-\,250$  & $0$  \\
			& $\lambda_5$, $\lambda_{10}$   & $0.2$   & $0.225$  \\
			& $Y_{10}$   & --  & $0$  \\
			& $\widetilde{m}^2$   & $-1.2\times 10^6$  & $-6\times 10^5$ \\
			& $\widetilde{M}$   & $2$  & $0$   \\
			\hline
			\rule{0pt}{2.5ex}   & $\mathrm{tan} \beta$   & $40$   & $7.5$  \\
			& $\lam_{D,2}$, $\lam_{U,2}$   & $0.47$  & $0.59$  \\
			& $\lam_{Qu,2}$ / $\lam_{Qd,2}$    & -- & $0.56$ / $0.76$  \\
			& $\lam_{L,2}$, $\lam_{E,2}$   & $0.24$  & $0.31$  \\
			\small{\textbf{SUSY scale}}  & $\mu$   & $1680$   & $3100$  \\
			& $B_{\mu}$   & $5.5\times 10^4$ & $1\times 10^6$  \\
			& $M_1$   & $546$ & $377$   \\
			& $M_2$   & $984$  & $633$   \\
			& $M_3$   &  $2561$ & $1757$   \\
			& $M_D$, $M_U$   & $2125$  & $810$   \\
			& $M_L$, $M_E$   & $1352$  & $298$  \\
			& $\widetilde{M}_{D,2}$,  $\widetilde{M}_{U,3}$   & $3.5$  & $3.2$   \\
			\hline
			\rule{0pt}{2.5ex}   & $m_{h}$ & $124.4$  &  $126.2$   \\
                           & $m_{A,H}$ & $1084$  &  $2570$   \\
			& $m_{\chi_1^0}$ & $539$  &  $372$   \\
			& $m_{\charone}$ & $1013$  &  $675$   \\
			& $m_{\glu}$ & $2700$  &  $1990$   \\
			\small{\textbf{Pole Masses}} & $m_{\tilde{e}_1}$ & $651$  &  $374$   \\
			& $m_{\tilde{e}_2}$ & $704$  &  $930$   \\
			& $m_{\tilde{\nu}_1}$ & $710$  &  $1290$   \\
			& $m_{\tilde{t}_R}$ & $2130$  &  $2210$   \\
			& $m_{E}$ & $1370$  &  $302$   \\
			& $m_{B}$ & $2260$  &  $1210$   \\
			\hline
			\rule{0pt}{2.5ex}  & $\deltagmtwomu$   & $2.2 \times 10^{-9}$ & $1.8 \times 10^{-9}$   \\
			\small{\textbf{Low Energy} }  & $\brbutaunu$   & $1.28 \times 10^{-4}$ & $1.24 \times 10^{-4}$   \\
			& \abunchi   & $0.119$ & $0.113$   \\
			\hline
			\hline
		\end{tabular}
		\caption{Benchmark points for the models \mfivem\ and \mtenm. Dimensionful quantities are given in~GeV and $\gev^2$,
and $\epsilon=\alpha=0$.}
		\label{tab:benchm} 
	\end{center}
\end{table}
%%%%%%%%%%%%%%%%%%%%%%%%%%%%%%%%%%%%%%%%%%%%%%%%%%%%%%%%%%%%%%

\begin{figure}[t]
\centering
\subfloat[]{%
%\label{fig:c}%
\includegraphics[width=0.47\textwidth]{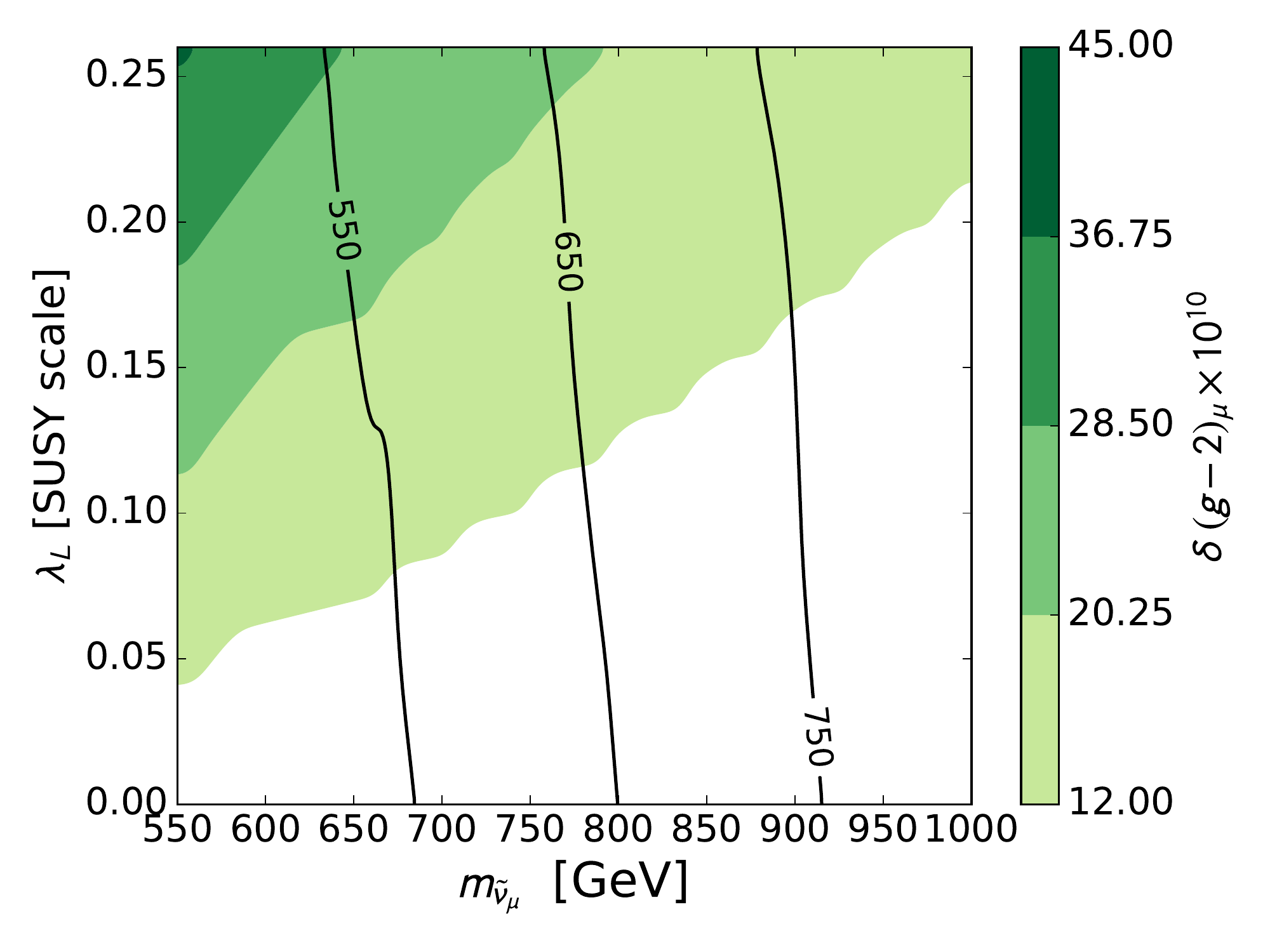}
}%
\hspace{0.02\textwidth}
\subfloat[]{%
%\label{fig:a}%
\includegraphics[width=0.47\textwidth]{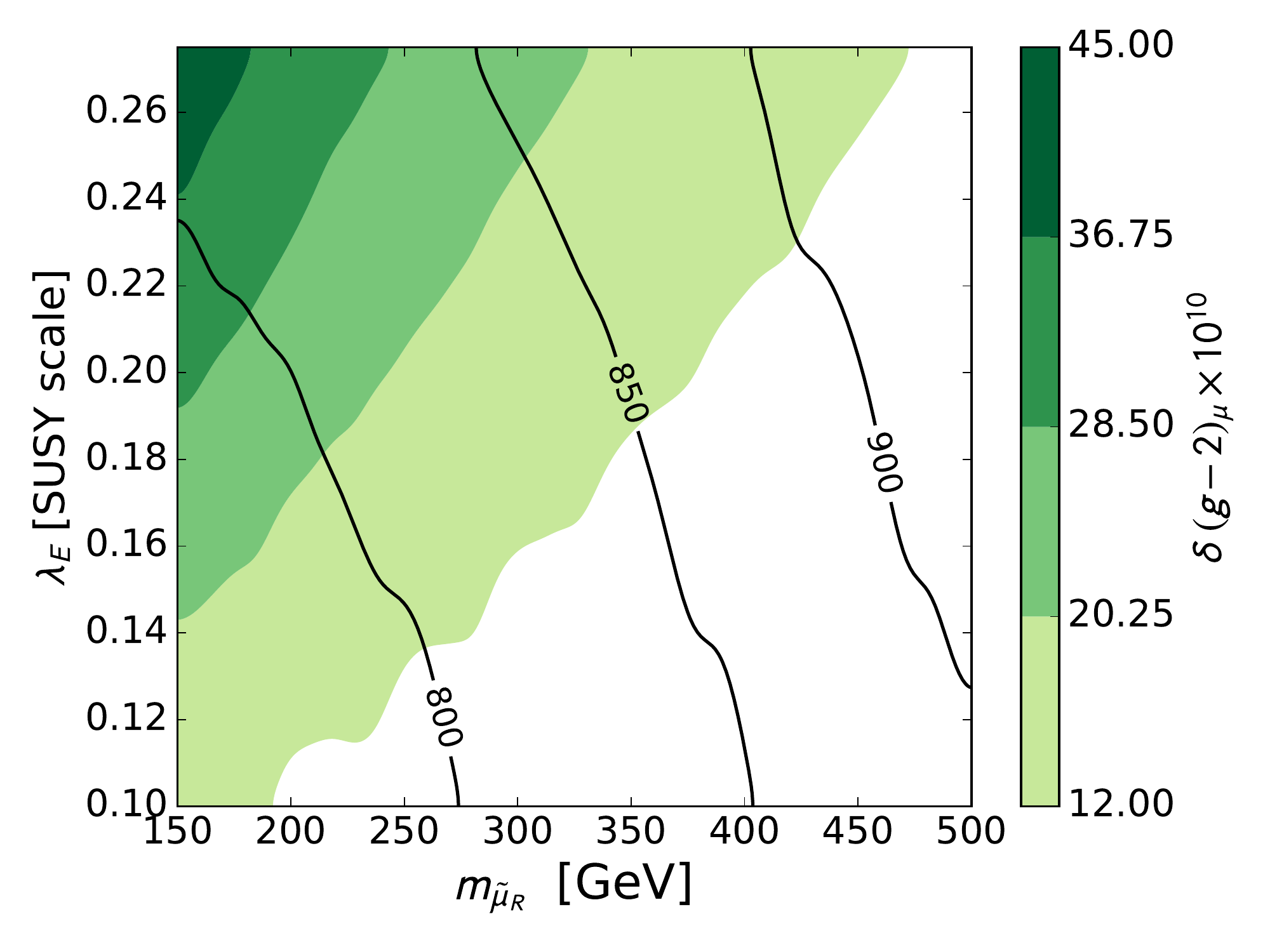}
}%
%}%
\caption{(a) \deltagmtwomu\ in the \mfivem\ model as a function of the pole mass of the first sneutrino eigenstate 
(a mixed muon sneutrino/VL sneutrino state) and of the Yukawa $\lam_L$ at the SUSY scale. 
The remaining parameters are chosen as in the benchmark point of \reftable{tab:benchm}. 
Black lines show contours of the lightest slepton mass. (b) 
\deltagmtwomu\ in the \mtenm\ model as a function of the pole mass of the first selectron eigenstate 
(a mixed smuon/ VL slepton state) and of the Yukawa $\lam_E$ at the SUSY scale. 
Black solid lines show contours of the second-lightest slepton mass.}
\label{fig:gm2m12}
\end{figure}

The parametric dependence of \deltagmtwomu\ around the benchmark point of the \mfivem\ model
given in \reftable{tab:benchm} is presented in \reffig{fig:gm2m12}(a). One can easily read out how the size of the observable 
depends on the new Yukawa 
couplings and the sneutrino mass. The parametric dependence of \deltagmtwomu\ around the benchmark point of the \mtenm\ model is 
given in \reffig{fig:gm2m12}(b).

%%%%%%%%%%%%%%%%%%%%%%%%%%%%%%%%
%%%%%%%%%%%%%%%%%%%%%%%%%%%%%%%%

%%%%%%%%%%%%%%%%%%%%%%%%%%%%%%%%%%%%%%
\begin{figure}[b]
	\begin{center}
		\includegraphics[width=0.6\textwidth]{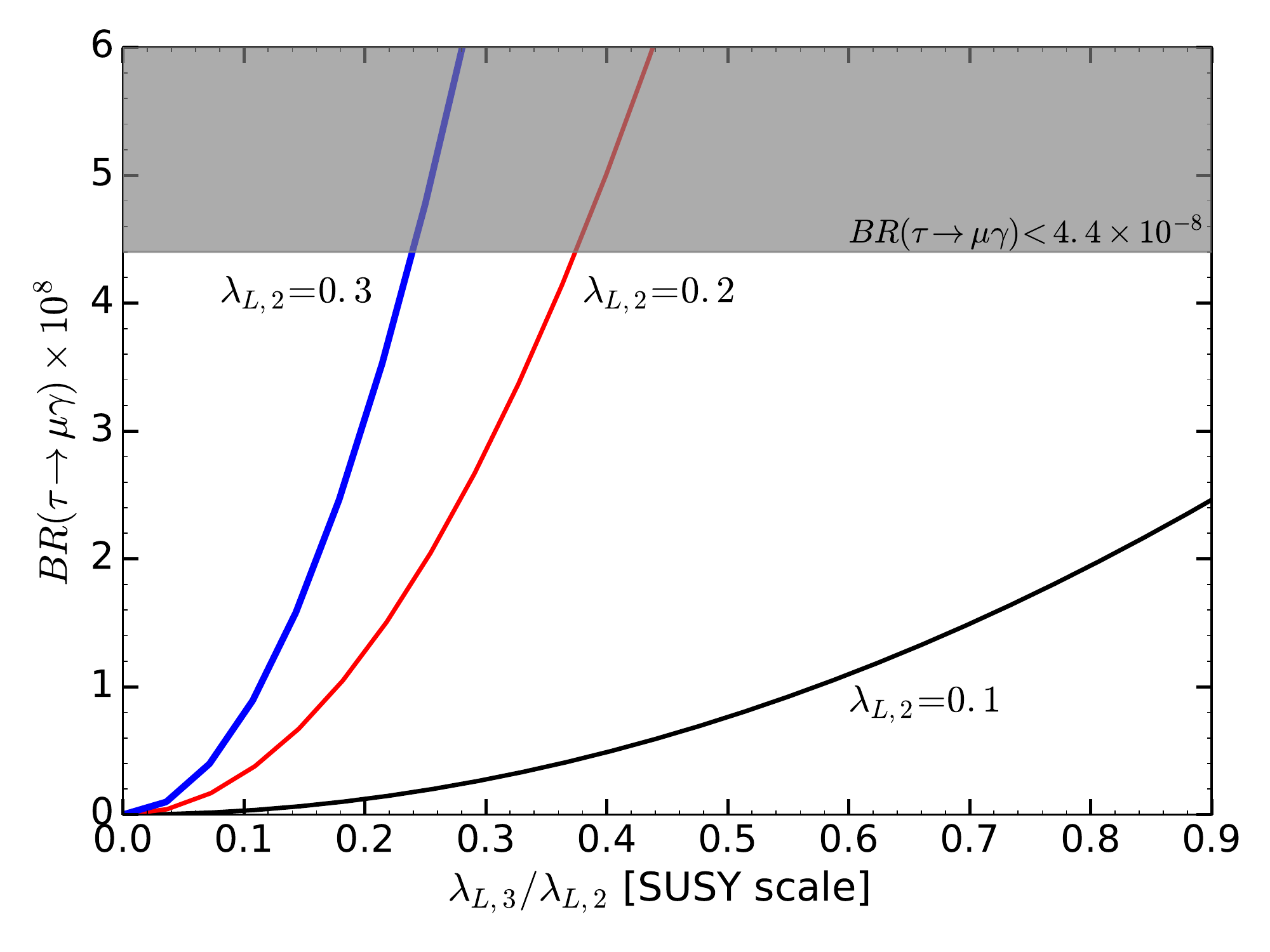}
		\caption{$\textrm{BR}(\tau \rightarrow \mu \gamma) \times 10^8$ as function of the ratio $\lambda_{L,3} / \lambda_{L,2}$ at the SUSY scale in the \mfivem\ model. From top to bottom, the lines correspond to $\lambda_{L,2} \approx 0.3$, 0.2, and 0.1 
(which leads to $\deltagmtwomu = 12 \times 10^{-10},\,18 \times 10^{-10}$, and $27 \times 10^{-10}$). 
All other parameters are fixed as in the benchmark point in \reftable{tab:benchm}.
%All other parameters are taken from the benchmark point of Table~\ref{tab:benchm}.
}
		\label{fig:epsgm2}
	\end{center}
\end{figure}
%%%%%%%%%%%%%%%%%%%%%%%%%%%%%%%%%%%%%%%%

While in the semi-analytical treatment of the previous sections we have often assumed that the mixing in the Yukawa and soft sectors 
only involve the second generation and VL particles, the scans include a coupling to the third generation, 
controlled by a small parameter $\epsilon$. While this parameter is not directly relevant for \deltagmtwomu\ 
it will affect for instance the collider phenomenology. The strongest constraint on this parameter comes from the flavor-violating decay, 
$\textrm{BR}(\tau \rightarrow \mu \gamma)< 4.4 \times 10^{-8}$. 
This is illustrated in \reffig{fig:epsgm2}, where we show the evolution of $BR(\tau \rightarrow \mu \gamma) \times 10^8$ 
as function of the ratio between third and second generation Yukawa couplings.

The \mtenm\ model points with a sizable \deltagmtwomu\ present an
interesting compressed spectrum, which can be seen in
\reftable{tab:benchm}. There is a bino-like neutralino, almost degenerate with a mixed
smuon/VL slepton and, with mass approximately twice their size, the first chargino. The rest
of the spectrum is heavier. A spectrum of this kind is likely to evade LHC
bounds due to the degeneracy between the slepton and neutralino, and simultaneously provides the
correct relic density (from smuon co-annihilation) and a good
\deltagmtwomu, as the smuon is relatively light. These
interesting properties come however at the expense of an
additional fine-tuning in the mass spectrum. We will therefore focus in the rest of the
paper on the more promising \mfivem\ model.

%%%%%%%%%%%%%%%%%%%%%%%%%%%%%%%%
\begin{figure}[t]
\centering
\subfloat[]{%
%\label{fig:c}%
\includegraphics[width=0.47\textwidth]{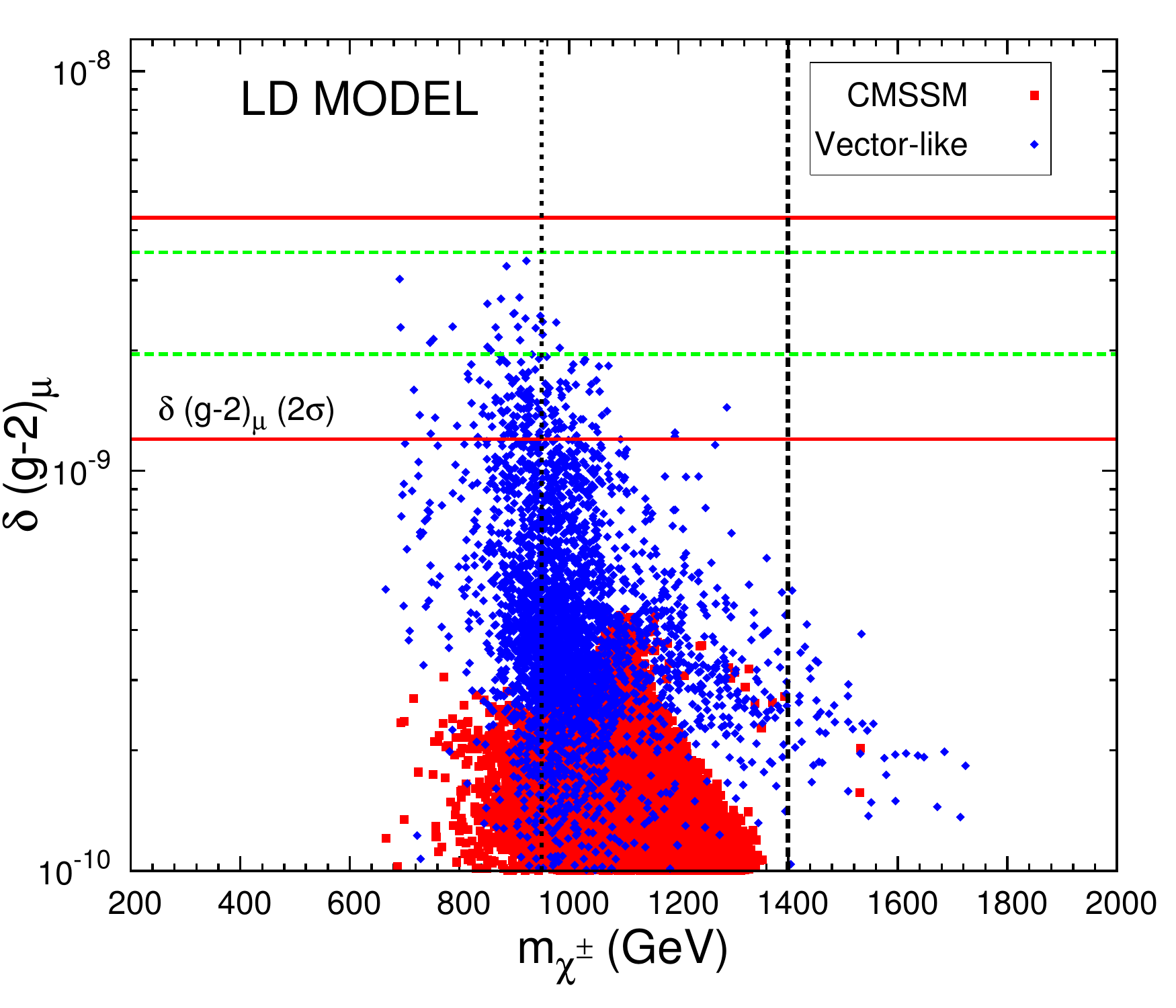}
}%
\hspace{0.02\textwidth}
\subfloat[]{%
%\label{fig:a}%
\includegraphics[width=0.47\textwidth]{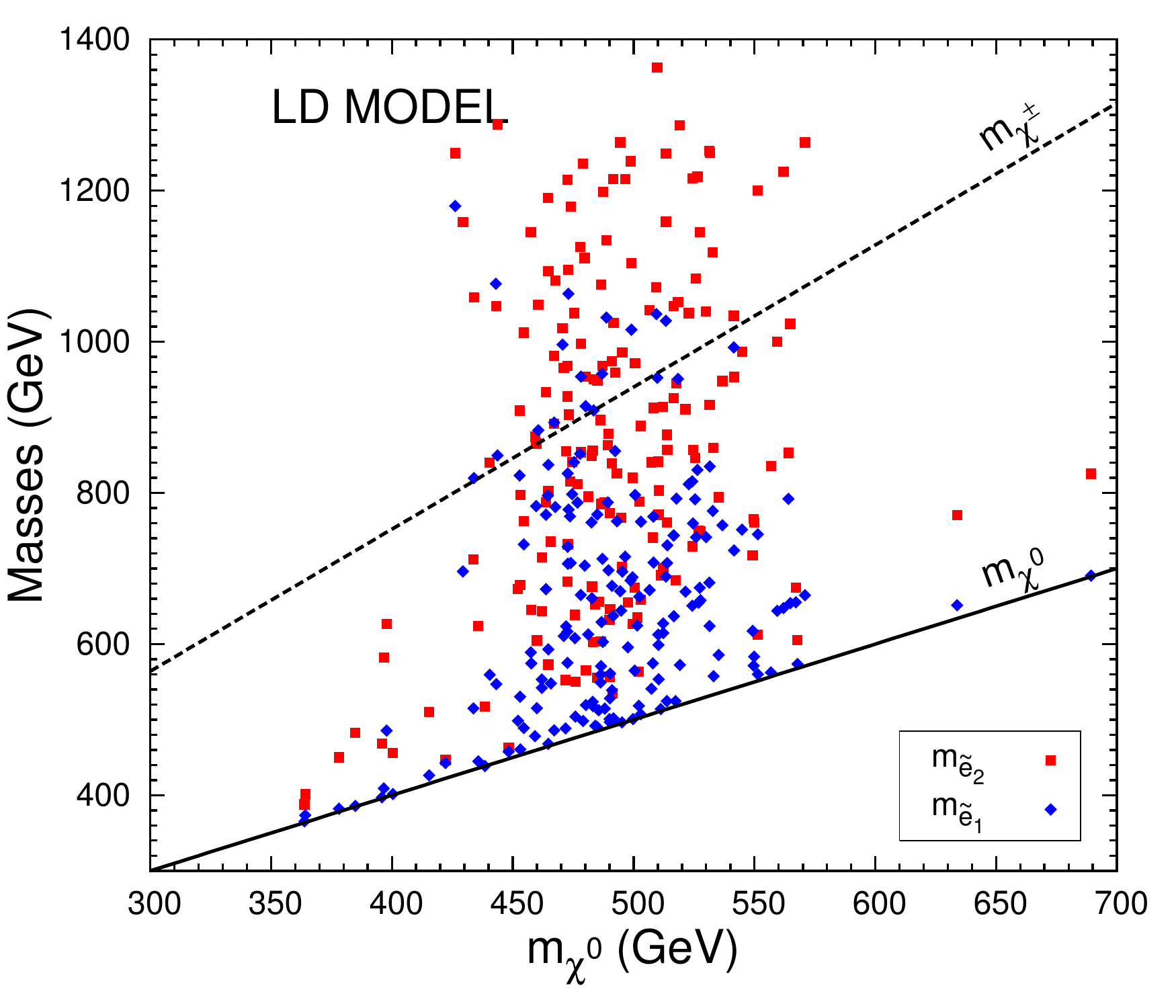}
}%
%\\
%}%
\caption{(a) Comparison of the calculation of \deltagmtwomu\ for the
  \mfivem\ model (blue diamonds) versus the CMSSM (red squares). The
  horizontal red solid lines indicate the current \deltagmtwomu\ $2\,\sigma$
  region, while the horizontal green dashed ones show the projected $2\,\sigma$
  region of the upcoming New Muon g-2 experiment, assuming the
  measured \deltagmtwomu\ remains unchanged. 
The vertical dashed lines show the current (thin) and projected (thick) bounds from 3-lepton 
searches\cite{ATLAS-CONF-2016-096,CMS:2016gvu} 
in the simplified-model interpretation. (b) Masses of the two
  lightest sleptons as function of the neutralino LSP mass. The thin
  dash line represents the mass of the first chargino.}
%The points can be tested by the LHC 13\tev. \es{add stuff!}}
\label{fig:gm2VL5scan}
\end{figure}
%%%%%%%%%%%%%%%%%%%%%%%%%%%%%%%%

In \reffig{fig:gm2VL5scan}(a) we show a plot of \deltagmtwomu\ versus the lightest chargino mass, $m_{\charone}$\,, for the points of 
the \mfivem\ model (blue diamonds).  The CMSSM case (red squares) is shown for comparison. 
One can clearly see the significant enhancement in \deltagmtwomu, which now allows one to easily find points that properly 
fit the experimental anomaly. For the points within $2\sigma$ of the \deltagmtwomu\ measurement, 
we show in \reffig{fig:gm2VL5scan}(b) the mass of the lightest slepton eigenstate (blue diamonds) and of the second slepton 
mass eigenstate (red squares).  
All the constraints of \reftable{tab:exp_constraints} are satisfied at the $3\sigma$ level in both plots, with the exception of the Higgs 
mass that is required to be within a $1\sigma\approx3\gev$ theoretical error.
The gluino mass lower bound\cite{ATLAS-CONF-2016-078} is satisfied for the points in the plot. 
We have also applied the bounds from direct LHC searches for VL quarks and leptons, which are also 
satisfied automatically by the points in the plot. The details of the latter bounds, along with corresponding projections 
for 14\tev, 300\invfb, are presented in Appendix~\ref{app:LHC}.

Because of the frequent presence of light sleptons of mass in between the chargino and the neutralino, 
the points in the figure are also subject to the most recent constraints from the ATLAS and CMS 
3-lepton searches for electroweakino pair production\cite{ATLAS-CONF-2016-096,CMS:2016gvu}. 
The thin dashed vertical line shows the current limit interpreted 
in the ``flavor-democratic'' simplified model with intermediate sleptons\cite{ATLAS-CONF-2016-096}. 
Note that the bound is not to be taken at face value.
We postpone a detailed LHC analysis, which requires a full numerical simulation, to future work, but
we point out here that we have checked several points characterized by $m_{\charone}\approx850\gev$, with an 
intermediate selectron at about $700\gev$, finding that in many cases the branching fraction of the chargino/neutralino decay 
chains are different from the simplified model considered by the experimental collaboration (for example, one often finds 
$\textrm{BR}( \neuttwo\rightarrow \tilde{\mu}\mu)\approx 50\%$ and $\textrm{BR}( \neuttwo \rightarrow \tilde{\tau}\tau)\approx 50\%$) so that the efficiency to the 3-lepton final state is reduced. Most of the points shown, even those below 900\gev, 
appear thus to be presently allowed, albeit some of them marginally.

On the other hand, the next round of data with increased luminosity is bound to deeply test the full 
parameter space that allows for a good \gmtwo\ fit.
We report in \reffig{fig:gm2VL5scan}(a), marked with a thick dashed line, the projected bound from 3-lepton searches in the 
flavor-democratic scenario at 14\tev\ and 300\invfb, which we take from Ref.\cite{Kowalska:2015zja}.
If the \gmtwo\ anomaly is real there will be unmistakable signatures at the LHC.
%%%%%%%%%%%%%%%%%%%%%%%%%%%%%%%%%%%%%%%%%%%%%%%%%%%%%%%%%

\bigskip
\textbf{Comments of flavor anomalies.} 
We conclude this subsection with some comments on the flavor observables. We have performed a survey of the values of the Wilson 
coefficients $C_7$, $C_8$, $C_9$, and $C_{10}$ in the \mfivem\ model. 
We observe for several points significant deviations from the SM in the coefficients $C_9$ and $C_{10}$, which 
may be useful to partially alleviate the current tensions between the SM predictions and the experimental measurements broadly related to the $b \rightarrow s$ transitions (see, e.g.,\cite{Altmannshofer:2014rta}). Indeed global fits for these two coefficients described in\cite{Altmannshofer:2014rta,DescotesGenon:2015uva} report best fit points around $(C_{9}^{\textrm{NP}},C_{10}^{\textrm{NP}}) \approx (-1.0,0.3)$, with the $2\sigma$ region extending for the former over the range $[-1.5,-0.3]$.
A certain number of scan points satisfying the constraints of \reftable{tab:exp_constraints} show $C_{9}^{\textrm{NP}} (\approx -0.3 C_{10}^{\textrm{NP}}) \approx -0.4$, thus being placed within the $2\sigma$ region from global fits. While it is unlikely that our model can explain all of the anomalies, it can reduce the pull compared to the SM. 
We point out, though, that we do not notice a significant correlation with the region of
parameter space that leads to good \deltagmtwomu\ and for this reason we refrain from further investigating this direction here.

%%%%%%%%%%%%%%%%%%%%%%%%%%%%%%%%%%%%%%%%%%%%%%%%%%%%%%%%%%%%%%%%%%%%%%%%%%%%%%
\subsection{Dark matter and direct detection\label{sec:dm}}
%%%%%%%%%%%%%%%%%%%%%%%%%%%%%%%%%%%%%%%%%%%%%%%%%%%%%%%%%%%%%%%%%%%%%%%%%%%%%%%

One of the most important phenomenological features of SUSY models is that they provide a natural DM candidate, 
which is typically the lightest neutralino. The DM relic density plays a crucial role in determining the allowed parameter space of such models since this constraint can be satisfied only if specific conditions characterizing the mass spectrum are met. By adding new VL 
fields to the model, we can modify the MSSM picture in basically two ways. One is by changing the position of the known regions in the parameter space in which the correct value of the DM relic density can be obtained, \textsl{e.g.}, due to modified RGE 
of the mass parameters from the GUT scale to the EW scale (see, \textsl{e.g.},\cite{Moroi:2011aa}). The other possibility is that new particles appearing in the model will be involved in additional annihilation channels for the lightest neutralino which can even open up new regions of the parameter space (see, \textsl{e.g.},\cite{Abdullah:2015zta,Abdullah:2016avr}).

As we focus on GUT-constrained scenarios, it is useful to compare our results with the ones obtained for the prototypical model of this kind, 
the CMSSM (see, \textsl{e.g.},\cite{Roszkowski:2014wqa} for an extensive discussion). 
The correct value of the DM relic density in the CMSSM for the region of the parameter space with bino-like neutralino (we do not treat here the promising region characterized by a higgsino-like neutralino with $\sim1\tev$ mass, as it requires a roughly SM-like value of \gmtwo) features either an approximate mass degeneracy between the lightest neutralino and the lightest slepton (slepton coannihilation), 
or the resonance condition for the $s$-wave pseudoscalar Higgs boson, $A$ ($A$-funnel region). 
 
Both the slepton-coannihilation and $A$-funnel regions are present in the \mfivem\ and \mtenm\ models that we analyze. 
The slepton-coannihilation region contains both CMSSM-like points characterized by a close mass degeneracy between the neutralino and the lightest stau, as well as points in which the lightest slepton is a mixture of an MSSM-like left chiral smuon and a VL slepton.
For the latter points the lightest neutralino is also mass-degenerate with the lightest sneutrino which, 
being lighter than the lightest charged slepton, can play the dominant role in the coannihilation mechanism responsible for reducing the
otherwise too large relic abundance of the bino-like neutralino. 

The slepton-coannihilation region in the \mtenm\ model is extended to contain points with larger mass difference between the neutralino 
and the lightest slepton (up to $\sim 160$ GeV) for which the correct value of the DM relic density is achieved partly thanks to coannihilations and partly due to efficient annihilations of the bino-like neutralino into the heavy VL leptons that avoid chirality 
suppression\cite{Abdullah:2015zta}. This effect is less pronounced in the~\mfivem~model since in the~\mtenm~model such annihilations are hypercharge enhanced for weak-isosinglet leptons.

\renewcommand{\arraystretch}{1.3}
\begin{table}[!t]
   \centering\footnotesize
   %\topcaption{Table captions are better up top} % requires the topcapt package
   \begin{tabular}{|c|c|c|c|} % Column formatting, @{} suppresses leading/trailing space
      \hline
      \textbf{ } & \textbf{ } & \textbf{SC} & \textbf{ } \\
      \textbf{Region} & \textbf{SC} & \textbf{$+$} & \textbf{AF} \\
      \textbf{ } & \textbf{ } & \textbf{bino ann.} & \textbf{ } \\
      \hline
      \textbf{Model} & \underline{\mfivem} / \mtenm & \mtenm & \underline{\mfivem} / \mtenm\\
      \hline
      \hline
      \mchi & $469.6$ GeV & $367.8$ GeV & $541.4$ GeV\\
%      \hline
      $m_{\neuttwo}\simeq m_{\charone}$ & $886$ GeV & $670$ GeV & $1013.7$ GeV\\      
%      \hline
      $\mlone$ & $479.6$ GeV & $455.8$ GeV & $992.4$ GeV \\      
%      \hline
      $m_{\tilde{\nu}_1}$ & $470.0$ GeV & $1163.7$ GeV & $988.0$ GeV\\      
%      \hline
      \ma & $1334.3$ GeV & $1543.1$ GeV & $1082.0$ GeV\\      
      \hline
%      \hline
      \abunchi & $0.123$ & $0.156$ & $0.128$\\
%      \hline
      \sigsip\ [$\textrm{cm}^2$] & $1.5\times 10^{-47}$ & $3.1\times 10^{-47}$ & $1.6\times 10^{-47}$\\
%      \hline
      \deltagmtwomu\ & $2.1\times 10^{-9}$ & $1.6\times 10^{-9}$ & $1.2\times 10^{-9}$\\
%      \hline
      \mh & $123.9$ GeV & $123.8$ GeV & $123.1$ GeV\\      
       \hline   
   \end{tabular}
 \caption{
Benchmark points for regions with the correct value of the DM relic density: the slepton-coannihilation (SC), 
$A$-funnel (AF) and a region of slepton-coannihilation with additional efficient bino annihilation into 
$4th$ generation leptons (SC+bino ann.). The model in which a given region is present in the favored parameter space is also denoted. 
In case of regions present in both models the underline indicates the model from which the benchmark point was taken. 
Note that the second benchmark point is most probably excluded by 3-lepton searches.}
\label{Tab:benchmarkpointsDM}
\end{table}

As was discussed in \refsec{sec:higgs}, in the \mfivem\ model the pseudoscalar Higgs mass, \ma, 
is sensitive to the value of the additional Yukawa coupling $\lambda_D$, giving more freedom to find points that fit the $A$-funnel 
condition $\ma\approx 2\,\mchi$. The last effect is not present in the \mtenm\ model which is, in addition, characterized by larger loop corrections to the Higgs boson mass that can easily lead to too large \mh. As a result, in the \mtenm\ model the allowed DM parameter space  in the (\mzero, \mhalf) plane is overall shrunk with respect to the \mfivem\ model.

The current and future direct detection limits introduce another constraint on the allowed regions of the parameter space. The actual value of the spin-independent scattering cross section, $\sigma_p^{\textrm{SI}}$, depends on how large is the bino-higgsino mixing of the lightest neutralino. In particular, in the slepton-coannihilation and $A$-funnel regions such a mixing is typically very small so that 
one easily satisfies the recent LUX exclusion bounds\cite{Akerib:2016vxi}. This scenario is also often beyond the reach of the Xenon1T experiment\cite{Aprile:2015uzo}, however, may be probed, \textsl{e.g.}, in its several-tonne extension Xenon-nT. 

In Table~\ref{Tab:benchmarkpointsDM} we present 3 benchmark points for the scenarios described above, 
in which the correct DM relic density can be obtained. For each point we present the masses of the particles relevant for the 
discussion of neutralino relic abundance, \textsl{i.e.}, \mchi, \ma, the mass of the lightest charged slepton, \mlone, 
and the lightest sneutrino, $m_{\tilde{\nu}_1}$, the mass of the second neutralino/lightest chargino, $m_{\neuttwo}/m_{\charone}$, 
as well as the basic observables.

%%%%%%%%%%%%%%%%%%%%%%%%%%%%%%%%%%%%%%%%%%%%%%%%%%
\section{Summary and conclusions\label{sec:summary}}
%%%%%%%%%%%%%%%%%%%%%%%%%%%%%%%%%%%%%%%%%%%%%%%%%%%

We have analyzed in this work two, minimal, supersymmetric models with vector-like matter: the \mfivem\ model, 
where the MSSM is enriched with one pair $\mathbf{5}+\mathbf{\bar{5}}$ of multiplets of $SU(5)$, and the \mtenm\ model, with instead one pair $\mathbf{10}+\mathbf{\overline{10}}$. Driven by minimality, we did not include any extra symmetry to prevent mixing 
between the new VL leptons and the SM ones, and did not consider additional singlets. 
Furthermore, we have imposed universal boundary conditions at GUT scale, thereby maintaining a relatively low number of parameters.

Our key finding is that, unlike the usual MSSM under similar constraints, these two models can accommodate the \deltagmtwomu\ measurement, while satisfying a large number of requirements. More precisely, we have imposed perturbativity of our couplings up to GUT scale, required physicality of our mass spectrum, confronted the models with various EW and flavor precision tests, and 
applied bounds from direct searches for SUSY particles. We have additionally ensured that one can find a DM candidate with the correct 
relic density and avoid bounds from direct detection experiments. Note that it was not \textsl{a priori} guaranteed that in the phenomenologically-driven extensions of the CMSSM that we discuss one could accommodate both the \deltagmtwomu\ measurement and these constraints since, given the minimality of GUT-constrained models, the modifications that we introduce have an impact on both physicality and many observational constraints, \textsl{e.g.}, via modified RGE running.

Enhancing the Higgs boson mass has been in the last few years one of the top reasons for introducing in SUSY models new 
colored VL matter. However, we showed in \refsec{sec:higgs} that additional colored fields, as found in the $\mathbf{10}+\mathbf{\overline{10}}$ multiplet of the \mtenm\ model, can also make the Higgs boson too heavy in broad regions of the parameter space, particularly once the current LHC bounds are taken into account. 
Parameter space in good agreement with the experimental value for the Higgs mass can nonetheless be easily found, especially if the gluino
is found just above the current LHC bounds. 

As pertains to \gmtwo, while most of the good points in the parameter space currently escape LHC bounds from direct 
electroweakino searches, the entire viable parameter space will be probed by the end of LHC 14\tev\ run. 
In case the \deltagmtwomu\ measurements is confirmed in the next few years, a more complete analysis of the collider constraints in the precise case of our models will be crucial.

\bigskip
%%%%%%%%%%%%%%%%%%%%%%%%%%%%%%%%%%%%%%%%%%%%%%%%%%%%%%%%%%%%%%%%%%%%%%%%%%%%%%%%
\noindent \textbf{Acknowledgments}
\medskip

\noindent We would like to thank K.~Kowalska for discussions and inputs on the most recent LHC bounds. AC would like to thank S. Mondal for helpful discussions. LD and EMS would like to thank M.~Kazana for useful discussions. ST would like to thank S.~Iwamoto for helpful discussions.  AC and LR are supported by the Lancaster-Manchester-Sheffield Consortium for Fundamental Physics under STFC Grant No.\ ST/L000520/1. LD, LR, EMS and ST are supported in part by the National Science Council (NCN) research grant No.~2015-18-A-ST2-00748. ST is supported in part by the Polish Ministry of Science and Higher Education under research grant 1309/MOB/IV/2015/0 and by NSF Grant No. PHY-1620638. The use of the CIS computer cluster at the National Centre for Nuclear Research in Warsaw is gratefully acknowledged.
%%%%%%%%%%%%%%%%%%%%%%%%%%%%%%%%%%%%%%%%%%%%%%%%%%%%%%%%%%%%%%%%%%%%%%%%%%%%%%%%
\newpage
\appendix
%%%%%%%%%%%%%%%%%%%%%%%%%%
\section{Soft Lagrangian and mass matrices\label{app:soft}}
%%%%%%%%%%%%%%%%%%%%%%%%%%

\subsection*{The 5-plet LD model}

The superpotential of the model with a pair of VL $\mathbf{5}+\mathbf{\bar{5}}$ multiplets (\mfivem) is given in \refeq{superpot5}.
One can write down the soft terms
\bea
\mathcal{L}_{\textrm{soft}}&=&-\left[\tilde{q}^{\dag}\mathbf{m_q^2}\tilde{q}+\tilde{d}^{\dag}\mathbf{m_d^2}\tilde{d}
+\tilde{u}^{\dag}\mathbf{m_u^2}\tilde{u}+\tilde{l}^{\dag}\mathbf{m_l^2}\tilde{l}+\tilde{e}^{\dag}\mathbf{m_e^2}\tilde{e}\right.\nonumber\\
 & &\left.+m_{H_u}^2|H_u|^2+m_{H_d}^2|H_d|^2+m_L^2|\tilde{L}|^2+m_{L'}^2|\tilde{L}'|^2
+m_D^2|\tilde{D}|^2+m_{D'}^2|\tilde{D}'|^2\right.\nonumber\\
 & &\left.+\left(\widetilde{m}_L^2\tilde{l}^{\dag}\tilde{L}
+\widetilde{m}_D^2\tilde{d}^{\dag}\tilde{D}+\textrm{h.c.}\right)\right]
-\frac{1}{2}\left(M_1\,\bar{\lam}_1\lam_1+M_2\,\bar{\lam}_2\lam_2+M_3\,\bar{\lam}_3\lam_3\right)\nonumber\\
 & &+\left(\mathbf{T_u}\,\tilde{q}H_u\tilde{u}^{\dag}+\mathbf{T_d}\,\tilde{q}H_d\tilde{d}^{\dag}
+\mathbf{T_e}\,\tilde{l}H_d\tilde{e}^{\dag}+T_D\,\tilde{q}H_d\tilde{D}^{\dag}
+T_L\,\tilde{L}H_d\tilde{e}^{\dag}+\textrm{h.c.}\right)\nonumber\\
 & &-\left(B_{\mu}H_u H_d+B_{M_L}\tilde{L}\tilde{L}'+B_{\widetilde{M}_L}\tilde{l}\tilde{L}'
+B_{M_D}\tilde{D}\tilde{D}'+B_{\widetilde{M}_D}\tilde{d}\tilde{D}'+\textrm{h.c.}\right),\label{soft5full}
\eea
where the generation indices, as well as the $SU(2)$ indices, are considered as summed over and suppressed from the notation.

Using the notation $v_u=v\sin\beta$ and $v_d=v\cos\beta$ 
with $v=174\,\textrm{GeV}$, one can construct the quark mass matrices, which in the basis $\{(\bar{d}_L,\bar{s}_L,\bar{b}_L,\bar{D}'),\,(d_R,s_R,b_R,D)^{T}\}$ read
\[
   M_d=
  \left( {\begin{array}{cc}
 Y_d v_d & \lam_D v_d \\
 \widetilde{M}_D & M_D
 \end{array} } \right),
\]
and the charged lepton mass matrix, which in the basis
$\{(\bar{e}_L,\bar{\mu}_L,\bar{\tau}_L,\bar{E}_L),\,(e_R,\mu_R,\tau_R,E_R)^{T}\}$ is
\[
   M_e=
  \left( {\begin{array}{cc}
 Y_e v_d & -\widetilde{M}_L \\
 \lam_L v_d & -M_L 
 \end{array} } \right),
\]
where we have explicitly indicated the doublet $L$ as $L=(N_L, E_L)^T$ and $L'$ as $L'=(N_R, E_R)^T$. 

We use the lepton mass matrix above to give an explicit form of the tree-level mass of the muon and the VL lepton.
By using the simplified notation $\widetilde{M}_L\equiv\widetilde{M}_{L,2}$, $\lam_L\equiv\lam_{L,2}$ and defining
\begin{align}
 \overline{M}_L^2 = \widetilde{M}_L^2 + M_L^2+ (Y_{e,22}^2 +\lam_L^2 ) v_d^2
\end{align}
one can write
\begin{equation}
m_{e^{\pm}_2,e^{\pm}_4}=\frac{1}{\sqrt{2}}\left(\overline{M}_L^2 
\mp\sqrt{\overline{M}_L^4-4 v_d^2 \left(Y_{e,22}^2 M_L-\lam_D \widetilde{M}_L\right)^2}~\right)^{1/2}.\label{muon5}
\end{equation}
Analogous formulas apply to the leptons of the other generations and to the quarks.

For completeness we also write down the mass matrix of the smuons in the 
$(\tilde{\mu}_L,\tilde{\mu}_R,\tilde{E}_L,\tilde{E}_R)$ basis, and under the assumption of only second-generation mixing with VL
matter, i.e., $\widetilde{M}_L\equiv\widetilde{M}_{L,2}$, $\widetilde{m}_L^2\equiv\widetilde{m}^2_{L,2}$, $T_L\equiv T_{L,2}$, and $B_{\widetilde{M}_L}\equiv B_{\widetilde{M}_{L,2}}$. 
For compactness, we neglect all the terms proportional to the gauge and Yukawa couplings, except for the new VL Yukawas. We get
 \begin{multline}
 M^2_{\tilde{\mu}}=\\
  \left( {\begin{array}{cccc}
 m_{\tilde{\mu}_L}^2+\widetilde{M}_L^2+m_{\mu}^2 & m_{\mu}( A_{\mu}-\mu\tanb)  & M_L\widetilde{M}_L+\widetilde{m}_L^2+\lam_L v_d m_{\mu} & -B_{\widetilde{M}_L} \\[0.5em]
 m_{\mu}( A_{\mu}-\mu\tanb)  & m_{\tilde{\mu}_R}^2+\lam_L^2 v_d^2+m_{\mu}^2 & \lam_{L}v_d(A_L-\mu\tanb ) &  -\lam_L v_d M_L-m_{\mu} \widetilde{M}_L\\[0.5em] 
  M_L\widetilde{M}_L+\widetilde{m}_L^2+\lam_L v_d m_{\mu} & \lam_{L}v_d(A_L-\mu\tanb) & M_L^2+m_L^2+v_d^2 |\lam_L|^2 & -B_{M_L}\\[0.5em]
-B_{\widetilde{M}_L} & -\lam_L v_d M_L-m_{\mu} \widetilde{M}_L & -B_{M_L} & M_L^2+\widetilde{M}_L^2+m_{L'}^2
 \end{array} } \right),\label{eq:smumass}
\end{multline}
where we have used the tree-level mass of the muon $m_\mu=v_d Y_{e,22}$, 
$A_\mu = T_{e,22}/Y_{e,22}$ and similarly $A_L = T_L / \lam_L$.

It is also useful to explicitly write down the mass matrix of the muon sneutrinos, under the assumption of second generation/VL mixing. 
In the basis $(\tilde{\nu}_{\mu},\tilde{N}_L,\tilde{N}_R)$
 \begin{equation}
M^2_{\tilde{\nu}_{\mu}}=
  \left( {\begin{array}{ccc}
m_{\tilde{\nu}_{\mu}}^2+\widetilde{M}_L^2  & M_L\widetilde{M}_L+\widetilde{m}_L^2  & B_{\widetilde{M}_L} \\[0.5em]
M_L\widetilde{M}_L+\widetilde{m}_L^2 &  M_L^2+m_L^2 & B_{M_L} \\[0.5em] 
B_{\widetilde{M}_L} & B_{M_L} & M_L^2+\widetilde{M}_L^2+m_{L'}^2  
 \end{array} } \right).\label{eq:snumass}
\end{equation}
 
%%%%%%%%%%%%%%%%%%%%%%%%%%%%%%%%%%%%%%%%%%%%%
\subsection*{The 10-plet QUE model}

The superpotential of the \mtenm\ model, in which we add a pair of VL fields $\mathbf{10}+\mathbf{\overline{10}}$ to the MSSM, 
is given in \refeq{superpot10}.
The soft terms of the MSSM fields have the same form as in \refeq{soft5full}. 
The additional soft terms proper of the VL fields are in this case:
\bea
\mathcal{L}_{\textrm{soft}}&=&-\left[m_Q^2|\tilde{Q}|^2+m_{Q'}^2|\tilde{Q}'|^2
+m_U^2|\tilde{U}|^2+m_{U'}^2|\tilde{U}'|^2+m_E^2|\tilde{E}|^2+m_{E'}^2|\tilde{E}'|^2\right.\nonumber\\
 & &\left.+\left(\widetilde{m}_Q^2\tilde{q}^{\dag}\tilde{Q}
+\widetilde{m}_U^2\tilde{u}^{\dag}\tilde{U}+\widetilde{m}_E^2\tilde{e}^{\dag}\tilde{E}+\textrm{h.c.}\right)\right]\nonumber\\
 & &+\left(T_{Qu}\,\tilde{Q}H_u\tilde{u}^{\dag}+T_{Qd}\,\tilde{Q}H_d\tilde{d}^{\dag}+T_U\,\tilde{q}H_u\tilde{U}^{\dag}
+T_{E}\,\tilde{l}H_d\tilde{E}^{\dag}+T_{10}\,\tilde{Q}H_u\tilde{U}^{\dag}+T_{10}'\,\tilde{Q}'H_d\tilde{U}'^{\dag}
\right.\nonumber\\
 & &-\left.B_{M_Q}\tilde{Q}\tilde{Q}'-B_{\widetilde{M}_Q}\tilde{q}\tilde{Q}'
-B_{M_U}\tilde{U}\tilde{U}'-B_{\widetilde{M}_U}\tilde{u}\tilde{U}'-B_{M_E}\tilde{E}\tilde{E}'
-B_{\widetilde{M}_E}\tilde{e}\tilde{E}'+\textrm{h.c.}\right).\label{soft10}
\eea 

Similarly one can construct the fermion and scalar mass matrices, as was done for the \mfivem\ model. The extra quarks and leptons also mix with their SM counterparts. 

Let us write down in particular the mass matrix for the five up-type quarks in the basis $\{(\bar{u}_L,\bar{c}_L,\bar{t}_L,\bar{T},\bar{U}'),\,(u_R,c_R,t_R,U,T')^{T}\}$:
 \be
 M_u=
  \left( {\begin{array}{ccc}
 Y_u v_u & \lam_U v_u & \widetilde{M}_Q \\
 \lam_{Qu} v_u & Y_{10} v_u & M_Q \\ 
 \widetilde{M}_U  & M_U & Y_{10}' v_d   
 \end{array} } \right),
\ee
where we have explicitly written the doublets as $Q=(T,B)^T$ and $Q'=(T',B')^T$. 

The smuon mixing matrix in the $(\tilde{\mu}_L,\tilde{\mu}_R,\tilde{E}',\tilde{E})$ basis reads
\begin{multline}
 M^2_{\tilde{\mu}}=\\
  \left( {\begin{array}{cccc}
 m_{\tilde{\mu}_L}^2+\lam_E^2 v_d^2+m_{\mu}^2 & m_{\mu}( A_{\mu}-\mu\tanb)  & \lam_{E}v_d(A_E-\mu\tanb ) & \lam_E v_d M_E+m_{\mu} \widetilde{M}_E \\[0.5em]
 m_{\mu}( A_{\mu}-\mu\tanb)  & m_{\tilde{\mu}_R}^2+\widetilde{M}_E^2+m_{\mu}^2 & M_E\widetilde{M}_E+\widetilde{m}_E^2+\lam_E v_d m_{\mu} &  B_{\widetilde{M}_E} \\[0.5em] 
  \lam_{E}v_d(A_E-\mu\tanb ) & M_E\widetilde{M}_E+\widetilde{m}_E^2+\lam_E v_d m_{\mu} & M_E^2+m_E^2+v_d^2 |\lam_E|^2 & B_{M_E}\\[0.5em]
\lam_E v_d M_E+m_{\mu} \widetilde{M}_E & B_{\widetilde{M}_E} & B_{M_E} & M_E^2+\widetilde{M}_E^2+m_{E'}^2
 \end{array} } \right).\label{eq:smumass2}
\end{multline}

%%%%%%%%%%%%%%%%%%%%%%%%%%%%%%%%%%%%%%%%%%%%%
\section{Leptonic rotation matrices and electroweak precision observables\label{app:prec}}
%%%%%%%%%%%%%%%%%%%%%%%%%%%%%%%%%%%%%%%%%%%%%%

We briefly investigate here the consequences of mixing in the leptonic sector. 
We define $c_W $ and $s_W$ as the sine and cosine of the Weinberg angle, and use the rotation matrices $L^{E}$, $R^{E}$, $L^{N}$, 
and $R^{N}$, such that
\begin{align*}
L^{E}  \mathbf{M_e} R^{E\dagger}=\textrm{diag}(m_{e_1},\dots) &&
L^{N}  \mathbf{M_N} R^{N\dagger}=\textrm{diag}(m_{\nu_1},\dots).
\end{align*}

Since the lepton mass eigenstates $e_2$ and $\nu_2$ now contain a fraction of VL lepton, their gauge coupling to $Z$ and $W$ bosons are
\begin{align}
 \scr{L} \supset  Z_{\mu} \bar{e}_{2} \gamma^\mu (P_L \glzmu + P_R \grzmu) e_2 + 
\left[ W_\mu \bar{\nu}_{2} \gamma^\mu P_L \glwmu  e_2 + \textrm{h.c.} \right].
\end{align}
We define 
\begin{align}
 \delta\glwmu\equiv\glwmu - \glwmuSM ~&=~ \begin{cases}  \displaystyle \frac{g}{2\,c_W} (-1+L^{N}_{22} L^{E\dagger}_{22} + L^{N}_{24} L^{E\dagger}_{42})    &(\mfivem) \\[0.5em] \displaystyle \frac{g}{2\,c_W}  L^{E\dagger}_{22}  & (\mtenm)\end{cases}\label{app2_a}
\end{align}
and 
\begin{align}
\delta\glzmu\equiv \glzmu - \glzmuSM~&=~ \begin{cases} 0   &(\mfivem) \\ \displaystyle \frac{g}{2 c_W} |L^E_{24}|^2  & (\mtenm)\end{cases}\\
\delta\grzmu\equiv \grzmu - \grzmuSM~&=~ \begin{cases} \displaystyle -\frac{g}{2 c_W} |L^R_{24}|^2   &(\mfivem) \\  0 & (\mtenm)\end{cases}.\label{app2_b}
\end{align}
The SM contributions are, as usual,
\begin{align}
 \glzmuSM = \displaystyle \frac{g}{2 c_W} (s_W^2-1/2) , \qquad   \grzmuSM = \displaystyle \frac{g }{2 c_W} s_W^2 , \qquad \  \glwmuSM = \displaystyle \frac{g }{\sqrt{2}}\,.
\end{align}

In the limit of \refeq{eq:masshierc}, we can write explicitly the form of the mixing matrices 
used in Eqs.~(\ref{app2_a})-(\ref{app2_b}), up to a normalization factor. In the \mfivem\ model we have
\begin{align}
  L^{E}  &\sim \begin{pmatrix}
1 && -\frac{\widetilde{M}_L}{M_L}  \\ - \frac{\widetilde{M}_L}{M_L} && -1
\end{pmatrix}
 &&&  R^{E}  &\sim  \begin{pmatrix}
1 && \frac{\lambda_L v_d}{M_L} \\ - \frac{\lambda_L v_d}{M_L} && 1
\end{pmatrix}
\end{align}
and 
\begin{align}
L^{N}  &\sim  \begin{pmatrix}
1 && -\frac{\widetilde{M}_L}{M_L}  \\ - \frac{\widetilde{M}_L}{M_L} && -1
\end{pmatrix}
&&&  R^{N}  &\sim  \begin{pmatrix}
1 && 0 \\ 0&& 1
\end{pmatrix} \ .
\end{align}

Equation~(\ref{app2_a}) then becomes in the \mfivem\ model
\begin{align}
\delta\glwmu &=  \displaystyle \frac{g}{2\,c_W} \left(-1 +  L^{N\dagger}_{22} L^{E}_{22} + L^{N}_{24} L^{E}_{24}\right)\\
			&\approx \displaystyle \frac{g}{2\,c_W} \left[1 - \left(1- \frac{L^{N\,2}_{24}}{2}\right)
\left(1-  \frac{L^{E\,2}_{24}}{2}\right)+ L^{N}_{24}\,L^{E}_{24}  \right]\\
			&= \displaystyle \frac{g}{2\,c_W} \left[ - \frac{1}{2} \left(L^{N}_{24} - L^{E}_{24} \right)^2\right] \\
			&\approx 0  \ ,
\end{align} 
where the second equality follows from the unitarity of $L^N$ and $L^E$, while the last line holds up to terms of the fourth order in 
$\lambda_L v_d/M_L$ and $ \widetilde{M}_L/M_L$\,. 
This interesting cancellation arises since, at the leading order, the mixing between left-handed neutrinos and VL neutrinos, 
and between left-handed leptons and VL leptons proceeds through the same superpotential mixing term $\widetilde{M}_L$, and leads to 
the mild constraint that follows \refeq{prec_bound2}.

In the \mtenm~model, we have instead no new neutrinos and
\begin{align}
  L^{E}  &\sim \begin{pmatrix}
1 && -\frac{\widetilde{M}_E}{M_E}  \\ - \frac{\widetilde{M}_E}{M_E} && -1
\end{pmatrix}
 &&&  R^{E}  &\sim  \begin{pmatrix}
1 && - \frac{\lambda_L v_E}{M_E} \\ \frac{\lambda_L v_E}{M_E} && 1
\end{pmatrix}
\end{align}
can be used to derive the bounds of Eqs.~(\ref{prec_bound}) and (\ref{prec_bound2}).

%%%%%%%%%%%%%%%%%%%%%%%%%%%%%%%%%%%%%%%%
\section{Approximate formulas for {\boldmath \gmtwo}\label{app:gm2}}
%%%%%%%%%%%%%%%%%%%%%%%%%%%%%%%%%%%%%%%%%

We derive in this appendix some of the formulas in \refsec{gm2analytic}. 

Our starting point will be \refeq{eq:amuchar}. 
The explicit form of the couplings 
$c_{jk}^L$, and $c_{jk}^R$ is 
\bea
c_{jk}^L&=&-g_2\,V_{k1} Y_{1j}\\
c_{jk}^R&=&\left(y_{\mu} Y_{1j}+\lam_L Y_{2j}\right)U_{k2},
\eea
where the equations above are expressed in terms of the eigenvectors of the chargino and sneutrino 
mass matrices. 
One has (we limit ourselves to real parameters)
\bea
UM_{\chi^{\pm}}V^T&=&\textrm{diag}(m_{\chi_1^{\pm}},m_{\chi_2^{\pm}})\\
YM^2_{\tilde{\nu}}Y^T&=&\textrm{diag}(m_{\tilde{\nu}^2_1},m_{\tilde{\nu}^2_2},m_{\tilde{\nu}^2_3}).
\eea
For the chargino mass matrix we follow the convention of\cite{Martin:2001st}, 
whereas the sneutrino mass squared matrix is given in \refeq{eq:snumass}.

We can now derive the explicit form of the dominant chargino/sneutrino contribution.
We assume that all $B\mu$ terms in \refeq{eq:snumass} are negligible, 
so that effectively we just need to diagonalize the upper left $2\times 2$ minor of the sneutrino mass matrix or, in other words, 
we only consider that the 2 lightest sneutrinos produce the dominant contributions.
Then,
\begin{multline} 
\Delta a_{\mu}^{\chi^{\pm}}\approx \frac{m_{\mu}}{24\,\pi^2}\left[M_2 \left(c_{11}^L c_{11}^R \frac{\mathcal{F}_C(z_{11})}{m_{\tilde{\nu}_{\mu,1}}^2}+c_{21}^L c_{21}^R \frac{\mathcal{F}_C(z_{21})}{m_{\tilde{\nu}_{\mu,2}}^2}\right)\right.\\
\left.+\mu \left(c_{12}^L c_{12}^R \frac{\mathcal{F}_C(z_{12})}{m_{\tilde{\nu}_{\mu,1}}^2}+c_{22}^L c_{22}^R \frac{\mathcal{F}_C(z_{22})}{m_{\tilde{\nu}_{\mu,2}}^2}\right)\right].\label{app1}
\end{multline}

From the explicit form of the chargino mass matrix\cite{Martin:2001st} one can see 
\be
M_2 V_{11} U_{12}\equiv-\frac{\mu\,M_2\sqrt{2}M_W\sin\beta}{m_{\chartwo}^2-m_{\charone}^2}=-\mu V_{21} U_{22},\label{cha_rot}
\ee
which, in the limit of one sneutrino and $\lam_L=0$, leads to \refeq{eq:bino_char}.

Note that from \refeq{cha_rot} approximate relations follow:
\be
M_2\,c_{11}^L c_{11}^R\approx \left(-M_2 V_{11} U_{12}\right)g_2 \left(y_{\mu}+Y_{11}Y_{21}\lam_{L}\right)\approx
\left(\mu V_{21} U_{22}\right)g_2 \left(y_{\mu}+Y_{11}Y_{21}\lam_{L}\right)\approx-\mu\,c_{12}^Lc_{12}^R\label{app2}
\ee
and
\be
M_2\,c_{21}^L c_{21}^R\approx \left(-M_2 V_{11} U_{12}\right)g_2 Y_{12}Y_{22}\lam_{L}\approx
\left(\mu V_{21} U_{22}\right)g_2 Y_{12}Y_{22}\lam_{L}\approx-\mu\,c_{22}^Lc_{22}^R\,.\label{app3}
\ee
Moreover,
\be
Y_{12}Y_{22}=-Y_{11}Y_{21}=\frac{M_L\widetilde{M}_L+\widetilde{m}_L^2}{m^2_{\tilde{\nu}_{\mu,2}}-m^2_{\tilde{\nu}_{\mu,1}}}.
\label{app4}
\ee

By combining Eqs.~(\ref{app1})-(\ref{app4}) one obtains Eqs.~(\ref{eq:VL5tot}) and (\ref{eq:VL5char}).

%%%%%%%%%%%%%%%%%%%%%%%%%%%%%%%%%%%%%%%%%
\section{Scan Range\label{app:scan}}
%%%%%%%%%%%%%%%%%%%%%%%%%%%%%%%%%%%%%%%%%%

\begin{table}[t]
\begin{center}
\begin{tabular}{|c|c|c|c|}
\hline
\rule{0pt}{1.25em}
\textbf{Parameter} &  \textbf{Description} & \textbf{Range} &  \textbf{Prior}\\[0.15em]
\hline
\hline
$\lam_5,\lam_{10}$ & Universal VL Yukawa coupling & $-\,0.5,\,0.5$ (\mfivem)& Linear \\
 &  & $-\,0.3,\,0.3$ (\mtenm)& Linear \\
\hline
$\epsilon$  & Yukawa hierarchy factor & $-\,0.5,\,0.5$ (\mfivem)&  Linear \\
 &  & $-\,0.25,\,0.25$ (\mtenm) &  Linear \\
\hline
$M_V$  &  Universal superpotential mass VL fields & $50,\,1500$  (\mfivem) &  Log \\
  &  & $100,\,1500$ (\mtenm) &  Log \\
\hline
$\widetilde{M}$  &  Universal superpotential mass mixing & $-\,20,\,20$ &  Linear \\
\hline
$\alpha$  &  Mass mixing hierarchy factor & $0.01,\,1$ &  Log \\
\hline
$Y_{10}$  & $10+\overline{10}$ Yukawa coupling & $-0.7,0.7$ &  Linear \\
\hline
\mzero\ &  Universal scalar mass & $100,\,4000$ (\mfivem)&   Log \\
 &  & $50,\,4000$ (\mtenm) &   Log \\
\hline
\mhalf &  Universal gaugino mass & $300,\,4000$ (\mfivem)&   Log \\
 &  & $1500,\,4000$ (\mtenm) &   Log \\
\hline
$\widetilde{m}^2$  &  Universal soft mass mixing & $-\,5\times 10^6,\,5\times 10^6$ (\mfivem)& Linear \\
 &  & $-\,2\times 10^6,\,2\times 10^6$ (\mtenm) & Linear \\
\hline
\azero\ &  Universal trilinear coupling & $-\,4000,\,4000$ &   Linear \\
\hline
$B_0$  &  Universal soft bilinear term VL fields & $-\,1500,\,1500$ &  Linear \\
\hline
\tanb\ &  Ratio of the Higgs vevs & $1,\,60$ & Linear\\
\hline
\signmu\ &  Sign of the Higgs mass parameter $\mu$ & $+1$ & \\
\hline
\hline
\multicolumn{4}{|l|}{\textbf{Nuisance parameters}}   \\
\hline
$m_t$  & Top quark pole mass  & $173.34\pm0.76$ & Gaussian \\
\hline
$m_b$  & Bottom quark mass ($\overline{\textrm{MS}}$)  & $4.18\pm0.03$ & Gaussian \\
\hline
\end{tabular}
\caption{Parameters of the models analyzed in this work. All soft SUSY-breaking masses are defined at the GUT scale.
Dimensionful quantities are given in GeV and $\gev^2$.}
\label{tab:models}
\end{center}
\end{table}%

We summarize our independent parameters and the scanned ranges and priors in \reftable{tab:models}.

%%%%%%%%%%%%%%%%%%%%%%%%%%%
\section{Collider constraints\label{app:LHC}}
%%%%%%%%%%%%%%%%%%%%%%%%%%%

We have confronted our models with the most recent LHC searches for VL matter, 
highlighting the parameter space that survives the collider bounds. 
Let us start by commenting on direct searches for VL leptons. 
The current bounds are approximately twice as strong as the 
LEP constraints\cite{Achard:2001qw}, excluding new leptons up 
to about $200\gev$ from multilepton searches\cite{Kumar:2015tna, Dermisek:2014qca}.
The most severe limits are derived under the assumption of 
a small mixing allowing decays of the new lepton to taus, e.g.,
$e_4^{\pm} \ra W^{\pm}\nu_\tau, Z\tau^{\pm},  h\tau^{\pm}$ (see\cite{Kumar:2015tna}) 
or to muons instead: $e_4^{\pm} \ra W^{\pm}\nu_\mu, Z\mu^{\pm},  h\mu^{\pm}$ 
(see\cite{Dermisek:2014qca}).\footnote{It may be noted that the ATLAS Collaboration 
	has excluded VL leptons with masses roughly up to 170\gev\ from direct 
	searches for VL leptons, decaying to a Z boson and an electron or a muon, 
	using LHC Run-I data\cite{Aad:2015dha}.}

On the other hand, in the case of models unified at the GUT scale the mass of colored VL particles is correlated to the VL lepton mass. 
As a consequence, the searches for VL quarks will lead to much stronger bounds that the direct lepton searches. 
Indeed, in the \mfivem\ model we have approximately the ratios
\begin{align}
m_{e_4}:m_{d_4}:M_V \simeq 1:1.65:0.74
\end{align}
where the subscript ``4'' indicates here a VL lepton or quark. For reference, we also include the ratio to the GUT-scale parameter $M_V$. 
Similarly we have
\begin{align}
m_{e_4}:m_{u_4}:m_{u_5}:m_{d_4}:M_V \simeq 1:2.7:3.5:3.5:0.8 \ ,
\end{align}
in the \mtenm\ model.

Let us focus on the \mfivem\ model, as this is the most promising regarding the \deltagmtwomu\ anomaly. 
For the points satisfying all the constraints summarized in \reftable{tab:exp_constraints} 
the dominant decay mode of a VL down-type quark
is to $W^{\pm} c$, with contributions from $Hs$ and $Zs$. Hence most relevant 
limits will come from searches of $d_4$ pair production with $W^{+}W^{-}q\bar q$ 
($q = u,c$) final states. Both ATLAS and CMS have looked for VL quark pair production, 
where the VL quark dominantly decays into a $W$ and a light quark jet. 
In the absence of any excess, upper bounds have been set on production 
cross section times branching ratios at 95\%~C.L.\cite{Aad:2015tba,CMS:2016pul,CMS:2014dka}. 
It may be noted that, with the assumption $\textrm{BR}(d_4 \ra W^{\pm}q)=1$, 
ATLAS has excluded new quarks below $690\gev$ at the 95\%~C.L.\cite{Aad:2015tba}, while CMS gives an even 
stronger bound, 845\gev\cite{CMS:2016pul,CMS:2014dka}.
As the branching ratio is often smaller than 100\% in the \mfivem\ model, a direct comparison of the quantity  
$\sigma \times  \textrm{BR}(d_4 \rightarrow W^{\pm}q)^2$ between model points 
and experimental upper limit is needed to derive the mass limits. 

\begin{figure}[t]
	\centering
	\subfloat[]{%
		%\label{fig:c}%
		\includegraphics[width=0.47\textwidth]{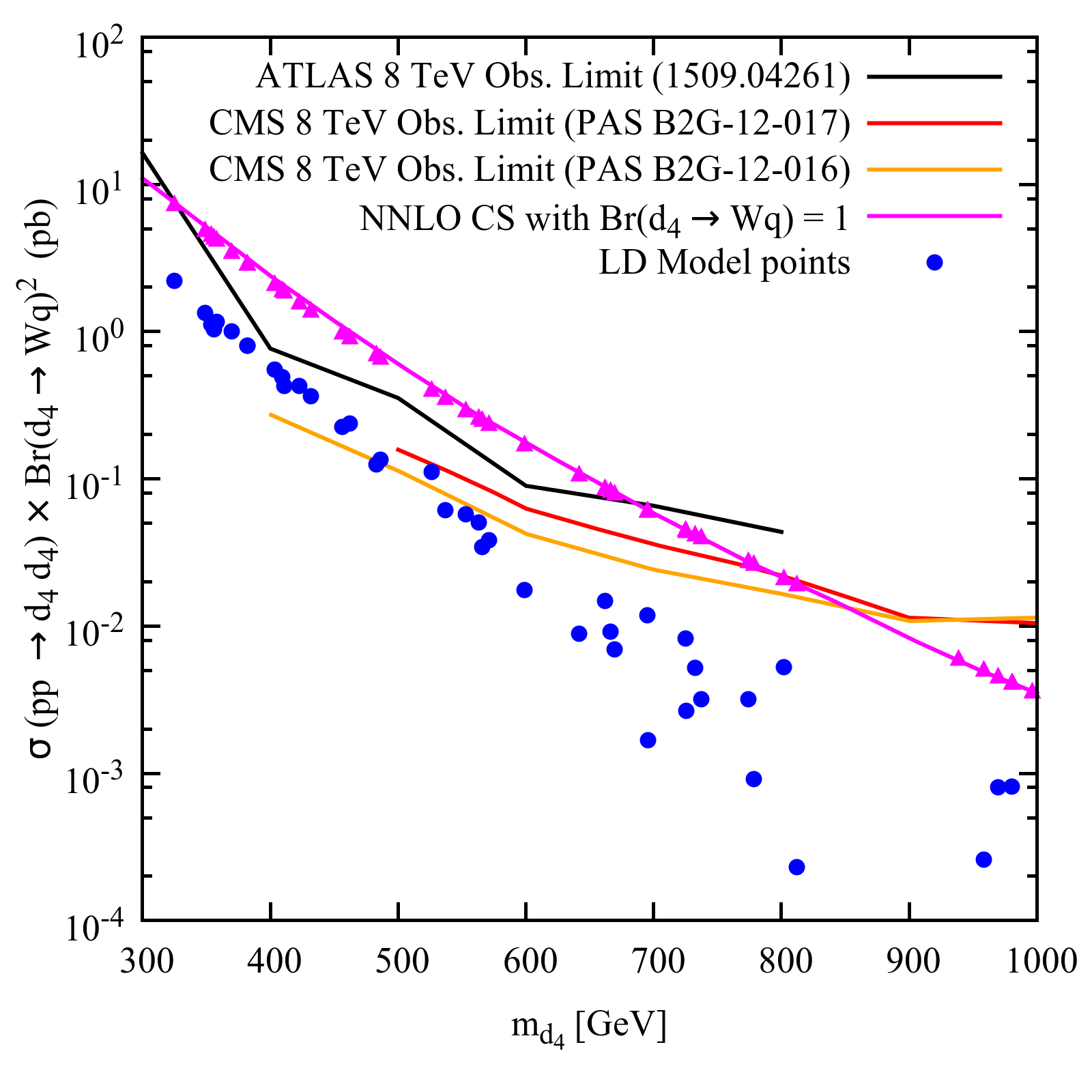}
	}%
	\hspace{0.02\textwidth}
	\subfloat[]{%
		%\label{fig:a}%
		\includegraphics[width=0.47\textwidth]{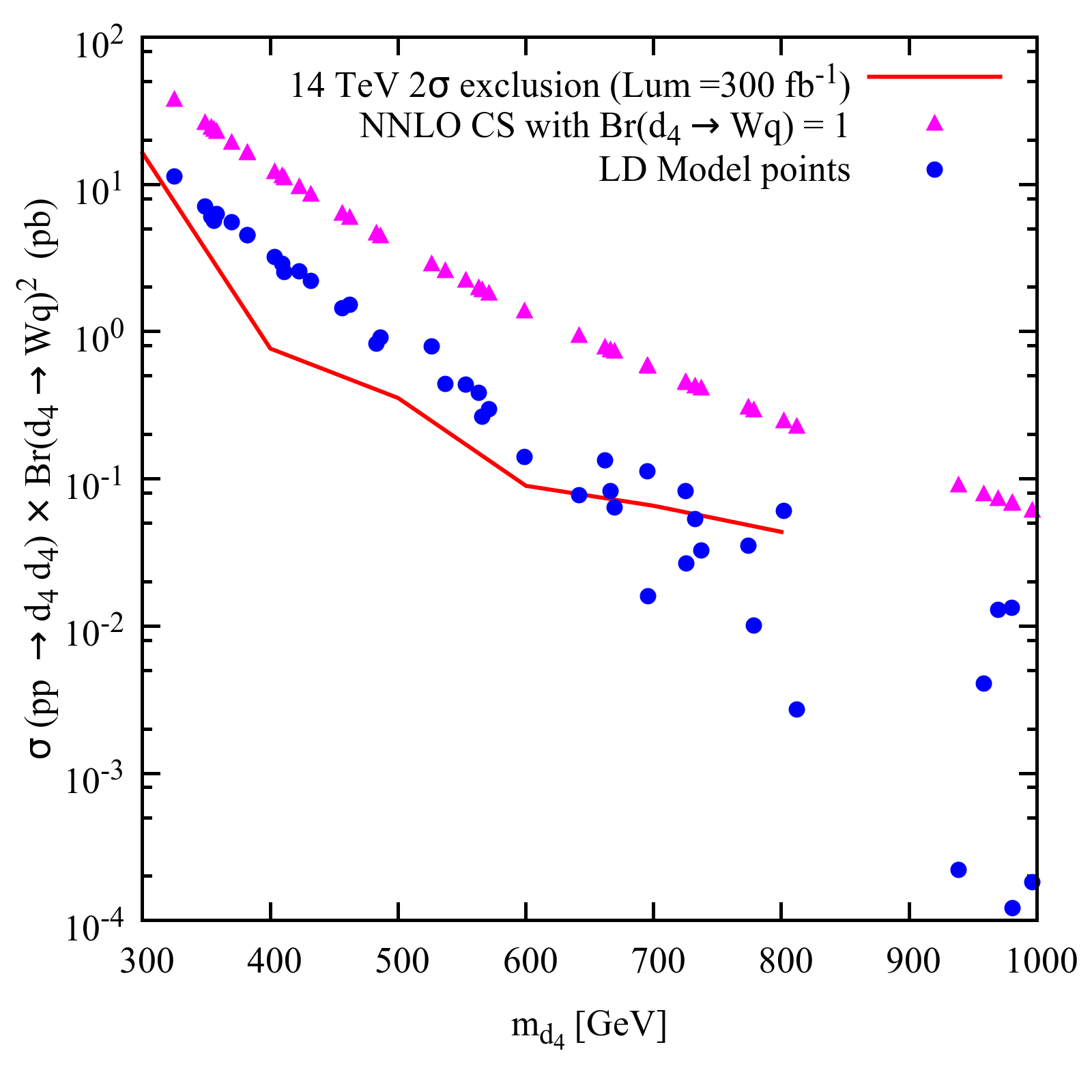}
	}%
	%\\
	%\subfloat[]{%
	%\label{fig:b}%
	%\includegraphics[width=0.47\textwidth]{Plots/HinoNeutrino.eps}
	%}%
	\caption{(a) A survey of collider constraints for the \mfivem\ model in the plane ($d_4$, $\sigma \times  \textrm{BR}(d_4 \rightarrow W^{\pm}q)^2$). Points are excluded above the solid yellow line. (b) Solid red line gives our projected exclusion bound at 14\tev, 300\invfb.}
	\label{fig:mssm5_decays2}
\end{figure}

In \reffig{fig:mssm5_decays2}(a), we show the 
quantity $\sigma \times  \textrm{BR}(d_4 \rightarrow W^{\pm}q)^2$
for a subset of points satisfying the constraints summarized in \reftable{tab:exp_constraints} and compare them with 
the upper bounds by both 
CMS \cite{CMS:2016pul,CMS:2014dka} and ATLAS \cite{Aad:2015tba} 
from LHC Run-I data. We calculate the $d_4$ pair-production cross section 
using \madgr\cite{Alwall:2014hca} with UFO model files generated by SARAH.
Branching ratios have been evaluated using the SARAH-produced SPheno code. 
To match with the NNLO cross sections provided by ATLAS and CMS 
in Ref\cite{Aad:2015tba,CMS:2016pul,CMS:2014dka} we have assumed 
an overall k-factor of 1.35. The black line 
in \reffig{fig:mssm5_decays2}(a) corresponds to the 
observed limit obtained by ATLAS\cite{Aad:2015tba} and the CMS limits 
from two similar but slightly different analyses\cite{CMS:2016pul,CMS:2014dka} 
are presented by 
red and yellow lines. The blue points represents the quantity 
$\sigma \times  \textrm{BR}(d_4 \rightarrow W^{\pm}q)^2$ for the \mfivem\  model 
and the magenta line is obtained for $\textrm{BR}(d_4 \ra W^{\pm}q)=1$. 

Overall, we see that the current LHC searches restrict the new 
quark $d_4$ to be above $550 \gev$. As we expected, this translates 
in our constrained models to a bound on the VL lepton pole mass of 
$\sim330\gev$, a bound almost two times more stringent than the one from direct searches. Note that, interestingly, although the mass range $300 - 400 \gev$ is not allowed 
	for $\textrm{BR}(d_4 \ra Wq)=1$, our points in these regions are not excluded by these searches due to the branching ratio 
	suppression (as $\textrm{BR}(d_4 \rightarrow W^{\pm}q)^2 \approx 0.25$ here).
	
%	this region seems to be still allowed (CMS has interpreted the results 
%	from 400 \gev).}

%%%%%%%%%%%%%%%%%%%%%%%%%%%%%%%%
%%%%%%%%%%%%%%%%%%%%%%%%%%%%%%%%

In \reffig{fig:mssm5_decays2}(b), we present the future projection 
limit for the 14\tev\ LHC with luminosity 300\invfb. 
To obtain an approximate future $2\sigma$ exclusion projection, 
we have used the ATLAS 8\tev\ results\cite{Aad:2015tba} 
with few simplifying assumptions.  We consider that the background 
events at 14\tev\ will be increased by a factor 2 compared to the 8\tev\ data  
for the same luminosity. From the ATLAS analysis\cite{Aad:2015tba}, 
we have evaluated the signal cut efficiencies for ${d_4}$ mass 
range 300 to 800\gev, and we assume here that these efficiencies  
remain the same for the 14\tev\ search. We assume a background systematic uncertainty $\delta B$
of about 30\% and, in the approximation of normally distributed statistics, we find the exclusion bound by applying the condition
\be
\frac{S}{\sqrt{B+(B\cdot\delta B)^2}}>2\,,
\ee 
where $B$ is the new number of background events and $S$ is the calculated signal.

We calculate the $d_4$ pair production cross section using MadGraph5 at 14\tev\ 
(multiplied with a k-factor of 1.35) and present the quantity $\sigma \times  Br(d_4 \rightarrow W^{\pm}q)^2$ 
in \reffig{fig:mssm5_decays2}(b). 
It appears that VL quark masses can be excluded up to around 700\gev\ with 14\tev\ LHC data and luminosity 
300\invfb, even with $\textrm{BR}(d_4 \ra W^{\pm}q)\approx 0.4 - 0.5$.

\bibliographystyle{JHEP}

\bibliography{VL}

\end{document}

%% file: macros_VL.tex
\newcommand{\newc}{\newcommand*}

\long\def\begincomment#1\endcomment{%
        \begingroup\sf\baselineskip12pt#1\endgroup}

\newc{\etal}{\textrm{et al.}} 
\newc{\eg}{\textrm{e.g.}} 
\newc{\ie}{\textrm{i.e.}}
\newc{\etc}{\textrm{etc}}
\newc\vs{\textrm{vs.}}
\newc{\cl}{\rm {C.L.}}
\newc{\ev}{\ensuremath{\,\mathrm{eV}}}
\newc{\kev}{\ensuremath{\,\mathrm{keV}}}
\newc{\mev}{\ensuremath{\,\mathrm{MeV}}}
\newc{\gev}{\ensuremath{\,\mathrm{GeV}}}
\newc{\tev}{\ensuremath{\,\mathrm{TeV}}}
\newc{\MeV}{\mev} 
\newc{\TeV}{\tev}
\newc{\invpb}{\ensuremath{/\text{pb}}}
\newc{\invfb}{\ensuremath{\,\text{fb}^{-1}}}
\newc\nb{\ensuremath{\,\mathrm{nb}}} \newc\pb{\ensuremath{\,\mathrm{pb}}} \newc\fb{\ensuremath{\,\mathrm{fb}}}
\newc\pc{\ensuremath{\,\mathrm{pc}}}
\newc\kpc{\ensuremath{\,\mathrm{kpc}}}
\newc\mpc{\ensuremath{\,\mathrm{Mpc}}}
\newc\ps{\ensuremath{\,\mathrm{ps}}} 
\newc\cmeter{\ensuremath{\,\mathrm{cm}}} 
\newc\meter{\ensuremath{\,\mathrm{m}}} 
\newc\kmeter{\ensuremath{\,\mathrm{km}}}
\newc\second{\ensuremath{\,\mathrm{s}}}
\newc\msecond{\ensuremath{\,\mathrm{ms}}}
\newc\nsecond{\ensuremath{\,\mathrm{ns}}}
\newc\psecond{\ensuremath{\,\mathrm{ps}}}
\newc{\chisqmin}{\ensuremath{\chi^2_{\mathrm{min}}}}
\newc{\Delchisq}{\ensuremath{\Delta\chi^2}}
\newc{\chisq}{\ensuremath{\chi^2}}
\newc{\like}{\ensuremath{\mathcal{L}}}
\newc\lsim{\ensuremath{\mathrel{\rlap{\lower4pt\hbox{\hskip1pt$\sim$}}\raise1pt\hbox{$<$}}}}
\newc\gsim{\ensuremath{\mathrel{\rlap{\lower4pt\hbox{\hskip1pt$\sim$}}\raise1pt\hbox{$>$}}}}
\newc{\VEV}[1]{\ensuremath{\langle #1 \rangle}}
\newc{\dl}{\ensuremath{\stackrel{\leftarrow}{D}}}
\newc{\dr}{\ensuremath{\stackrel{\rightarrow}{D}}}
\newc{\scr}[1]{\ensuremath{\mathcal{#1}}}

\newc{\bcenter}{\begin{center}}    \newc{\ecenter}{\end{center}}
\newc{\bfl}{\begin{flushleft}}    \newc{\efl}{\end{flushleft}}
\newc{\bfr}{\begin{flushright}}    \newc{\efr}{\end{flushright}}

\newc{\bi}{\begin{itemize}}
\newc{\ei}{\end{itemize}}
\newc{\bed}{\begin{description}}
\newc{\eed}{\end{description}}
\newc{\ben}{\begin{enumerate}}
\newc{\een}{\end{enumerate}}

\newc{\be}{\begin{equation}}
\newc{\ee}{\end{equation}}
\newc{\bea}{\begin{eqnarray}}
\newc{\eea}{\end{eqnarray}}
\newc{\ra}{\rightarrow}
\newc{\alphas}{\ensuremath{\alpha_s}}
\newc{\alphatwo}{\ensuremath{\alpha_2}}
\newc{\alphaone}{\ensuremath{\alpha_1}}
\newc{\alphai}[1]{\ensuremath{\alpha_{#1}}}
\newc{\alphaem}{\ensuremath{\alpha_{\mathrm{em}}}}
\newc{\alphaeff}{\ensuremath{\alpha_{\mathrm{eff}}}}
\newc{\sineff}{\ensuremath{\sin \theta_{\mathrm{eff}}}}
\newc{\sinsqeff}{\ensuremath{\sin^2 \theta_{\mathrm{eff}}}}
\newc{\dalphahad}{\ensuremath{\Delta \alpha_{\mathrm{had}}}}
\newc{\yt}{\ensuremath{h_t}} \newc{\yb}{\ensuremath{h_b}} \newc{\ytau}{\ensuremath{h_{\tau}}}
\newc\mz{\ensuremath{M_Z}} 
\newc\mw{\ensuremath{m_W}}
\newc\mZ{\mz}        \newc\mW{\mw}
\newc\mhsm{\ensuremath{ m_{H_{\mathrm{SM}}}}}
\newc{\mtop}{\ensuremath{ m_t}}               \newc{\mtpole}{\ensuremath{ M_t}}
\newc{\mbottom}{\ensuremath{ m_b}} 
\newc{\mtau}{\ensuremath{ m_{\tau}}}
\newc{\mt}{\mtpole}
\newc{\mb}{\mbottom} 
\newc{\rgg}{\ensuremath{R_{h}(\gamma\gamma)}}
\newc{\rzz}{\ensuremath{R_{h}(ZZ)}}
\newc{\rtwogg}{\ensuremath{R_{h_2}(\gamma\gamma)}}
\newc{\rtwozz}{\ensuremath{R_{h_2}(ZZ)}}
\newc{\ronegg}{\ensuremath{R_{h_1}(\gamma\gamma)}}
\newc{\ronezz}{\ensuremath{R_{h_1}(ZZ)}}
\newc{\rsiggg}{\ensuremath{R_{h_\textrm{sig}}(\gamma\gamma)}}
\newc{\rsigzz}{\ensuremath{R_{h_\textrm{sig}}(ZZ)}}
\newc{\llbar}{\ensuremath{\ell\bar{\ell}}}
\newc{\tauptaum}{\ensuremath{ \tau^+\tau^-}}
\newc{\qqbar}{\ensuremath{ q\bar{q}}} \newc{\ppbar}{\ensuremath{ p\bar{p}}}
\newc{\bbbar}{\ensuremath{ b\bar{b}}} \newc{\ttbar}{\ensuremath{ t\bar{t}}}
\newc{\ffbar}{\ensuremath{ f\bar{f}}} \newc{\tautaubar}{\ensuremath{ \tau\bar{\tau}}}
\newc{\mchi}{\ensuremath{m_{\chi}}}
\newc{\squark}{\ensuremath{\tilde{q}}}
\newc{\slepton}{\ensuremath{\tilde{l}}}
\newc{\gluino}{\ensuremath{\tilde{g}}} 
\newc{\mgluino}{\ensuremath{{m_{\gluino}}}}
\newc{\tone}{\ensuremath{{\tilde{t}_1}}}
\newc{\sthw}{\ensuremath{ \sin\theta_W}}              \newc{\cthw}{\ensuremath{\cos\theta_W}}
\newc{\tanthw}{\ensuremath{ \tan\theta_W}}              \newc{\cotthw}{\ensuremath{\cot\theta_W}}
\newc{\ssqthw}{\ensuremath{\sin^2 \theta_W}}
\newc{\msbar}{\ensuremath{\overline{MS}}} \newc{\drbar}{\ensuremath{\overline{DR}}}
\newc{\mtmtsmmsbar}{\ensuremath{ m_t(m_t)^{\msbar}_{{\mathrm{SM}}}}}
\newc{\mtmtsmdrbar}{\ensuremath{ m_t(m_t)^{\drbar}_{{\mathrm{SM}}}}}
\newc{\mtmtmssmdrbar}{\ensuremath{ m_t(m_t)^{\drbar}_{{\mathrm{SUSY}}}}}
\newc{\mbmbmsbar}{\ensuremath{ m_b(m_b)^{\msbar} }}
\newc{\mbmbsmmsbar}{\ensuremath{ m_b(m_b)^{\msbar}_{{\mathrm{SM}}}}}
\newc{\mbmzsmmsbar}{\ensuremath{ m_b(\mz)^{\msbar}_{{\mathrm{SM}}}}}
\newc{\mbmzsmdrbar}{\ensuremath{ m_b(\mz)^{\drbar}_{{\mathrm{SM}}}}}
\newc{\mbmzmssmdrbar}{\ensuremath{ m_b(\mz)^{\drbar}_{{\mathrm{SUSY}}}}}
\newc{\mtaumzsmmsbar}{\ensuremath{ m_{\tau}(\mz)^{\msbar}_{{\mathrm{SM}}}}}
\newc{\mtaumzsmdrbar}{\ensuremath{ m_{\tau}(\mz)^{\drbar}_{{\mathrm{SM}}}}}
\newc{\mtaumzmssmdrbar}{\ensuremath{ m_{\tau}(\mz)^{\drbar}_{{\mathrm{SUSY}}}}}
\newc{\alphasmzms}{\ensuremath{\alpha_s(M_Z)^{\overline{MS}}}}
\newc{\alphaimzms}[1]{\ensuremath{\alpha_{#1}(M_Z)^{\overline{MS}}}}
\newc{\alphaemmz}{\ensuremath{\alpha_{\mathrm{em}}(M_Z)^{\overline{MS}}}}
\newc{\mzero}{\ensuremath{{m_0}}}
\newc{\mhalf}{\ensuremath{ m_{1/2}}}
\newc{\tanb}{\ensuremath{\tan\beta}}
\newc{\azero}{\ensuremath{ A_0}}
\newc{\bzero}{\ensuremath{ B_0}}
\newc{\signmu}{\ensuremath{\rm{sgn}\,\mu}}
\newc{\mueff}{\ensuremath{\mu_{\rm{eff}}}}
\newc{\lam}{\ensuremath{{\lambda}}}
\newc{\kap}{\ensuremath{{\kappa}}}
\newc{\alam}{\ensuremath{{A_{\lambda}}}}
\newc{\akap}{\ensuremath{{A_{\kappa}}}}
\newc{\hs}{\ensuremath{ H_s}}      
\newc{\mhs}{\ensuremath{ m_{H_s}}} 
\newc{\mgut}{\ensuremath{ M_{\rm GUT}}}
\newc{\mplanck}{\ensuremath{ M_{\rm P}}}      \newc{\mpl}{\ensuremath{ M_{\rm Pl}}}
\newc{\msusy}{\ensuremath{ M_{\rm SUSY}}}      \newc{\ms}{\ensuremath{ M_{\rm S}}}
 \newc{\mhl}{\ensuremath{m_\hl}} 
 \newc{\mhone}{\ensuremath{m_{h_1}}} 
 \newc{\mhtwo}{\ensuremath{m_{h_2}}} 
 \newc{\mglu}{\ensuremath{m_{\tilde g}}} 
 \newc{\mul}{\ensuremath{m_{\tilde{u}_L}}} 
 \newc{\mtone}{\ensuremath{m_{\tilde{t}_1}}} 
 \newc{\ma}{\ensuremath{m_A}} 
 \newc{\maone}{\ensuremath{m_{a_1}}} 
 \newc{\matwo}{\ensuremath{m_{a_2}}}
 \newc{\hone}{\ensuremath{h_1}}
 \newc{\htwo}{\ensuremath{h_2}}
 \newc{\aone}{\ensuremath{a_1}}
 \newc{\atwo}{\ensuremath{a_2}}
 \newc{\mhu}{\ensuremath{ m_{H_u}}}       
 \newc{\mhd}{\ensuremath{ m_{H_d}}}
 \newc{\mhusq}{\ensuremath{ m_{H_u}^2}}       
 \newc{\mhdsq}{\ensuremath{ m_{H_d}^2}}
 \newc{\mhuew}{\ensuremath{ m^{\ast}_{H_u}}}       
 \newc{\mhdew}{\ensuremath{ m^{\ast}_{H_d}}}
 \newc{\mhuewsq}{\ensuremath{ m^{\ast\, 2}_{H_u}}}       
 \newc{\mhdewsq}{\ensuremath{ m^{\ast\, 2}_{H_d}}}
 \newc{\hu}{\ensuremath{ H_u}}       
 \newc{\hd}{\ensuremath{ H_d}}
 \newc{\barmhu}{\ensuremath{ \bar{m}_{H_u}}}
 \newc{\barmhd}{\ensuremath{ \bar{m}_{H_d}}}

 \newc{\mqthree}{\ensuremath{m_{\widetilde{Q}_3}^2}}
 \newc{\muthree}{\ensuremath{m_{\tilde{u}_3}^2}}
 \newc{\mdthree}{\ensuremath{m_{\tilde{d}_3}^2}}
 \newc{\mlthree}{\ensuremath{m_{\widetilde{L}_3}^2}}
 \newc{\methree}{\ensuremath{m_{\tilde{e}_3}^2}}
 \newc{\mqtwo}{\ensuremath{m_{\widetilde{Q}_2}^2}}
 \newc{\mutwo}{\ensuremath{m_{\tilde{u}_2}^2}}
 \newc{\mdtwo}{\ensuremath{m_{\tilde{d}_2}^2}}
 \newc{\mltwo}{\ensuremath{m_{\widetilde{L}_2}^2}}
 \newc{\metwo}{\ensuremath{m_{\tilde{e}_2}^2}}
 \newc{\mqone}{\ensuremath{m_{\widetilde{Q}_1}^2}}
 \newc{\muone}{\ensuremath{m_{\tilde{u}_1}^2}}
 \newc{\mdone}{\ensuremath{m_{\tilde{d}_1}^2}}
 \newc{\mlone}{\ensuremath{m_{\widetilde{L}_1}^2}}
 \newc{\meone}{\ensuremath{m_{\tilde{e}_1}^2}}
 \newc{\msmul}{\ensuremath{m_{\tilde{\mu}_L}}}
 \newc{\msmur}{\ensuremath{m_{\tilde{\mu}_R}}}
 \newc{\msneumu}{\ensuremath{m_{\tilde{\nu}_{\mu}}}}
 \newc{\mone}{\ensuremath{M_1}}
 \newc{\monesq}{\ensuremath{M_1^2}}
 \newc{\mtwo}{\ensuremath{M_2}}
 \newc{\mtwosq}{\ensuremath{M_2^2}}
 \newc{\mthree}{\ensuremath{M_3}}
 \newc{\mthreesq}{\ensuremath{M_3^2}}
 \newc{\atau}{\ensuremath{{A_{\tau}}}}
 \newc{\at}{\ensuremath{{A_{t}}}}
 \newc{\ab}{\ensuremath{{A_{b}}}}
 \newc{\atausq}{\ensuremath{{A_{\tau}^2}}}
 \newc{\atsq}{\ensuremath{{A_{t}^2}}}
 \newc{\absq}{\ensuremath{{A_{b}^2}}}
 \newc{\dmzero}{\ensuremath{\Delta{_{m_0}}}}
 \newc{\dmhalf}{\ensuremath{\Delta{_{m_{1/2}}}}}
 \newc{\dmu}{\ensuremath{\Delta{_{\mu}}}}

 \newc{\pten}{\ensuremath{\psi_{10}}}
 \newc{\ffive}{\ensuremath{\phi_{5}}}
 \newc{\hfive}{\ensuremath{h_{5}}}
 \newc{\hbfive}{\ensuremath{h_{\bar{5}}}}
 \newc{\thet}{\ensuremath{\theta_{50}}}
 \newc{\thetb}{\ensuremath{\theta_{\,\overline{50}}}}
 \newc{\ptenhat}{\ensuremath{\hat{\psi}_{10}}}
 \newc{\ffivehat}{\ensuremath{\hat{\phi}_{5}}}
 \newc{\hfivehat}{\ensuremath{\hat{h}_{5}}}
 \newc{\hbfivehat}{\ensuremath{\hat{h}_{\bar{5}}}}
 \newc{\thethat}{\ensuremath{\hat{\theta}_{50}}}
 \newc{\thetbhat}{\ensuremath{\hat{\theta}_{\,\overline{50}}}}
 \newc{\si}{\ensuremath{\Sigma}}
 \newc{\mfive}{\ensuremath{m_5^2}}
 \newc{\mten}{\ensuremath{m_{10}^2}}
 \newc{\dfive}{\ensuremath{\Delta^2_5}}
 \newc{\dbfive}{\ensuremath{\Delta^2_{\bar{5}}}}
 \newc{\dfifty}{\ensuremath{\Delta^2_{50}}}
 \newc{\dfiftyb}{\ensuremath{\Delta^2_{\,\overline{50}}}}
 \newc{\msi}{\ensuremath{m_{\Sigma}^2}}
 \newc{\lamh}{\ensuremath{\lambda_{H}}}
 \newc{\lamhb}{\ensuremath{\lambda_{\bar{H}}}}
 \newc{\ah}{\ensuremath{A_{H}}}
 \newc{\ahb}{\ensuremath{A_{\bar{H}}}}
 \newc{\lams}{\ensuremath{\lambda_{S}}}
 \newc{\as}{\ensuremath{A_{S}}}
 \newc{\lamsig}{\ensuremath{\lambda_{\si}}}
 \newc{\asig}{\ensuremath{A_{\si}}}

 \newc{\msten}{\ensuremath{m_{16}^2}}
 \newc{\mhun}{\ensuremath{m_{126}^2}}
 \newc{\mhunb}{\ensuremath{m_{\bar{126}}^2}}
 \newc{\mthun}{\ensuremath{m_{210}^2}}
 \newc{\ahun}{\ensuremath{A_{\bar{126}}}}
 \newc{\yhun}{\ensuremath{Y_{\bar{126}}}}
 \newc{\aten}{\ensuremath{A_{10}}}
 \newc{\yten}{\ensuremath{Y_{10}}}
 \newc{\alone}{\ensuremath{A_{\lambda_1}}}
 \newc{\altwo}{\ensuremath{A_{\lambda_2}}}
 \newc{\althree}{\ensuremath{A_{\lambda_3}}}
 \newc{\althreeb}{\ensuremath{A_{\bar{\lambda_3}}}}
 \newc{\lone}{\ensuremath{\lambda_1}}
 \newc{\ltwo}{\ensuremath{\lambda_2}}
 \newc{\lthree}{\ensuremath{\lambda_3}}
 \newc{\lthreeb}{\ensuremath{\bar{\lambda_3}}}

\newc{\sigsip}{\ensuremath{\sigma^{\rm SI}_{p}}}	\newc{\sigsin}{\ensuremath{\sigma^{\rm SI}_{n}}}
\newc{\sigsdp}{\ensuremath{\sigma^{\rm SD}_{p}}}	\newc{\sigsdn}{\ensuremath{\sigma^{\rm SD}_{n}}}
\newc{\sigsi}{\ensuremath{\sigma^{\rm SI}}}	\newc{\sigsd}{\ensuremath{\sigma^{\rm SD}}}
\newc{\sigv}{\ensuremath{\sigma v}}
\newc{\abund}{\ensuremath{ \Omega h^2}}
\newc{\omegadm}{\ensuremath{ \Omega_{{\rm DM}}}}     \newc{\abunddm}{\ensuremath{ \Omega_{{\rm DM}} h^2}} 
\newc{\omegam}{\ensuremath{ \Omega_{{\rm m}}}}       \newc{\abundm}{\ensuremath{ \Omega_{{\rm m}} h^2}}
\newc{\omegab}{\ensuremath{ \Omega_{{\rm b}}}}	\newc{\abundb}{\ensuremath{ \Omega_{{\rm b}} h^2}}
\newc{\omegatot}{\ensuremath{ \Omega_{{\rm TOT}}}}
\newc{\omegacdm}{\ensuremath{ \Omega_{{\rm CDM}}}}   \newc{\abundcdm}{\ensuremath{ \Omega_{{\rm CDM}} h^2}}
\newc{\omegalambda}{\ensuremath{ \Omega_{\Lambda}}} \newc{\abundlambda}{\ensuremath{ \Omega_{\Lambda} h^2}}
\newc{\omegarad}{\ensuremath{ \Omega_{{\rm rad}}}}  \newc{\abundrad}{\ensuremath{ \Omega_{{\rm rad}} h^2}}
\newc{\rhocrit}{\ensuremath{ \rho_{\rm crit}}}
\newc{\rhochi}{\ensuremath{ \rho_{\chi}}}
\newc{\abunchi}{\ensuremath{\Omega_\chi h^2}}
\newc{\abundlsp}{\ensuremath{\Omega_{\rm LSP}h^2}}
\newcommand*{\abundchi}{\ensuremath{\Omega_\chi h^2}}% For multiple citations with one key
\newc{\amu}{\ensuremath{ a_{\mu}}}        \newc{\amususy}{\ensuremath{ a_{\mu}^{\mathrm{SUSY}}}}
\newc{\amuexpt}{\ensuremath{ a_{\mu}^{\mathrm{expt}}}}        \newc{\amusm}{\ensuremath{ a_{\mu}^{\mathrm{SM}}}}
\newc\deltaamu{\ensuremath{\Delta a_{\mu}}} \newc{\deltaamususy}{\ensuremath{\delta a_{\mu}^{\mathrm{SUSY}}}}
\newc\gmtwo{\ensuremath{ (g-2)_{\mu}}} 
\newc{\deltagmtwomususy}{\ensuremath{\delta\left(g-2\right)_{\mu}^{\mathrm{SUSY}}}}
\newc{\deltagmtwomu}{\ensuremath{\delta\left(g-2\right)_{\mu}}}
\newc\BR{\ensuremath{\textrm{BR}}}
\newc\bsgamma{\ensuremath{ b\rightarrow s \gamma }}
\newc\bxsgamma{\ensuremath{\overline{B}\rightarrow X_{s}\gamma}}
\newc\brbsgamma{\ensuremath{\BR\left(\bsgamma\right)}}
\newc\brbxsgamma{\ensuremath{\BR\left(\bxsgamma\right)}}
\newc\bsmumu{\ensuremath{B_s\to\mu^+\mu^-}}
\newc\brbsmumu{\ensuremath{\BR\left(B_s\to\mu^+\mu^-\right)}}
\newc\bdmmumu{\ensuremath{\overline{B}_d\to\mu^+\mu^-}}
\newc\bbbarmix{\ensuremath{\overline{B}_s\mbox{-}B_s}}      % B_s mixing
\newc\delmbs{\ensuremath{\Delta M_{B_s}}}
\newc{\butaunu}{\ensuremath{B_u \rightarrow \tau \nu}}
\newc{\brbutaunu}{\ensuremath{\BR\left(B_u \rightarrow \tau \nu\right)}}

\newc{\brmuegamma}{\ensuremath{\BR\left(\mu^{\pm}\rightarrow e^{\pm}\gamma\right)}}
\newc{\brtauegamma}{\ensuremath{\BR\left(\tau^{\pm}\rightarrow e^{\pm}\gamma\right)}}
\newc{\brtaumugamma}{\ensuremath{\BR\left(\tau^{\pm}\rightarrow \mu^{\pm}\gamma\right)}}
\newc{\brmuthreee}{\ensuremath{\BR\left(\mu^{\pm}\rightarrow e^{\pm}e^+e^-\right)}}
\newc{\brtauthreee}{\ensuremath{\BR\left(\tau^{\pm}\rightarrow e^{\pm}e^+e^-\right)}}
\newc{\brtauthreemu}{\ensuremath{\BR\left(\tau^{\pm}\rightarrow \mu^{\pm}\mu^+\mu^-\right)}}

\newcommand*{\reftable}[1]{Table~\ref{#1}}         
       
\newcommand*{\reffig}[1]{Fig.~\ref{#1}}
 
        \newcommand*{\refeq}[1]{Eq.~(\ref{#1})}
     \newcommand*{\refsec}[1]{Sec.~\ref{#1}}

\newcommand*{\neuttwo}{\ensuremath{\tilde{{\chi}}^0_2}}

\newcommand*{\charone}{\ensuremath{\chi^{\pm}_1}}
\newcommand*{\chartwo}{\ensuremath{\chi^{\pm}_2}}

\newcommand*{\glu}{\ensuremath{\tilde{g}}}

\newcommand*{\SARAH}{SARAH}

\let\oldcite\cite
\renewcommand*{\cite}{~\oldcite}

\newcommand*{\hl}{\ensuremath{h}}

\newcommand*{\mh}{\ensuremath{m_h}}

\newcommand*{\mtenm}{\textbf{QUE}}
\newcommand*{\mfivem}{\textbf{LD}}

\newc{\glzmu}{\ensuremath{{g^{Z\mu \mu}_{L}}}}
\newc{\grzmu}{\ensuremath{{g^{Z\mu \mu}_{R}}}}
\newc{\glwmu}{\ensuremath{{g^{W\mu \nu_\mu}_{L}}}}
\newc{\grwmu}{\ensuremath{{g^{W\mu \nu_\mu}_{R}}}}
\newc{\glzmuSM}{\ensuremath{{g^{Z\mu \mu}_{L,\textrm{SM}}}}}
\newc{\grzmuSM}{\ensuremath{{g^{Z\mu \mu}_{R,\textrm{SM}}}}}
\newc{\glwmuSM}{\ensuremath{{g^{W\mu \nu_\mu}_{L,\textrm{SM}}}}}
\newc{\grwmuSM}{\ensuremath{{g^{W\mu \nu_\mu}_{R,\textrm{SM}}}}}

\newcommand*{\madgr}{\texttt{MadGraph5\_aMC@NLO}}